\documentclass[fleqn,usenatbib]{mnras}

\usepackage{newtxtext,newtxmath}

\usepackage[T1]{fontenc}
\usepackage{ae,aecompl}

\usepackage[caption=false]{subfig}

\usepackage{graphicx}	
\usepackage{amsmath}	
\usepackage{amssymb}	
\usepackage{lscape}

\title[X-ray Irradiation in Black Hole X-ray Binaries]{Thermally-Driven Disc Winds as a Mechanism for X-ray Irradiation Heating in Black Hole X-ray Binaries: The Case Study of GX339$-$4}

\author[B.E. Tetarenko et al.]{
B.E. Tetarenko,$^{1}$\thanks{E-mail: btetaren@umich.edu}
G. Dubus$^{2}$,
G. Marcel$^{3}$,
C. Done$^{4}$,
M. Clavel$^{2}$ 
\\
$^{1}$Department of Astronomy, University of Michigan, 1085 South University Avenue, Ann Arbor, MI 48109, USA\\
$^{2}$Univ. Grenoble Alpes, CNRS, IPAG, 38000, Grenoble, France\\
$^{3}$Villanova University, Department of Physics, Villanova, PA 19085, USA\\
$^{4}$Department of Physics, University of Durham, South Road, Durham, DH1 3LE, UK \\
}

\date{Accepted XXX. Received YYY; in original form ZZZ}

\pubyear{2020}

\begin{document}
\label{firstpage}
\pagerange{\pageref{firstpage}--\pageref{lastpage}}
\maketitle

\begin{abstract}
X-ray irradiation heating of accretion discs in black hole X-ray binaries (BHXBs) plays a key role in regulating their outburst cycles. However, despite decades of theoretical and observational efforts, the physical mechanism(s) responsible for irradiating these discs remains largely unknown. 
We have built an observationally-based  methodology to estimate the strength of irradiation of BHXB discs by combining multiwavelength X-ray and optical/IR data throughout transient outbursts. We apply this to 
 $\sim15$ yrs of activity in the Galactic BHXB GX339$-$4. 
Our findings suggest that the irradiation heating
required by the optical data is large in this system. Direct illumination of the outer disc
does not produce sufficient irradiation, but this
should also produce a thermal-radiative wind
which adds to the irradiation heating by scattering flux down onto the disc. 
However, analytic estimates of X-ray illumination from scattering in the wind is still not sufficient to produce the observed heating, even in combination with direct illumination. Either the analytic thermal-radiative wind models are underestimating the effect of the wind, or there are additional scattering mechanisms at work, such as magnetically-driven outflows, acting to increase the optical/IR flux.
While wind-driven irradiation is likely a common feature among long-period BHXBs, fully understanding the driving mechanism(s) behind such a wind will require radiation-hydrodynamic simulations.

\end{abstract}




\section{Introduction}\label{sec:intro}

Illumination of the outer accretion disc, by X-rays produced in the inner accretion flow, plays a key role in regulating the outburst cycles of X-ray binary systems \citep{vanparadijs1994,vanparadijs1996}. This X-ray irradiation determines the temperature over most of the accretion disc during outburst
and thus, is a major contributor to the thermal balance in the accretion flow of these binary systems. It controls the outburst decay towards quiescence (and thus the overall outburst duration) and sets the limit on the amount of mass able to be accreted during outburst (thus impacting the overall outburst recurrence timescales) \citep{king1998,dubus2001}. As a result, the light curves of X-ray binary outbursts display characteristic profile shapes, that encode within them distinct observational signatures of the X-ray irradiation source heating the disc in the system \citep{king1998,kim1999,dubus2001,tetarenko2018b}.

Among X-ray binary systems, black hole low-mass X-ray binaries (BH-LMXBs), offer ideal laboratories in which to understand the mechanism behind the irradiation heating of X-ray binary accretion discs. They undergo bright X-ray ($L_{\rm X,peak}\sim10^{36}-10^{39} \, \rm{erg/s}$; \citealt{chen1997,tetarenko2016}) and optical \citep{vanparadijs1996} outbursts, indicative of episodes of rapid mass-transfer from 
a low-mass ($<1M_{\odot}$) companion star onto a stellar-mass ($5-30 M_{\odot}$) black hole, that recur frequently on observable month to year timescales \citep{mcclintock2006,tetarenko2016}. Moreover, the majority of the optical/infrared (OIR) light emitted by the accretion discs in BH-LMXBs comes from reprocessed X-rays \citep{vanparadijs1994,vanparadijs1996}, making the OIR regime the only direct probe of the X-ray irradiating flux we have.

The mechanism behind the bright (X-ray and optical) outbursts observed in BH-LMXBs can be understood using the disc-instability model (DIM; \citealt{osaki1974,meyer1981,smak1983,smak1984,cannizzo1985,mineshige1989,cannizzo1993}), with the addition of irradiation and evaporation (``truncation'') of the inner thin disc to a radiatively-inefficient flow (DIM+irradiation; \citealt{dubus2001}). The DIM+irradiation predicts that the outburst light curve of a BH-LMXB will show an exponential+linear shaped decay profile. With the exponential-shaped decay, attributed to a viscously accreting irradiated disc, transitioning to a linear-shaped decay when the temperature drops low enough in the outer disc, resulting in the formation and propagation of a cooling front inward at a speed controlled purely by the decaying X-ray irradiating flux.


Despite being the subject of extensive theoretical and observational work for decades, how X-ray binary discs are irradiated is not well understood. Both the mechanism by which the discs are heated, and the fraction of the X-ray flux that is intercepted and reprocessed in the outer disc (hereafter referred to as $\cal C$), remain open questions in the field. The reason for this stems from the fact that main factors determining the intercepted fraction remain largely unknown. Such factors include: geometry of the disc and irradiating source, X-ray albedo of the disc, and effect the illuminating spectrum has on thermal properties of the disc itself. Moreover,  whether such quantities vary as functions of time and/or disc radius, has not yet been studied extensively.

Note that, throughout this work, $\cal C$ is defined using the formulation in \citealt{dubus1999}. Here the irradiation temperature is defined as a function of $\dot{M}$ (see Equation \ref{eq:temp_irr_eq}). In the literature, irradiation temperature may also be defined in terms of luminosity (e.g., \citealt{dubus2001}). To directly compare values of $\cal C$ derived via the luminosity formulation, to the $\cal C$ computed in this paper, one must multiply by an additional factor of accretion efficiency.

Recently, \cite{tetarenko2018b} analyzed the X-ray lightcurves for a large sample of BH-LMXB outbursts. They derived estimates for $\cal C$ by comparing the observed X-ray lightcurve profiles to the predictions of the DIM+irradiation, assuming a source of X-ray irradiation proportional to the central mass accretion rate ($\dot{M}_{\rm in}$) throughout outburst. These authors are able to show that an initial exponential-shaped decline after the outburst peak is a robust feature of a fully irradiated disc accreting on a viscous timescale. However, they also find that the predictions of the DIM+irradiation do not adequately describe the later stages of BH-LMXB outburst light curves. 

As a result, they derive values of $\cal C$ from the X-ray lightcurves significantly in excess of $\sim5\times10^{-3}$. This is the typical value assumed in theoretical work \citep{vrtilek1990,dejong1996}. This value is also consistent with the amount of X-ray heating required to stabilize persistent X-ray binary systems against the thermal-viscous instability \citep{vanparadijs1996,coriat2012,tetarenko2016}. \cite{tetarenko2018b} postulate that this suggests that BH-LMXB X-ray light curve profiles, beyond the initial exponential decay, are shaped by a variety of physical mechanisms, for which irradiation is only one of them. Examples of such mechanisms include: mass loss through either inner disc evaporation to a radiatively-inefficient structure or mass loss from an accretion disc wind.

X-ray light-curves alone may be insufficent to understand how the X-ray irradiating source heats the discs through the course of BH-LMXB outbursts \citep{tetarenko2018b}. However, the use of simultaneous multi-wavelength data sets does provide a promising alternative approach. By modelling the irradiated discs in BH-LMXBs, assuming a constant irradiation geometry, values of ${\cal C}\sim6\times10^{-4}-7\times10^{-3}$ have been found to sufficiently explain the observed multi-wavelength outburst behaviour in a small sample of systems (e.g., \citealt{hynes2002,suleimanov2008,lipunova2017}). 
However, evidence also exists suggesting the possibility that $\cal C$ may change between the hard and soft accretion states (e.g., \citealt{gierlinski2009,kimura2019}). Moreover, a handful of BH-LMXBs show complex light-curve morphology. Here,  variability on a range of timescales (e.g., flaring episodes) and extended plateau phases are observed. These temporal features are suggestive of a non-constant irradiation geometry, where a temporal and/or spatially varying X-ray irradiation source heats the disc (e.g., \citealt{esin2000a,esin2000}).


As such, in this paper we focus on building a numerical methodology that can track the time-series evolution of the X-ray irradiation heating the discs in BH-LMXB systems using a combination of observed X-ray, optical, and infrared lightcurves. 
The paper is organized as follows: Section \ref{apply_gx_lcs} describes the multi-wavelength observations, and binary system characteristics, of Galactic BH-LMXB GX339$-$4 that we make use of in this work. Section \ref{sec:model_disc0} summarizes our methodology, while a detailed account describing the development of our methodology is provided in Appendix \ref{sec:model_disc}. Section \ref{results} describes the application of our methodology to the multi-wavelength observations of GX339$-$4. Section \ref{discuss} explores the physical mechanism(s) responsible for the irradiation heating of the accretion disc in GX339$-$4, 
and Section \ref{conclusion} summarizes this work.   

\section{The Galactic BHXB GX339$-$4}\label{apply_gx_lcs}

GX339$-$4 is a Galactic LMXB that was discovered during an X-ray outburst in 1972 \citep{markert1973}. No dynamical mass estimate currently exists for this source. However, both the known mass function \citep{hynes2003,heida2017}, as well as X-ray spectral and temporal properties \citep{zdziarski1998,sunyaev2000}, are indicative of the black hole nature of the compact object in the system. 
Over the past near half-century, this system has undergone more than 20 individual outbursts. As a result, GX339$-$4 is one of the most frequently recurring, and in-turn one of the most extensively studied, transient X-ray binaries in our Galaxy. During its multitude of outbursts, GX339$-$4 has been observed to display the entire array of X-ray spectral accretion states, as well as a range of morphology in its X-ray and optical lightcurves, including combination exponential+linear shaped profiles, extended plateau phases, and multiple flaring episodes during the outburst decay. 
See Table 14 of \cite{tetarenko2016} for a complete list of references. For these reasons, GX339$-$4 is an ideal source for study.

\subsection{Observational Data}\label{sec:data}

\begin{figure*}
  \center
  \includegraphics[width=0.97\linewidth,height=1.2\linewidth]{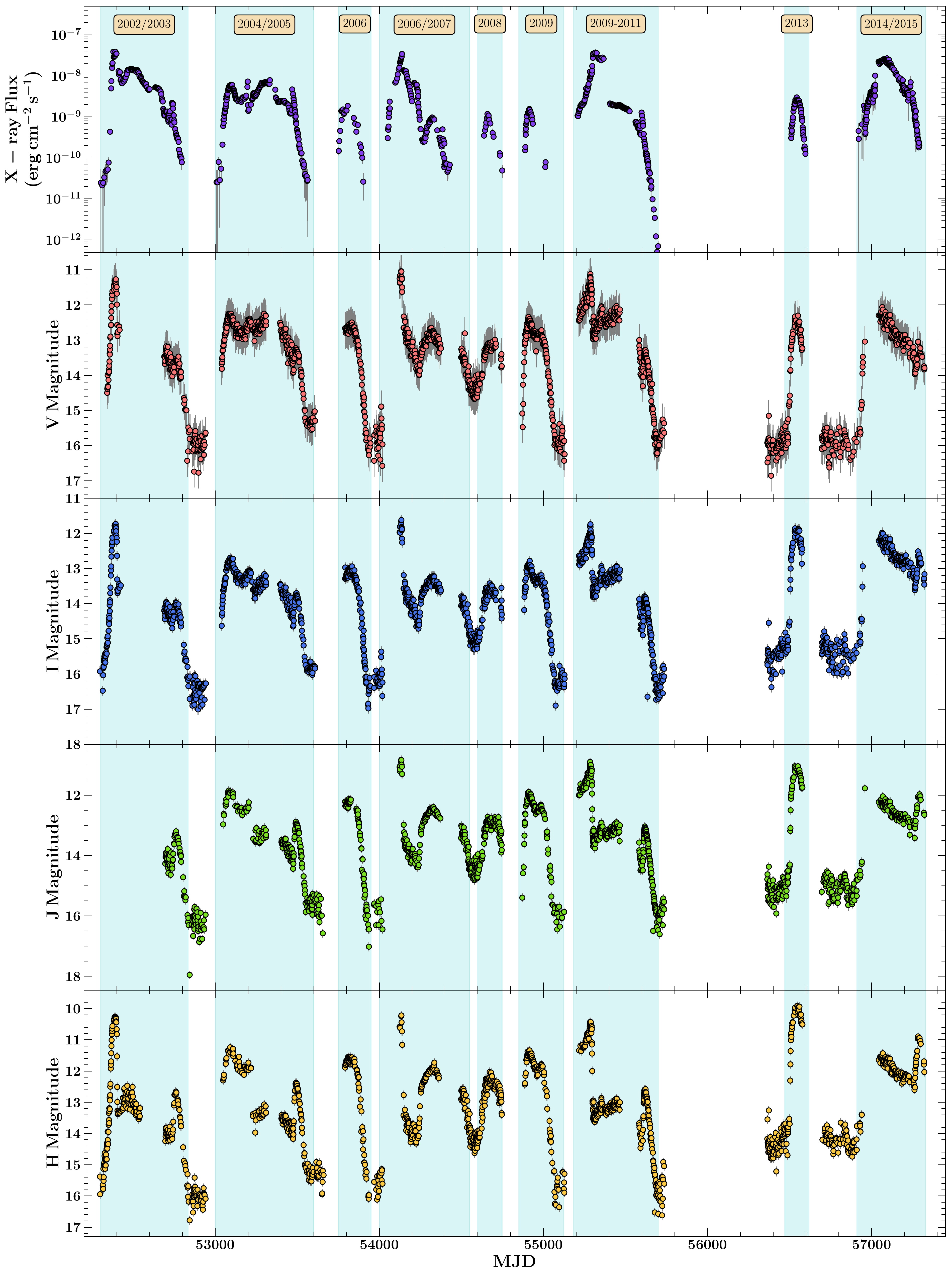}
  \caption{Multi-wavelength light-curve of GX339$-$4 between 2002$-$2015. X-ray flux is a combination of RXTE/PCA, Swift/XRT and MAXI/GSC data (see Section \ref{sec:xraydata} for details). All OIR data is from SMARTS/ANDICAM and has been dereddened. Shaded cyan regions are meant to guide the eye, marking individual outbursts. See Table \ref{tab:outinfo} for the definition of individual outbursts.}%
  \label{fig:longterm_lc}%
\end{figure*}

GX339$-$4 has been extensively observed over the past two decades, at X-ray, optical, and infrared wavelengths, with a combination of the Rossi X-ray Timing Explorer (RXTE), the Neil Gehrels Swift Observatory, the Monitor of All-sky X-ray Image (MAXI) telescope, and the Small \& Moderate Aperture Research Telescope System (SMARTS; \citealt{subasavage2010}) 1.3m telescope in Cerro Tololo, Chile. We have collected all X-ray through infrared data available for GX339$-$4 during the time period of 2002$-$2015 from the (i) Proportional Counter Array (PCA) aboard RXTE, (ii) X-ray Telescope (XRT) aboard Swift, (iii) MAXI telescope, and (iv) A Novel Dual Imaging CAMera (ANDICAM; \citealt{depoy2003}) aboard the SMARTS 1.3m telescope. This data set covers 9 individual outbursts of GX339$-$4. See Table \ref{tab:outinfo} and Figure \ref{fig:longterm_lc} for outburst information.

\subsubsection{X-ray Light Curves}\label{sec:xraydata}

We obtained all RXTE/PCA and MAXI/GSC data from the WATCHDOG project \citep{tetarenko2016}. This includes all good pointed PCA observations (i.e. no scans or slews) available (over the RXTE mission) in the HEASARC archive and publicly available data from the MAXI online archive\footnote{\url{http://maxi.riken.jp/top/}}. All Swift/XRT data,

\noindent including both windowed-timing and photon-counting mode observations, were obtained from the Swift/XRT online product builder\footnote{\url{http://www.swift.ac.uk/user\_objects/index.php}} \citep{evans2009}.

All X-ray light curves were originally extracted in the 2$-$10~keV band. These lightcurves were then converted from instrument specific count-rate to band-limited flux by using crabs as a baseline unit and calculating approximate count rate equivalences in the 2$-$10 keV band (see \citealt{tetarenko2016} for details on this method). 

Next, band-limited flux was converted to bolometric flux by splitting each outburst into individual accretion states using the WATCHDOG project's Accretion-State-By-Day tool\footnote{This tool provides accretion state information on daily time-scales during outbursts of the population of BH-LMXBs in our Galaxy.}, and applying the accretion state specific bolometric corrections estimated by \cite{migliari2006}.
Lastly, by applying an accretion efficiency defined as \citep{coriat2012}\footnote{However, also see Marcel et al. 2020 (in prep) for a thorough discussion on how accretion efficiency may vary in a more complex way during BHXB outburst cycles.},
\[ \eta=\begin{cases} 
      0.1\left(\frac{\dot{M}}{0.01 \dot{M}_{\rm edd}} \right) & L_X<0.01L_{\rm edd} \\
      0.1 & L_X\geq0.01L_{\rm edd}, 
   \end{cases}
\]
where the Eddington accretion rate is defined as $L_{\rm edd}=0.1\dot{M}_{\rm edd}c^2$, bolometric flux was converted to an observed $\dot{M}_{\rm in}$ via,
\begin{equation}
\dot{M}_{\rm in}=\frac{F_{\rm X,bol}(4 \pi D^2)}{\eta c^2}.
\end{equation}

\subsubsection{Optical/IR Light Curves}\label{sec:oirdata}

We have collected all available optical and IR observations of GX339$-$4 from SMARTS/ANDICAM in the V, I, J, and H bands. Data from 2002$-$2012 were obtained from \cite{buxton2012}. Additional data covering the time period of 2013$-$2015 were collected separately. For the reduction procedure used for this data, see \cite{buxton2012}.

Following \cite{buxton2012}, all data were corrected for interstellar extinction according to \cite{odonnell1994} (for V and I bands) and \cite{cardelli1989} (for J and H bands). Magnitudes were dereddened using the {\sc specutils} package in {\sc python} and $E(B-V)=1.2\pm0.1$ \citep{zdziarski1998}. The uncertainty on the dereddened magnitudes were calculated by adding the photometric and interstellar reddening errors in quadrature. Lastly, dereddened magnitudes were converted to flux density (in Jy) using the appropriate filter zero points obtained from the SVO filter service\footnote{\url{http://svo2.cab.inta-csic.es/theory/fps/}}. 
Note that, it is possible the reddening correction used here is over-estimated. See Figure \ref{fig:hs_seds}, and Section \ref{discuss}, for a discussion of the effect this would have on our results. See also \citet{buxton2012}, for a discussion of other reddening estimates present in the literature for GX339$-$4. Also see \citet{kosenkov2020} (which we became aware of after this manuscript was submitted). They use a similar SMARTS dataset, though focus on the non-disc components in the OIR regime.

\begin{table}
	\centering
	\setlength{\tabcolsep}{3pt}
	\caption{Outburst Activity in GX339$-$4 between 2002-2015}
	\medskip
	\label{tab:outinfo}
	\begin{tabular}{lcccc} 
		\hline
		Outburst & $t_{\rm begin}$ & $t_{\rm end}$ & Data &Outburst\\
		Year & (mjd) & (mjd) &Available& Type \\
		\hline
 2002$-$2003 &52350& 52750 &PCA,ANDICAM&C\\[0.045cm]
 2004$-$2005& 53054 &53515 &PCA,ANDICAM&C\\[0.045cm]
 2006& 53751 &53876 &PCA,ANDICAM&F\\[0.045cm]
 2006$-$2007& 54053 &54391 &PCA,XRT,ANDICAM&C\\[0.045cm]
2008 &54624 &54748 &PCA,XRT,ANDICAM&F\\[0.045cm]
2009 &54875 &55024 &PCA,XRT,ANDICAM&F\\[0.045cm]
2009$-$2011& 55182 &55665 &PCA,XRT,ANDICAM&C\\[0.045cm]
2013 &56505 &56608 &XRT,ANDICAM&F\\[0.045cm]
2014$-$2015 &56936& 57311&XRT,GSC,ANDICAM&C\\[0.2cm]
\hline
\multicolumn{5}{p{0.82\columnwidth}}{\hangindent=1ex NOTE. -- The outburst year, and mjd of the beginning ($t_{\rm begin}$) and end ($t_{\rm end}$) of the outbursts are from the WATCHDOG catalogue \citep{tetarenko2016}. Outburst type refers to the outburst classifications defined in \cite{tetarenko2016}: C - canonical, cycles through hard and soft accretion states during outburst, and F - failed, remains in the hard accretion state for the duration of the outburst. } 
	\end{tabular}
\end{table}

\subsubsection{X-ray Spectra}\label{sec:xrayspectra}

We make use of the spectral fitting and analysis done by \cite{clavel2016}, who have fit a two-component disc blackbody plus power-law model to all available RXTE/PCA spectra in the 3$-$40 keV band\footnote{Note that, in the case of the 2009$-$2011 outburst, the soft state spectral data was re-fit, allowing the photon index parameter ($\Gamma$) to only be sampled in the 2$-$2.5 interval. This was done to limit the wide dispersion  in $\Gamma$ initially obtained in the soft state spectral fits of this outburst by \cite{clavel2016}.}, to compute the (i) bolometric luminosity, $L_{\rm bol}(t)$, and (ii) Compton temperature, $T_{\rm IC}(t)$, as a function of time
during outbursts of GX339$-$4 occurring between 2002$-$2012. These two quantities are essential input parameters needed to model the evolution, and observable properties, of a thermally-driven (Compton-heated) wind in this system. See Section \ref{discuss} for further discussion. We compute $L_{\rm bol}(t)$ from the 3$-$200 keV flux, estimated by \cite{marcel2019} from the best-fits to all available RXTE/PCA spectra.

We follow the procedure outlined in \citet{shidatsu2019} to compute $T_{\rm IC}(t)$. 
Considering each time $t_i$ in which spectral information is available,
we integrate the observed (best-fit) spectral

\begin{table*}
	\centering
	\setlength{\tabcolsep}{3pt}
	\caption{Binary Orbital Parameters used for GX339$-$4}
	\medskip
	\label{tab:binary_params}
	\begin{tabular}{cccccccc} 
		\hline
		black hole mass & mass ratio & orbital period  & distance&$R_{\rm g}$&$ R_{\rm circ}$ &$R_1$&$R_{\rm out}$\\
		$M_{1}$ ($M_{\odot}$) & ($q=M_2/M_1)$ &$P_{\rm orb}$  (hrs) &$D$ (kpc)& (cm)& ($\times 10^{10} {\rm cm}$)&($\times 10^{10} {\rm cm}$)&($\times 10^{10} {\rm cm}$) \\
		\hline
$N(7.8,1.2)$&$U(0.024,0.45)$& 42.1 &$8\pm2$&$\left(1.15_{-0.19}^{+0.17} \right) \times 10^{6}$&$\left( 26.62_{-1.45}^{+1.42}\right)$&$\left(63.01_{-3.43}^{+3.35}\right)$&$\left(44.34_{-12.93}^{+12.64}\right)$\\
\hline
\end{tabular}
\end{table*}

\noindent energy distribution (SED) as follows:
\begin{equation}
 T_{\rm IC}(t_i)=\frac{\int h\nu F_{\nu} d \nu}{4k \int F_{\nu} d\nu },
\end{equation}
where $k$ is the Boltzmann constant. Here we set the lower limit of the integral at 0.1 keV. The dependence of the Compton temperature on the high-energy cutoff (for a power-law) saturates at $>100$ keV as a result of the rollover in the Klein-Nishima cross-section compared to the constant cross-section assumed in Thomson scattering \citep{done2018}. Thus, we set the upper limit of the integral at 100 keV. The uncertainty in $T_{\rm IC}$ is propagated via a Monte-Carlo technique from errors in the best-fit spectral model parameters (see details in \citealt{clavel2016}), inner disc radius ($R_{\rm in}$; see Section \ref{sec:inner_disc} and \citealt{marcel2019}), and our chosen distributions for the binary orbital parameters of black hole mass and binary mass ratio (see Table \ref{tab:binary_params} and the following Section).
In addition, an absolute minimum for $T_{\rm IC}$ is also applied in this method, whereby $T_{\rm IC}$ will not be sampled below $(T_{\rm IC}/10^{8}\, {\rm K})=0.06 kT_{\rm in}$, the absolute minimum $T_{\rm IC}$ for a pure disc black-body spectrum.

Note that, while the above method works well for the simple power law spectrum of the hard state, the complex (combination disc $+$ power law) spectra existing during the intermediate and soft states must be handled more carefully. The steeper power-law component here tends to dominate the low energy spectral region in an nonphysical way, causing an artificially low estimate for $T_{\rm IC}$. To combat this problem, for intermediate and soft state observations, we take the power law spectral component to start from the $E_{\rm min}=kT_{\rm in}$ keV, rather than $E_{\rm min}=0.1$ keV, in the above integral.

Lastly, we note that our estimates for $T_{\rm IC}$ in the soft state, derived using this method, are typically cooler than those estimated by \citet{shidatsu2019} for the Galactic BHXB H1743$-$322 (which based on outburst frequency, is thought to have a similar disc size to GX339$-$4). However, the soft state spectra of BHXBs tend to be much softer for face-on discs (gravitational redshift dominates) than edge-on discs (doppler blueshift dominates) as a result of doppler/general-relativistic effects (e.g., see \citealt{munozdarias2013}). Thus, this is likely only an inclination effect.

\subsection{Binary System Characteristics}

\subsubsection{Orbital Parameters}

While the mass function, orbital period ($P_{\rm orb}=42.1$ hrs; \citealt{hynes2003,heida2017}), and stellar companion (K-type star based on detected absorption lines in the near IR spectrum; \citealt{heida2017}), are known in GX339$-$4, no dynamical mass estimate or constrained estimate for mass ratio currently exist. Thus, we instead follow the procedure of \cite{tetarenko2018b}, and sample these quantities from the Galactic distributions of \cite{ozel2010} and 
\cite{tetarenko2016}, respectively.

The distance to the source is still a matter of debate. \cite{hynes2004} suggest GX339$-$4 is located beyond the Galactic tangent point (giving a lower limit of $>6$ kpc) based on optical spectra. This is consistent with the recent work of \cite{heida2017}, who derive a conservative lower limit of $>5$ kpc based on near IR spectra. \cite{zdziarski2004} on the other hand, prefer GX339$-$4 to be located in the Galactic bulge, estimating $D=8\pm2$ kpc based on OIR data. We adopt the distance estimate from \cite{zdziarski2004} in this work.

Lastly, no estimate of binary inclination for GX339$-$4 (via ellipsoidal variations)  currently exists (although see Section \ref{sec:xstar_sims} for a detailed discussion on this topic). Thus, we do not take into account inclination effects in this analysis. Instead, we average over all angles when computing the disc optical flux (Equation \ref{eq:disc_flux}). See Table \ref{tab:binary_params} for a summary of orbital parameters used in this work.

\subsubsection{Evolution of the Inner Disc Radius}\label{sec:inner_disc}

To define how $R_{\rm in}$ varies as a function of time during outbursts of GX339$-$4, we make use of two individual prescriptions (as described below) that employ very different methods. The first relies on modelling the reflection component in X-ray spectra, while the second models the continuum. 

The first prescription adopts the $R_{\rm in}$ estimated from X-ray reflection spectroscopy \citep{garcia2015,wangji2018}. These authors provide estimates of $R_{\rm in}$ for multiple hard state observations, taken during the 2002$-$2003, 2009$-$2011, 2013, and 2014$-$2015 outbursts of GX339$-$4, covering a luminosity range of $\sim0.6-23\%\,  L_{\rm edd}$ (assuming $M_{1}=N(7.8,1.2)M_{\odot}$; \citealt{ozel2010}). Using these results, we create a linearly interpolated function $R_{\rm in}(\dot{M}_{\rm in}/\dot{M}_{\rm edd})$, valid during the hard accretion state. To create the time-series evolution of $R_{\rm in}$ required for the methodology described in Section \ref{sec:model_disc0} and Appendix \ref{sec:model_disc}, we start by parsing through an individual outburst of GX339$-$4, and use the data from the WATCHDOG project's Accretion-State-By-Day tool \citep{tetarenko2016} to determine the accretion state evolution of the source. If the source is in the hard or intermediate states, we use our interpolated function, along with the observed $\dot{M}_{\rm in}(t)$, to determine $R_{\rm in}(t)$. If the source is in the soft state, we assume $R_{\rm in}(t)=R_g$. 

The second prescription uses the unified accretion-ejection paradigm for BH-LMXBs developed by Marcel et al. \citep{marcel2018a,marcel2018b,marcel2019}. These authors have developed a two-temperature plasma code to effectively model the spectral evolution (at X-ray and radio wavelengths) of BH-LMXB outbursts. They model the observed spectral evolution in a BH-LMXB as the interplay between two different regions of the accretion flow, an inner (jet-emitting) disc (JED) and an outer (standard) accretion disc (SAD)\footnote{The JED-SAD hybrid disc configuration involves (i) a geometrically thin, optically thick accretion disc extending down to the truncation radius, where (ii) a hot jet emitting disk (JED), threaded by a large scale vertical magnetic field transporting angular momentum vertically, exists down to the inner-most stable circular orbit (ISCO). See \cite{ferreira2006} and \cite{marcel2018a,marcel2018b,marcel2019} for details.}, that ultimately leads to changes in $\dot{M}_{\rm in}$ and the transition radius between two disc regions (i.e., $R_{\rm in}$) over an entire outburst cycle.
By applying this method to RXTE/PCA data of GX339$-$4 (see e.g., \citealt{marcel2019}), they have been able to obtain the time-series evolution of $\dot{M}_{\rm in}(t)$ and $R_{\rm in}(t)$, during outbursts of GX339$-$4 occurring between 2002$-$2012\footnote{Note that the 2009$-$2011 outburst results have already been published in \cite{marcel2019}. Results from the outbursts occurring between 2002$-$2009 will be published in a later paper (Marcel et al. 2020, submitted to A\&A).}, that together uniquely reproduce the X-ray (i.e., $\dot{M}_{\rm in}$ evolution) and (9 GHz) radio lightcurves of, and evolution of the spectral shape during, each outburst cycle.

\section{Constraining the X-ray Irradiation of BHXB Accretion discs with Observations}\label{sec:model_disc0}

\subsection{The Methodology}

By directly comparing BH-LMXB outburst lightcurves at optical and X-ray wavelengths, one can, in principle, track how properties of the X-ray irradiation heating the discs in these systems evolves over time.
While the X-ray light curve provides a measure of the bolometric luminosity in these systems, and thus a straightforward means to estimate the central mass-accretion rate onto the black hole ($\dot{M}_{\rm in}$; see Section \ref{sec:xraydata}), the optical lightcurve gives a direct measure of irradiation flux (under the assumption that reprocessing is the dominant source of emission in the optical regime; see Section~\ref{sec:emission_mechs} for discussion). Thus, building a relationship between the central mass accretion rate ($\dot{M}_{\rm in}$) and absolute magnitude in the optical bandpasses, valid during BH-LMXB outbursts, would allow one to place constraints on the fraction of X-ray emission needed to be reprocessed in the outer disc to explain the observed optical flux, and how this fraction changes over a complete outburst cycle. 

While the full details of this methodology are presented in Appendix \ref{sec:model_disc}, the basic idea is as follows: the absolute magnitude, in a particular bandpass, depends only on $\dot{M}_{\rm in}$, black hole mass, disc size, and the fraction of X-rays intercepted and reprocessed in the outer disc ($\cal C$). Thus, by assuming: (i) a constant outer disc radius ($R_{\rm out}$) during outburst, (ii) an inner disc radius that varies as a function of central mass accretion rate ($R_{\rm in}(\dot{M}_{\rm in}$); see Section \ref{sec:inner_disc}), and (iii) a disc temperature profile that is a combination of
viscous \citep{frank2002},
\begin{equation}
T_{\rm visc}^4=\frac{3G M_1 \dot{M}}{8 \pi \sigma R^3}\left[ 1-\left(\frac{R_{\rm in}}{R}\right)^{1/2} \right],
\end{equation}
and irradiated \citep{dubus1999},
\begin{equation}
T_{\rm irr}^4={\cal C} \frac{\dot{M} c^2}{4 \pi \sigma R^2},
\label{eq:temp_irr_eq}
\end{equation}
portions such that,
\begin{equation}
T_{\rm eff}^4(R)=T_{\rm visc}^4(R)+T_{\rm irr}^4(R),
\label{eq:temp}
\end{equation}
one can use numerical integration techniques to reconstruct $\cal C$ from an optical light curve, given the estimate of $\dot{M}_{\rm in}$ derived from a simultaneous X-ray light curve.

\subsection{Origin of the OIR Emission}\label{sec:emission_mechs}

In addition to reprocessed X-rays from the outer disc, OIR emission during BH-LMXB outbursts may also be produced by: 
(i) synchrotron emission from particles accelerated to very high energies (i.e., Lorentz factors of $\gamma\sim10^6$) in the collimated jets \citep{homan2005,russell2006}, and (ii) hot spots, created as a result of the accretion stream impacting the disc (see \citealt{maccarone2014} and references therein). As such, estimating the contribution of, and correcting for, these two emission mechanisms in the OIR lightcurves is essential to accurately derive 
how properties of the X-ray irradiation heating the disc in this system evolve with time (see Section \ref{results} for details).

\subsubsection{Tracking the Contributions of the Disc and Jet}

BH-LMXB jets produce a broad-band spectrum ($F_{\nu}\propto \nu^{\gamma}$ for spectral index $\gamma$) characterized by a flat to slightly inverted, optically thick component ($\gamma\gtrsim0$; \citealt{blandford1979,falcke1995,fender2001}) extending from the radio through OIR wavelengths \citep{corbel2002,homan2005,russell2006,chaty2011}, that breaks to a steep, optically thin component ($-0.7<\gamma<-0.5$; \citealt{russell2013}).

In contrast, for an accretion disc with a temperature profile of the form, $T(R)\propto R^{-n}$, an optical spectral index of $\gamma=3-2/n$ is expected in the spectral band corresponding to the summed multicolour disc blackbody. Hence, $\gamma=1/3$ is expected for a viscous disc ($n=3/4$). While a $\gamma$ ranging between $-5/3$ (irradiated isothermal disc following \citealt{cunningham1976} and \citealt{vrtilek1990}; $n=3/7$) and $-1$ (irradiated disc with $n=1/2$) would correspond to an irradiated disc. If the IR emission was purely from a disc (viscous or irradiated), this would correspond to the Rayleigh-Jeans regime, and thus a $\gamma=2$ \citep{frank2002}. 

\begin{table}
	\centering
	\setlength{\tabcolsep}{4.0pt}
	\caption{Disc Bright Spot Contributions in the OIR Regime}
	\medskip
	\label{tab:brightspots}
	\begin{tabular}{ccccc} 
		\hline
		Outburst & \multicolumn{4}{c}{$m_{\rm spot}$ in SMARTS/ANDICAM Bands}\\[0.05cm]
		Year & $V$ & $I$ &$J$& $H$ \\
		\hline
 2002$-$2003 &$15.6_{-1.1}^{+1.7}$&$15.7_{-1.1}^{+1.8}$&
 $16.0_{-1.1}^{+1.6}$&$15.9_{-1.2}^{+1.6}$\\[0.08cm]
 2004$-$2005 &$16.1_{-1.0}^{+1.7}$&$16.1_{-1.1}^{+1.7}$&
 $16.5_{-1.2}^{+1.8}$&$16.4_{-1.2}^{+1.7}$\\[0.08cm]
 2006 &$18.2_{-1.3}^{+1.8}$&$18.3_{-1.2}^{+1.7}$&
 $18.5_{-1.2}^{+1.6}$&$18.3_{-1.2}^{+1.6}$\\[0.08cm]
 2006$-$2007 &$15.7_{-1.1}^{+1.6}$&$15.8_{-1.1}^{+1.7}$&
 $16.0_{-1.1}^{+1.6}$&$15.8_{-1.1}^{+1.7}$\\[0.08cm]
 2008 &$18.7_{-1.1}^{+1.6}$&$18.8_{-1.1}^{+1.6}$&
 $18.9_{-1.2}^{+1.9}$&$18.8_{-1.2}^{+1.7}$\\[0.08cm]
 2009 &$18.5_{-1.2}^{+1.8}$&$18.6_{-1.1}^{+1.7}$&
 $18.8_{-1.1}^{+1.7}$&$18.6_{-1.1}^{+1.7}$\\[0.08cm]
 2009$-$2011 &$15.6_{-1.1}^{+1.8}$&$15.7_{-1.1}^{+1.7}$&
 $16.0_{-1.2}^{+1.6}$&$15.8_{-1.2}^{+1.8}$\\[0.08cm]
 2013 &$18.8_{-1.1}^{+1.7}$&$19.0_{-1.1}^{+1.6}$&
 $19.1_{-1.1}^{+1.7}$&$19.0_{-1.1}^{+1.8}$\\[0.08cm]
 2014$-$2015 &$15.8_{-1.0}^{+1.7}$&$15.9_{-1.1}^{+1.7}$&
 $16.1_{-1.0}^{+1.8}$&$16.0_{-1.1}^{+1.6}$\\[0.08cm]
 
\hline
	\end{tabular}
\end{table}

\noindent See \cite{hynes2002} and \cite{hynes2005} for a discussion on the temperature profile, and corresponding spectral energy distributions (SEDs), of LMXB accretion discs.

\begin{figure*}
\subfloat
  {\includegraphics[width=0.96\linewidth,height=.5\linewidth]{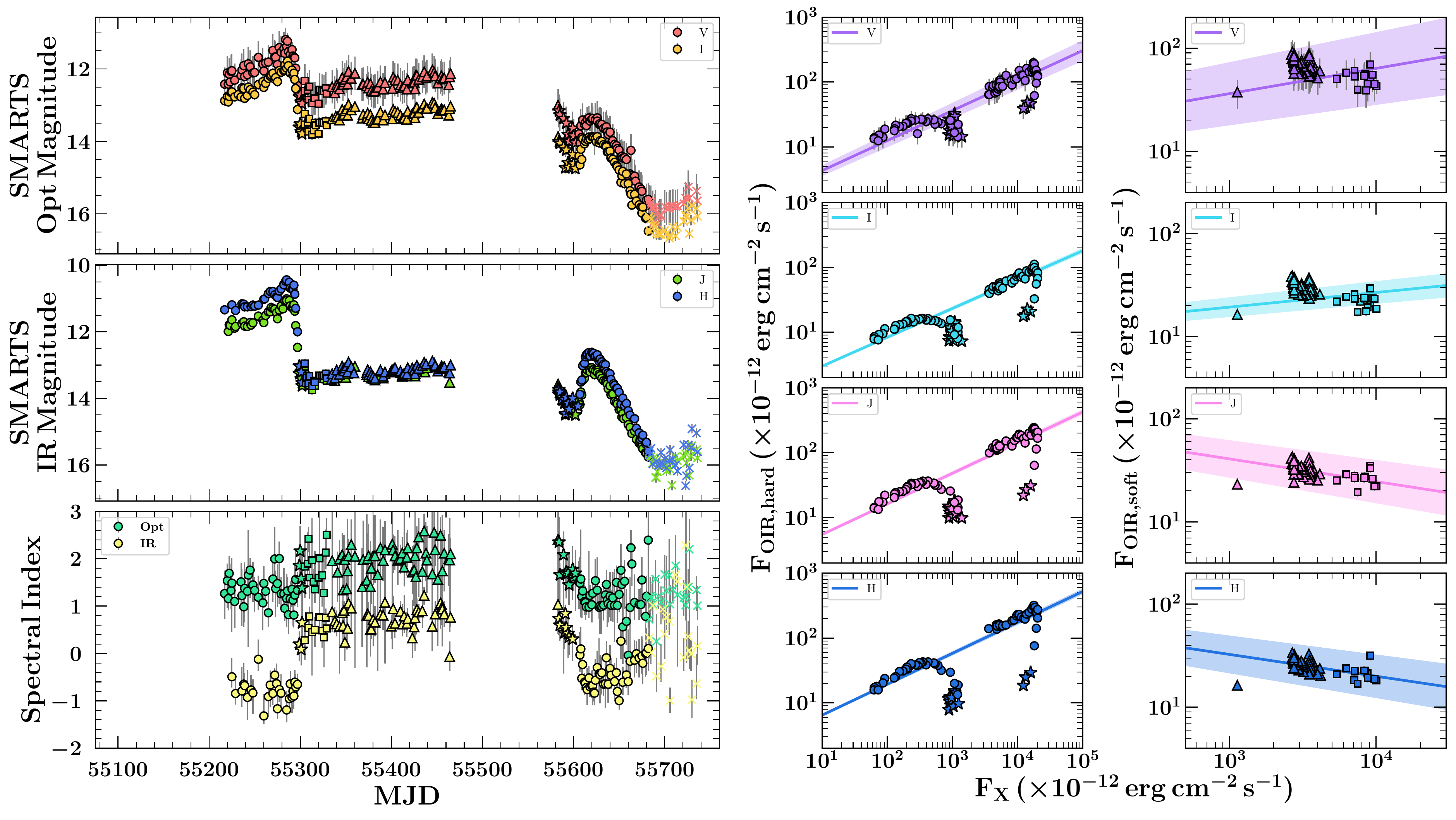}}
  \caption{Analysis of the OIR emission processes during the 2009$-$2011 outburst of GX339$-$4. \textbf{Left Panel:} dereddened SMARTS/ANDICAM lightcurves in the \textit{(top)} V and I, and \textit{(middle)} J and H, bands.
 Optical (V and I) and IR (J and H) spectral index \textit{(bottom)}, as a function of time. \textbf{Middle Panel:} OIR-X-ray correlations for the hard (and hard-intermediate) accretion states in the V-Band \textit{(top)}, I-Band \textit{(second)}, J-Band \textit{(third)}, and H-Band \textit{(bottom)}. \textbf{Right Panel:} OIR-X-ray correlations for the soft (and soft-intermediate) accretion states, in the V-Band \textit{(top)}, I-Band \textit{(second)}, J-Band \textit{(third)}, and H-Band \textit{(bottom)}. The bolometric (3$-$200 keV) flux, computed from all available RXTE/PCA spectra (see Section \ref{sec:xrayspectra} and \citealt{marcel2019} for details on the spectral fitting and analysis), is used as the X-ray data in all correlations.
 The shapes of the data points in all panels indicate the accretion state of the source as defined by Marcel et al.. Shapes are defined as follows: hard (circles), hard-intermediate (stars), soft (triangles), soft-intermediate (squares), and quiescence (X's). The best-fit power-law function (solid coloured lines), and $1\sigma$ confidence interval (coloured shaded regions) on the fit, are displayed in each correlation plot.}
  \label{fig:mw_outburst1}%
\end{figure*}

As the jet spectrum significantly differs from that of an accretion disc, tracking the changes in the (optical and IR) spectral index over time can be used to understand
how the OIR contribution of the disc and jet vary throughout BH-LMXB outbursts. We use a Bayesian Markov-Chain Monte-Carlo (MCMC) algorithm to compute both the optical (using V and I band data) and IR (using J and H band data) spectral indexes, along with corresponding $1\sigma$ error bars. The details of this algorithm are explained in Section \ref{sec:jet_mech}. 
The left panels of Figure \ref{fig:mw_outburst1} plot a combination of the (i) OIR lightcurve in the in the V, I, J, and H bands, and (ii) the optical (V and I bands) and IR (J and H bands) spectral indexes, used to determine which emission process dominates during the 2009$-$2011 outburst of GX339$-$4, as a function of time and X-ray accretion state.
 Figures \ref{fig:mw_outburst2}$-$\ref{fig:mw_outburst5} in Appendix \ref{sec:oir_app} show these results for the remaining 8 outbursts in our sample.

Bright spots in X-ray binaries are difficult to detect. To actually detect these bright spots, the X-ray binary would have to be: (i) in quiescence, where $\dot{M}_t$ into the outer disc $\gg$ $\dot{M}_{\rm in}$ onto the black hole, or (ii) in the radiatively inefficient accretion regime during outburst (i.e., early rise and late decay), where $\dot{M}_{\rm in}$ onto the black hole is low (e.g., \citealt{mcclintock1995,froning2011,maccarone2014}).
To compute the theoretical contribution of the bright spot to the V, I, J, and H bands, we follow the procedure outlined in \cite{dubus2018}, assuming a luminosity of the form,

\begin{equation}
L_{\rm spot}=\frac{G M \dot{M}_{t}}{2 R_{\rm disc}}\left(1-\frac{R_{\rm disc}}{R_1} \right),
\end{equation}
and a temperature of $T_{\rm eff,spot}=15,000$\,K \citep{groot2001}. Here, we estimate the mass-transfer rate from the companion ($\dot{M}_{t}$) by calculating the time-averaged $\dot{M}_{\rm in}$ over each outburst (see \citealt{tetarenko2016} for details on this method). This takes into account that the mass transfer rate from the companion may be increased during outburst from its time-averaged value over several outbursts. Thus, the estimated $\dot{M}_{t}$ is an upper limit to the average mass transfer. Uncertainties in $\dot{M}_{\rm in}$, $M_1$, $q$, $D$, and V-band magnitude (i.e., instrument uncertainty and error in interstellar reddening), are all taken into account when computing the bright spot contribution and its uncertainty (see Table \ref{tab:brightspots}). Comparing to the magnitudes in Figure \ref{fig:longterm_lc} confirms that the hot spot OIR contribution is small in outburst.

\begin{table}
	\centering
	\setlength{\tabcolsep}{3.5pt}
	\caption{Jet Contributions in the OIR Regime}
	\medskip
	\label{tab:jets_contrib}
	\begin{tabular}{ccccc} 
		\hline
		Outburst & \multicolumn{4}{c}{$F_{\rm jet}/F_{\rm tot}$ in SMARTS/ANDICAM Bands}\\[0.05cm]
		Year & $V$ & $I$ &$J$& $H$ \\
		\hline
 2002$-$2003 &$0.41_{-0.09}^{+0.08}$&$\cdots$&$0.99_{-0.0001}^{+0.0002}$&$\cdots$\\[0.08cm]

 2004$-$2005 &$0.38_{-0.07}^{+0.12}$&$\cdots$&$\cdots$&$0.56_{-0.02}^{+0.01}$\\[0.08cm]

 2006$-$2007 &$0.60_{-0.04}^{+0.03}$&$0.37_{-0.03}^{+0.02}$&$0.40_{-0.03}^{+0.02}$&$0.73_{-0.01}^{+0.01}$\\[0.08cm]

 2009$-$2011 &$0.81_{-0.01}^{+0.01}$&$0.84_{-0.002}^{+0.001}$&$0.99_{-0.00003}^{+0.00002}$&$0.99_{-0.00006}^{+0.00005}$\\[0.08cm]

 2014$-$2015 &$0.81_{-0.06}^{+0.11}$&$0.94_{-0.002}^{+0.003}$&$0.98_{-0.0006}^{+0.0007}$&$0.99_{-0.0002}^{+0.0001}$\\[0.08cm]  

\hline
\multicolumn{5}{p{0.92\columnwidth}}{\hangindent=1ex NOTE. -- The jet contribution cannot be estimated for the ``failed'' outbursts of GX339$-$4 (2006, 2008, 2009, and 2013), since there is no soft state data to draw from.} 
	\end{tabular}
\end{table}

\subsubsection{Quantifying the Jet Contribution}\label{sec:jet_mech}

Another instrumental tool we can use to study the emission processes in BH-LMXBs is the observed correlation between OIR flux and X-ray flux during outburst.
In a study of 33 LMXBs, \cite{russell2006} found individual global power-law correlations ($F_{\rm OIR}=N_{\rm pl} F_{X}^{\beta}$) valid during the hard and soft accretion states. The specific slope ($\beta$) of these correlations is expected to vary depending on the dominant emission mechanism \citep{vanparadijs1994,hynes2005,russell2006,coriat2009}. Thus, 
fitting power-law correlations to observed outburst data can help one determine the dominant emission mechanisms present, and also verify conclusions made from the multi-wavelength SEDs, as discussed above. 

If the disc temperature varies as $T=T_0 (R/R_0)^{-n}$ with $T_0\propto \dot{M}^{m}$ then the flux in the multicolour disc blackbody varies like $F_{\rm disc}\propto \dot{M}^{2m/n}$ while $F_{\rm RJ}\sim \dot{M}^{m}$ in the Rayleigh-Jeans (RJ) tail. The X-ray flux $F_{X}$ is either $\propto \dot{M}$ (soft state) or $\dot{M}^2$ (hard state)  depending on radiative efficiency (see Section \ref{sec:xraydata}). 

For a viscously-heated disc ($m=1/4$, $n=3/4$), the expected slope ranges from $\beta=0.13$ (RJ) to $\beta=0.33$ (disc) in the hard state, $0.26\leq\beta \leq 0.67$ in the soft state. For X-ray reprocessing with an isothermal disc ($m=2/7$, $n=3/7$) the slope ranges between $0.14\leq\beta \leq 0.67$ in the hard state, $0.28\leq\beta \leq 1.33$ in the soft state. The OIR flux is usually at the spectral transition between the RJ tail and the multicolour blackbody, given the outer disc temperature of 10,000\,K in outburst \citep{russell2006}. In contrast, under the assumption that the optically thick jet spectrum is flat from the radio through OIR regimes, a slope of $\beta\sim0.7$ is expected in the optical and IR regime \citep{corbel2003,gallo2003,russell2006}. 

As BH-LMXB jets are typically only observed in the hard state, they should contribute a negligible amount of OIR flux in the soft state. 
Thus, one can estimate the fraction of the total OIR flux (in a given bandpass) that comes from the jet vs the disc by computing the difference between the offset of hard and soft state data (i.e, comparing the power-law normalization parameters, $N_{\rm pl}$, of the hard and soft state correlations fits).
See \citet{russell2006} for a detailed description and application of this method to a large sample of LMXBs.

First, we use a Bayesian Markov-Chain Monte-Carlo (MCMC) algorithm \citep{foreman-mackey2013} to perform a linear fit in log-space, and estimate the slope ($s_{\nu}$) and intercept ($b_{\nu}$), for each individual correlation. To properly take into account uncertainties in both OIR flux (propagated from instrument and interstellar reddening errors) and X-ray flux\footnote{The bolometric (3$-$200 keV) flux used here has been computed by fitting all available RXTE/PCA spectra with a two-component disc blackbody plus power-law model. See Section \ref{sec:xrayspectra} and \citealt{marcel2019} for details on the spectral fitting and analysis.}, we use an alternative method to the standard linear formulation (i.e., $y=s_{\nu}x+b_{\nu}$). This method parametrizes the fit in terms of the $\theta$ parameter, defined as the angle that the linear function makes with the x-axis, and the $y$-intercept, $y_b$ \citep{hogg2010}. After likelihood maximization, posterior probability distributions (PDFs) of $s_{\nu}$ and $b_{\nu}$ are obtained by taking the tangent of the PDFs of $(\theta,y_b)$. See \citet{shaw2019} for details on this fitting algorithm. Table \ref{tab:fitinfo} displays the best-fit parameters $(s_{\nu},b_{\nu})$ for each correlation.

Second, we compute the hard state jet contribution to the V, I, J, and H bandpasses using the best-fit power-law normalization parameters ($N_{\rm pl}=10^{b_{\nu}}$) of the hard and soft state OIR-correlations. In this method, uncertainty in $D$ and V-band magnitude (a combination of instrument uncertainty and interstellar reddening errors) are taken into account when computing the jet contribution and its uncertainty. See Table \ref{tab:jets_contrib}, for jet contributions in each waveband, and Figure \ref{fig:hs_seds}, which shows select OIR SEDs, taken during the 2009-2011 outburst, before and after the OIR data was corrected for the jet contribution.

\begin{table}
	\centering
	\setlength{\tabcolsep}{2.5pt}
	\caption{OIR-X-ray Correlation Best Fits}
	\medskip
	\label{tab:fitinfo}
	\begin{tabular}{cccccc} 
		\hline
		Outburst&OIR &\multicolumn{2}{c}{Hard State} & \multicolumn{2}{c}{Soft State} \\
		ID&Band&$s_{\nu}$&$b_{\nu}$&$s_{\nu}$&$b_{\nu}$ \\
		\hline
2002$-$2003 &V&$0.37_{-0.02}^{+0.01}$&$0.37_{-0.06}^{+0.05}$&
        $0.27_{-0.08}^{+0.07}$&$0.60_{-0.25}^{+0.24}$\\[0.05cm]
         &I&$0.39_{-0.004}^{+0.005}$&$0.09_{-0.01}^{+0.01}$&
        $0.26_{-0.02}^{+0.01}$&$0.31_{-0.06}^{+0.07}$\\[0.05cm]
         &J&$0.25_{-0.01}^{+0.02}$&$0.59_{-0.03}^{+0.02}$&
        $-0.39_{-0.08}^{+0.09}$&$2.26_{-0.27}^{+0.23}$\\[0.05cm]
         &H&$0.48_{-0.004}^{+0.005}$&$0.16_{-0.01}^{+0.02}$&
        $0.38_{-0.01}^{+0.02}$&$-0.03_{-0.04}^{+0.05}$\\[0.3cm]
        2004$-$2005 &V&$0.41_{-0.01}^{+0.02}$&$0.25_{-0.04}^{+0.04}$&
        $0.73_{-0.09}^{+0.06}$&$-0.79_{-0.18}^{+0.29}$\\[0.05cm]
         &I&$0.40_{-0.004}^{+0.003}$&$0.07_{-0.01}^{+0.01}$&
        $0.37_{-0.02}^{+0.03}$&$0.06_{-0.08}^{+0.07}$\\[0.05cm]
         &J&$0.47_{-0.005}^{+0.004}$&$0.15_{-0.01}^{+0.02}$&
        $0.30_{-0.03}^{+0.04}$&$0.38_{-0.08}^{+0.09}$\\[0.05cm]
         &H&$0.49_{-0.005}^{+0.006}$&$0.17_{-0.01}^{+0.02}$&
        $0.21_{-0.03}^{+0.04}$&$0.58_{-0.09}^{+0.08}$\\[0.3cm]
 	2006 &V&$0.50_{-0.02}^{+0.01}$&$0.07_{-0.05}^{+0.04}$&$\cdots$&$\cdots$\\[0.05cm]
         &I&$0.51_{-0.01}^{+0.01}$&$-0.14_{-0.01}^{+0.02}$&
        $\cdots$&$\cdots$\\[0.05cm]
         &J&$0.67_{-0.01}^{+0.01}$&$-0.17_{-0.02}^{+0.01}$&
        $\cdots$&$\cdots$\\[0.05cm]
        &H&$0.65_{-0.01}^{+0.01}$&$-0.12_{-0.02}^{+0.03}$&
        $\cdots$&$\cdots$\\[0.3cm]
        2006$-$2007
        &V&$0.59_{-0.02}^{+0.01}$&$-0.36_{-0.06}^{+0.07}$&
        $0.30_{-0.06}^{+0.07}$&$0.47_{-0.27}^{+0.21}$\\[0.05cm]
         &I&$0.55_{-0.01}^{+0.02}$&$-0.40_{-0.02}^{+0.01}$&
        $0.25_{-0.02}^{+0.03}$&$0.30_{-0.06}^{+0.07}$\\[0.05cm]
         &J&$0.61_{-0.01}^{+0.01}$&$-0.27_{-0.02}^{+0.01}$&
        $0.25_{-0.02}^{+0.03}$&$0.34_{-0.07}^{+0.08}$\\[0.05cm]
         &H&$0.65_{-0.01}^{+0.01}$&$-0.29_{-0.02}^{+0.03}$&
        $0.16_{-0.02}^{+0.03}$&$0.62_{-0.07}^{+0.08}$\\[0.3cm]
        2008 &V&$0.28_{-0.09}^{+0.10}$&$0.78_{-0.28}^{+0.23}$&$\cdots$&$\cdots$\\[0.05cm]
         &I&$0.25_{-0.04}^{+0.03}$&$0.62_{-0.09}^{+0.10}$&
        $\cdots$&$\cdots$\\[0.05cm]
         &J&$0.45_{-0.03}^{+0.04}$&$0.43_{-0.09}^{+0.08}$&
        $\cdots$&$\cdots$\\[0.05cm]
         &H&$0.55_{-0.03}^{+0.02}$&$0.27_{-0.06}^{+0.08}$&
        $\cdots$&$\cdots$\\[0.3cm]
        2009 &V&$0.21_{-0.05}^{+0.04}$&$0.99_{-0.14}^{+0.16}$&$\cdots$&$\cdots$\\[0.05cm]
         &I&$0.30_{-0.02}^{+0.03}$&$0.49_{-0.08}^{+0.09}$&
        $\cdots$&$\cdots$\\[0.05cm]
         &J&$0.36_{-0.04}^{+0.03}$&$0.68_{-0.13}^{+0.12}$&
        $\cdots$&$\cdots$\\[0.05cm]
         &H&$0.47_{-0.03}^{+0.02}$&$0.44_{-0.07}^{+0.09}$&
        $\cdots$&$\cdots$\\[0.3cm]
        2009$-$2011 &V&$0.46_{-0.02}^{+0.01}$&$0.17_{-0.07}^{+0.06}$&
        $0.25_{-0.04}^{+0.05}$&$0.82_{-0.17}^{+0.16}$\\[0.05cm]
        &I&$0.45_{-0.01}^{+0.01}$&$0.03_{-0.02}^{+0.01}$&
        $0.14_{-0.01}^{+0.02}$&$0.86_{-0.05}^{+0.06}$\\[0.05cm]
         &J&$0.47_{-0.02}^{+0.01}$&$0.27_{-0.02}^{+0.01}$&
        $-0.22_{-0.03}^{+0.02}$&$2.27_{-0.10}^{+0.09}$\\[0.05cm]
         &H&$0.48_{-0.01}^{+0.01}$&$0.33_{-0.02}^{+0.02}$&
        $-0.21_{-0.03}^{+0.02}$&$2.15_{-0.09}^{+0.10}$\\[0.3cm]
        2013 &V&$0.30_{-0.16}^{+0.20}$&$0.72_{-0.63}^{+0.46}$&$\cdots$&$\cdots$\\[0.05cm]
         &I&$0.34_{-0.05}^{+0.05}$&$0.76_{-0.16}^{+0.15}$&
        $\cdots$&$\cdots$\\[0.05cm]
         &J&$0.39_{-0.05}^{+0.04}$&$0.98_{-0.12}^{+0.14}$&
        $\cdots$&$\cdots$\\[0.05cm]
         &H&$0.40_{-0.04}^{+0.05}$&$1.32_{-0.15}^{+0.13}$&
        $\cdots$&$\cdots$\\[0.3cm]
        2014$-$2015 &V&$0.23_{-0.04}^{+0.03}$&$0.72_{-0.12}^{+0.11}$&
        $2.09_{-0.72}^{+0.41}$&$-7.21_{-1.74}^{+3.10}$\\[0.05cm]
         &I&$0.11_{-0.01}^{+0.02}$&$1.23_{-0.05}^{+0.04}$&
        $0.71_{-0.13}^{+0.12}$&$-1.20_{-0.52}^{+0.56}$\\[0.05cm]
         &J&$-0.07_{-0.02}^{+0.01}$&$1.95_{-0.05}^{+0.04}$&
        $0.77_{-0.13}^{+0.12}$&$-1.48_{-0.54}^{+0.55}$\\[0.05cm]
         &H&$-0.23_{-0.01}^{+0.02}$&$2.69_{-0.03}^{+0.04}$&
        $0.82_{-0.12}^{+0.11}$&$-1.53_{-0.47}^{+0.50}$\\[0.3cm]

\hline
\multicolumn{6}{p{0.89\columnwidth}}{\hangindent=1ex NOTE. -- The best-fit linear function to the data in log-space, $\log_{10}(F_{\rm OIR})=s_{\nu}\log_{10}(F_{X})+b_{\nu}$ for a slope of $s_{\rm \nu}$ and a $y$-intercept of $b_{\rm \nu}$, is presented here. For comparison to the standard power-law correlations, $F_{\rm OIR}=N_{\rm pl}F_{X}^{\beta}$, the slope $\beta=s_{\nu}$ and the power-law normalization $N_{\rm pl}=10^{b_{\nu}}$.} 
	\end{tabular}
\end{table}  

The middle and right panels of Figure \ref{fig:mw_outburst1} plot the OIR-X-ray correlations in the V, I, J, and H bands, for each individual accretion state, used to estimate the
contributions of the jet at OIR wavelengths during the 2009$-$2011 outburst of GX339$-$4, as a function of X-ray accretion state and X-ray flux. The $\beta$ index is compatible with multicolour disc blackbody emission in the soft state, either from an irradiated or non-irradiated disc. The values of $\beta$ in the hard state are higher than in the soft state, and higher than expected from disc emission, supporting a significant contribution from the jet to the OIR emission. Figures \ref{fig:mw_outburst2}$-$\ref{fig:mw_outburst5} in Appendix \ref{sec:oir_app} show these results for the remaining 8 outbursts in our sample.

\begin{figure*}
\subfloat
  {\includegraphics[width=0.96\linewidth,height=.5\linewidth]{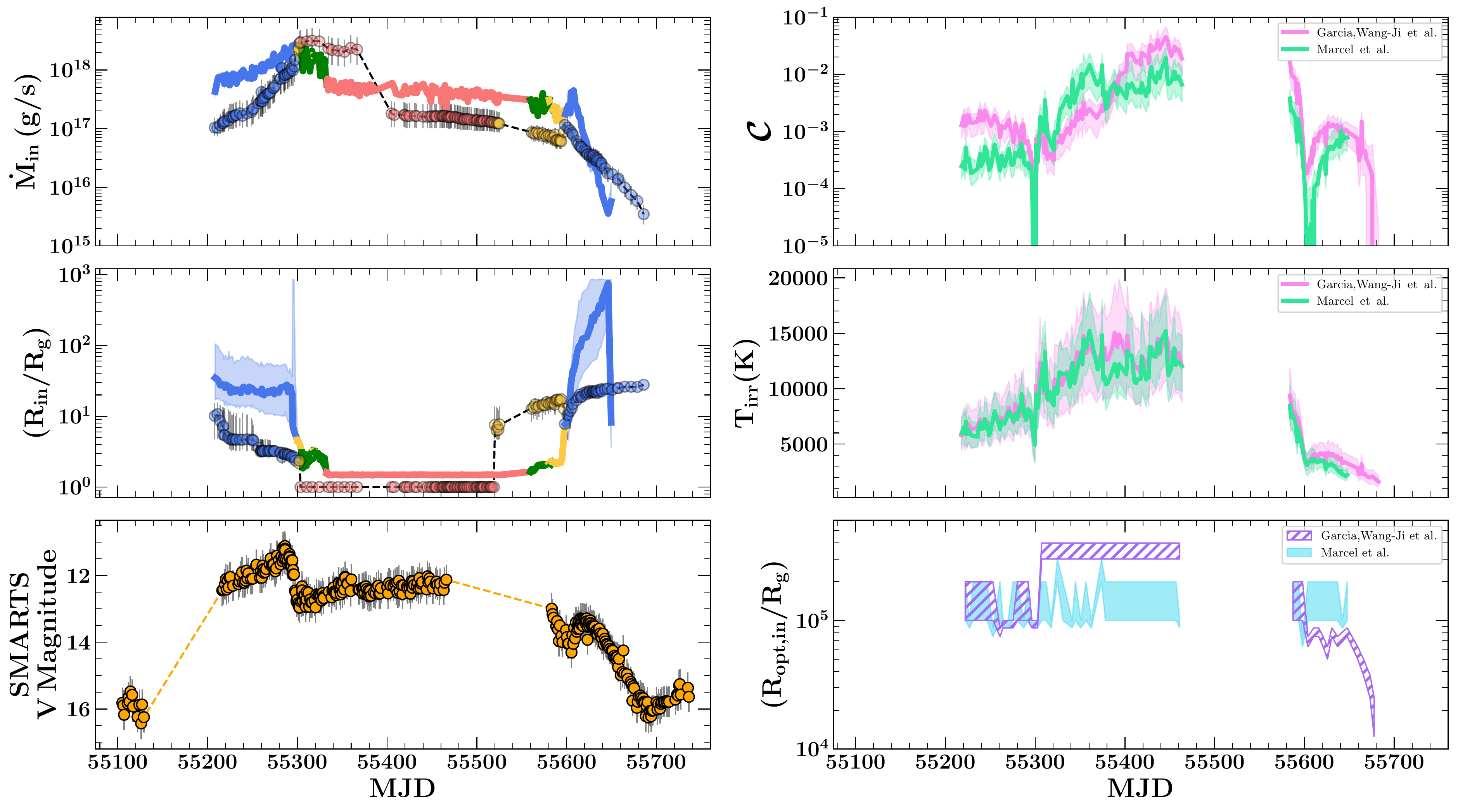}}
  
  \caption{The 2009$-$2011 outburst of GX339$-$4. \textbf{Left Panel:} \textit{(top)} central mass-accretion rate ($\dot{M}_{\rm in}$), \textit{(middle)} inner disc radius ($R_{\rm in})$ in units of $R_g$ (for $M_1=N(7.8,1.2)M_{\odot}$), and \textit{(bottom)} dereddened SMARTS/ANDICAM V-band magnitude, as a function of time. In the \textit{top} and \textit{middle} panels:
  (i) the translucent data points correspond to the $\dot{M}_{\rm in}$ calculated from the X-ray data and $R_{\rm in}$ interpolated from the Garcia,Wang-Ji et al. estimates using X-ray reflection spectroscopy, respectively. The solid lines are output from the Marcel et al. analysis. Both data points and solid lines are colour coded by accretion state. The colours of the data points and solid lines correspond to accretion state estimates from the WATCHDOG Project's Accretion-State-By-Day tool, and the accretion state definitions from Marcel et al., respectively. The colours are defined as follows: red (soft state), yellow (intermediate state; referred to as hard-intermediate state in Marcel et al.), blue (hard state), and green (soft intermediate state; Marcel et al. only). Uncertainties on the data points (represented as grey error bars) are propagated from errors in X-ray flux, distance, and black hole mass. The coloured shaded regions show the uncertainties in $(\dot{M}_{\rm in},R_{\rm in})$ derived in \citet{marcel2019}. The black \textit{(top and middle)} and orange \textit{(bottom)} dashed lines are only meant to guide the eye. 
  \textbf{Right Panel:} \textit{(top)} The fraction of X-rays intercepted and reprocessed in the outer disc ($\cal C$) as a function of time, calculated using: the $\dot{M}_{\rm in}$ computed from the X-ray data and the $R_{\rm in}$ interpolated from the Garcia,Wang-Ji et al. estimates using X-ray reflection spectroscopy, and the output of the Marcel et al. analysis. Coloured shaded regions show the $1\sigma$ confidence interval on $\cal C$, computed by taking into account uncertainty in X-ray flux, distance, black hole mass, binary mass ratio, inner disc radius, and V-band magnitude. \textit{(middle)} The irradiation temperature at the outer disc radius, $T_{\rm irr}(R_{\rm out})$, as a function of time. See Section \ref{sec:time_c} for details. Coloured shaded regions represent the $1\sigma$ confidence interval on $T_{\rm irr}(R_{\rm out})$. 
  (\textit{bottom}) The inner radius of the optically-emitting portion of the disc ($R_{\rm opt,in}$) as a function of time. $R_{\rm opt,in}$ is calculated using the two different prescriptions for $\cal C$ (see above) and considering a 3$\sigma$ deviation in $\cal C$ (see Section \ref{sec:time_c} for details).
  Note that optical V-band contributions from the disc bright spot and jet have been corrected for when computing $\cal C$$(t)$, $T_{\rm irr}(R_{\rm out})$, and $R_{\rm opt,in}(t)$. See Section \ref{sec:emission_mechs} for discussion.
  }%
  \label{fig:lc_outburst1}%
\end{figure*}

\begin{figure}
  \center
  \includegraphics[width=0.95\linewidth,height=0.6\linewidth]{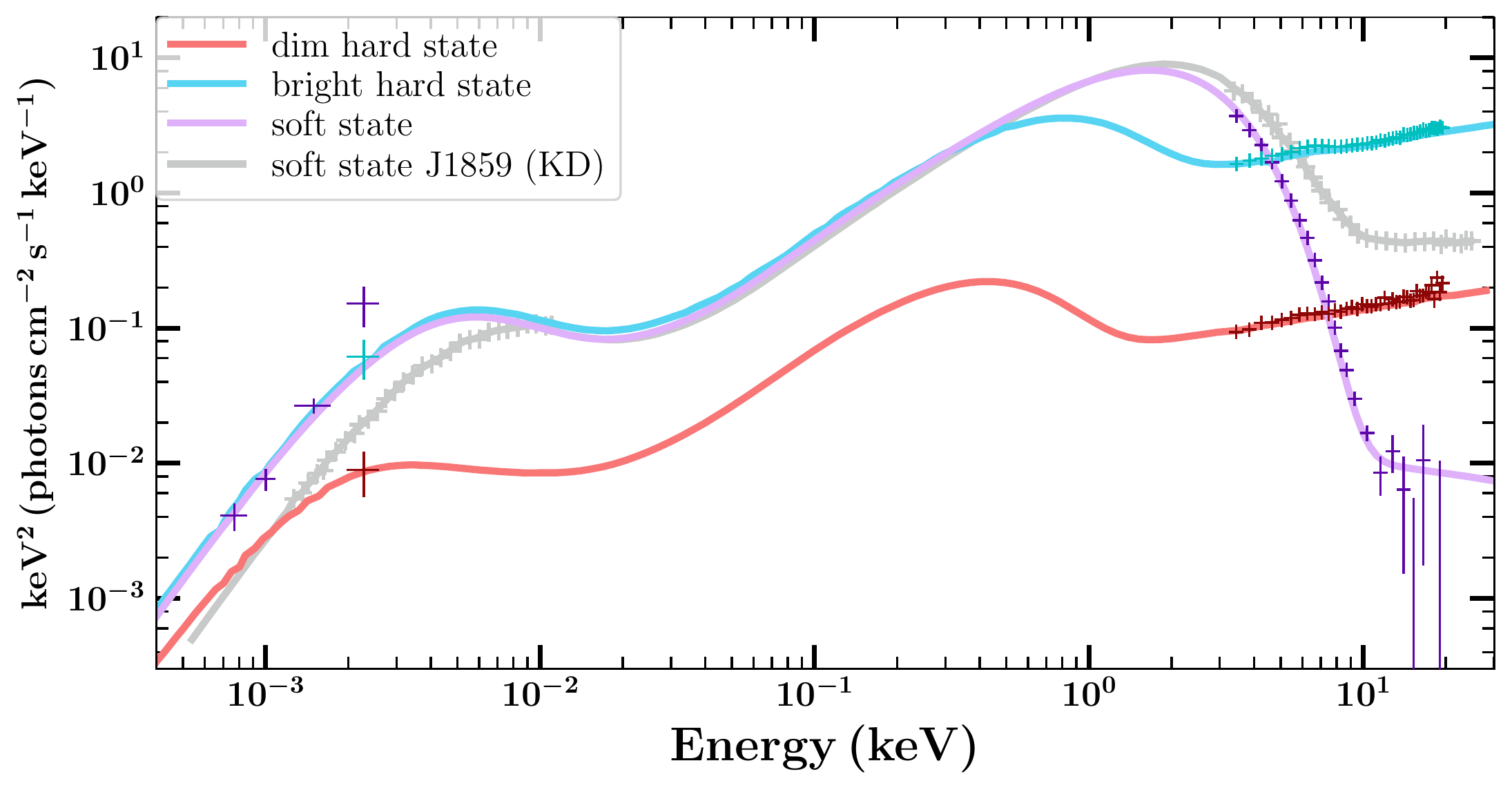}
  \caption{Example SEDs, for selected days near the peak (blue; bright hard-state - MJD 55282), in the soft-state plateau (purple; MJD 55464), and during the decay (red; dim hard-state - MJD 55610), of the 2009-2011 outburst. The available SMARTS/ANDICAM data, and 3-40 keV RXTE/PCA X-ray spectral data \citep{clavel2016}, are shown for the soft-state observation. To compare our method for computing irradiation properties to the broadband SED fitting of \citealt{kimura2019} (see discussion in Section \ref{discuss}), we: (i) plot the soft-state SED (i.e., data-set T3) of XTEJ1859+226, fit with the {\sc optxrplir} irradiated disc model of \citealt{kimura2019} (KD; grey), and (ii) the {\sc optxrplir} fit to the soft-state observation of GX339$-$4 (purple). The {\sc optxrplir} fit to the (jet corrected) V-band SMARTS/ANDICAM data, and 3-40 keV RXTE/PCA X-ray spectral data, used to compute the irradiation properties with our method,  are shown for the two hard-state observations (blue and red).}%
  \label{fig:hs_seds}%
\end{figure}

\section{Results}\label{results}

\subsection{The Time Series Evolution of the X-ray Irradiating Source in GX339$-$4}\label{sec:time_c}

We have applied the methodology, briefly summarized in Section \ref{sec:model_disc0} and thoroughly described in Appendix \ref{sec:model_disc}, to the X-ray (RXTE/PCA, Swift/XRT, and MAXI/GSC) and jet/bright spot corrected optical (SMARTS/ANDICAM) data available for GX339$-$4 (see Section \ref{sec:data} and Table \ref{tab:outinfo}).
In doing so, we have derived: (i) how the fraction of X-ray intercepted and reprocessed in the outer disc evolves with time, $\cal{C}$$(t)$, (ii) how the temperature of the irradiation at the outer disc radius evolves with time, $T_{\rm irr}(R_{\rm out},t)$, and (iii) placed constraints on the evolution of the inner radius of the optically-emitting portion of the disc, $R_{\rm opt,in}(t)$, throughout 9 individual outburst cycles.

$\cal C$$(t)$ is computed, for each time $t_i$ during which simultaneous X-ray and optical data are available, by starting with $\dot{M}_{\rm in}(t)$ (computed from the X-ray light-curve; see Section \ref{sec:xraydata}), then varying $\cal C$ until the observed V-band magnitude is obtained. 
The confidence interval on $\cal C$$(t)$ is propagated by taking into account errors in X-ray flux, $D$, $M_1$, $q$, $R_{\rm in}(t)$, and V-band magnitude (which itself is a combination of instrument uncertainty and error on interstellar reddening). Note that the V-band lightcurve used here, and thus the derived $\cal C$, has been corrected for both optical contributions from the jet and the disc bright spot (see Section \ref{sec:emission_mechs} for details). $T_{\rm irr}(t,R_{\rm out})$ is computed using $\cal C$$(t)$ and $\dot{M}_{\rm in}(t)$ in Equation \ref{eq:temp_irr_eq}. 

To compute $R_{\rm opt,in}(t)$, we start by computing $\dot{M}_{\rm in}-M_V$ relationships for a range of constant $R_{\rm in}$ between $R_g$ and $R_{\rm disc}\sim4\times10^{5}R_g$. Then for each $t_i$ during which simultaneous X-ray and optical data are available, we truncate the disc in increments of $R_g$, and compute $\cal{C}$$(t_i,R/R_g)$ in each case.
When the deviation between $\cal{C}$$(t_i,R_g)$ and $\cal{C}$$(t_i,R)$ reaches a particular threshold, we take this to indicate that the radius $R$ is within the optical emitting region of the disc. This procedure is performed for a threshold of $3\sigma$ for each outburst. The $R_{\rm opt,in}(t)$ we compute with this method can be thought of
  as a conservative upper limit on the inner radius of the outer, irradiated portion of the disc responsible for the optical emission.

Figure \ref{fig:lc_outburst1} displays $\cal{C}$, $T_{\rm irr}(R_{\rm out})$, and $R_{\rm opt,in}$ derived using two different prescriptions for the evolution of $\dot{M}_{\rm in}$ and $R_{\rm in}$ over the 2009$-$2011 outburst of GX339$-$4: (i) the observed $\dot{M}_{\rm in}(t)$ computed from the X-ray data and $R_{\rm in}(t)$ interpolated from X-ray reflection spectroscopy results \citep{garcia2015,wangji2018}, and (ii) the $\dot{M}_{\rm in}(t),R_{\rm in}(t)$ derived in \citet{marcel2018a,marcel2018b,marcel2019}. See Section \ref{sec:xraydata} and \ref{sec:inner_disc} for a detailed discussion on each prescription. The results for the remaining 8 outbursts in our sample can be found in Figures \ref{fig:lc_outburst2}$-$\ref{fig:lc_outburst5} in Appendix \ref{sec:xray_app}.

Note that Marcel et al. assume a black hole mass for GX339$-$4 of $M_1=5.8M_{\odot}$ and does not take into account accretion efficiency when computing $\dot{M}_{\rm in}$. To directly compare to the $\dot{M}_{\rm in}$ computed from the observational X-ray data and the $R_{\rm in}$ interpolated from the Garcia, Wang-Ji et al. estimates using X-ray reflection spectroscopy, we scale the ($\dot{M}_{\rm in},R_{\rm in}$) results from Marcel et al. to a black hole mass of $M_1=N(7.8,1.2)M_{\odot}$, and apply an accretion efficiency ($\eta$) as defined in Section \ref{sec:xraydata}.

\section{Discussion}\label{discuss}

The time-series evolution of the fraction of X-rays intercepted and reprocessed in the outer disc, $\cal C$, that we have derived from the observed X-ray and optical light curves,
varies in a complex way during the 9 outbursts of GX339$-$4 in our sample. However, the values of $\cal C$ do not exceed $\approx 10^{-2}$, in agreement with previous rough estimates of the reprocessed fraction based on the optical to X-ray ratio.  This confirms that using breaks in the X-ray lightcurve to constrain $\cal C$ is much less reliable and can lead to unphysical values \citep{tetarenko2018b}.

\begin{figure}
  \center
  \includegraphics[width=1.\linewidth,height=0.6\linewidth]{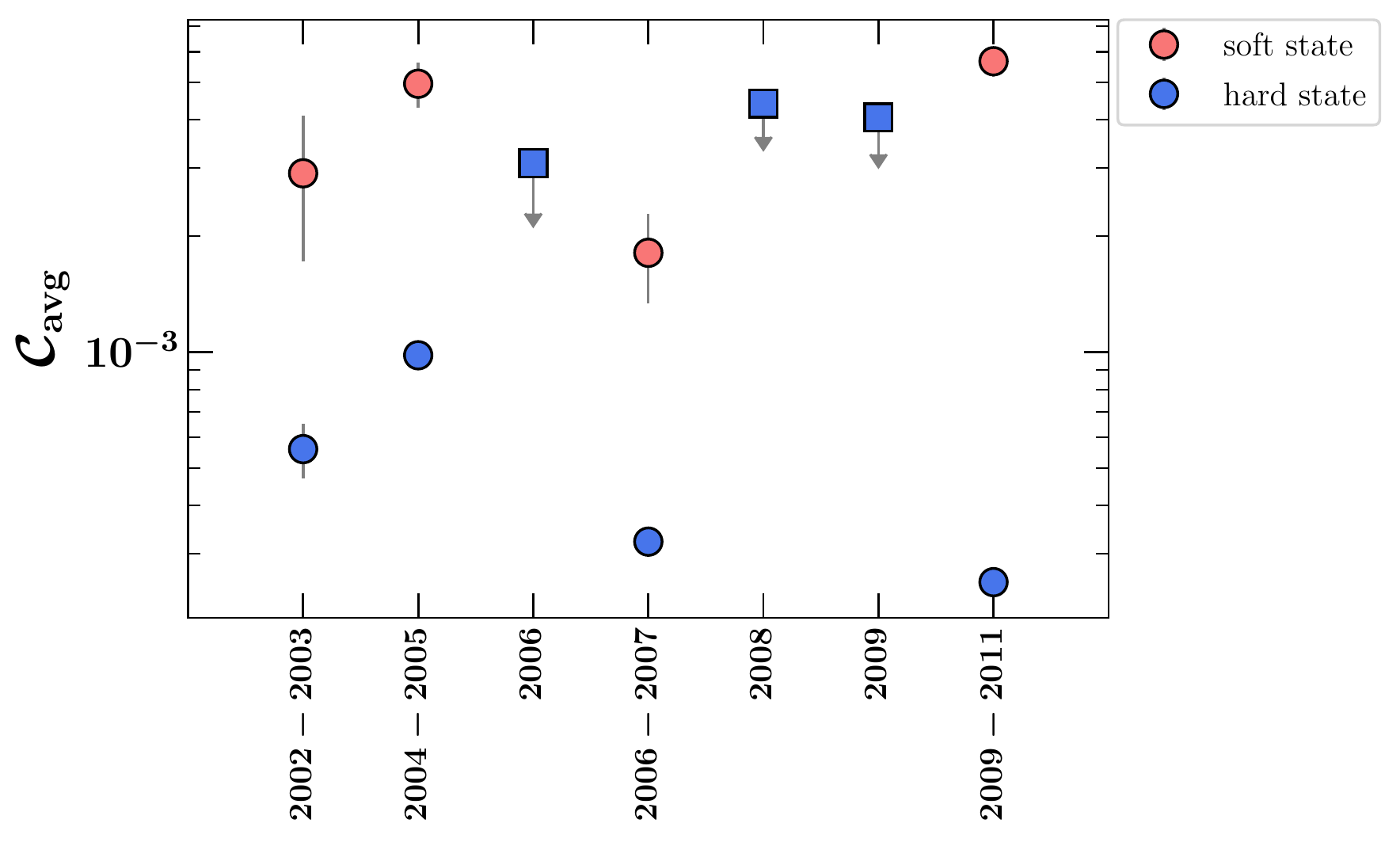}
  \caption{Comparison of the average value of $\cal C$, computed using the Marcel et al. analysis (see Table \ref{tab:Cirrinfo}), during the hard and soft accretion states of our GX339-4 outburst sample. Marker shape specifies outburst classification: canonical (circles) and failed (squares). Hard state $\cal C$ estimates during failed outbursts are only considered as upper limits, as the optical outburst light curves cannot be corrected for jet contribution (see Section \ref{discuss} for details).}%
  \label{fig:cavg_lc}%
\end{figure}

First, we find that the value of $\cal C$ in the soft accretion state (${\cal C}_{\rm soft}$) tends to be higher than $\cal C$ in the hard accretion state (${\cal C}_{\rm hard}$), at least during ``canonical'' outbursts (consistent with results from other BH-LMXB sources; see discussion below and Figure \ref{fig:cavg_lc}). However, we caution that the difference is sensitive to the OIR contribution from the jet and/or cyclosynchrotron emission from the hot flow itself. ${\cal C}_{\rm hard}$ during ``failed'' outbursts, when no jet contribution can be estimated due to the lack of soft state, is typically comparable to the ${\cal C}_{\rm soft}$ ``canonical'' outburst values. Thus, the larger values of ${\cal C}_{\rm hard}$ during ``failed'' outbursts should only be considered upper limits and are likely a consequence of not being able to correct the light curves for an optical jet contribution.  For comparison, ${\cal C}_{\rm hard}$ rises by a factor 10 to $\approx 2\times 10^{-3}$ for the 2009$-$2011 outburst if the optical jet contribution is not removed.

Second, we find some difference when comparing the values of ${\cal C}$ derived using the Marcel et al. and Garcia, Wang et al. prescriptions (e.g., see Table \ref{tab:Cirrinfo}). Of the seven outbursts for which we have estimated $\cal C$ using both prescriptions, typically the  ${\cal C}$ derived using Marcel et al. are smaller, when compared to the Garcia, Wang et al. prescription results for the same outburst. This difference is entirely the result of their different estimates for $\dot{M}_{\rm in}$. Their very different 
assumptions for $R_{\rm in}$ have no effect whatsoever for this as the innermost radius of the disc which contributes substantially to the optical emission, $R_{\rm opt,in}$ (see Figure \ref{fig:lc_outburst1} and Appendix \ref{sec:xray_app})
is around $10^5 R_g$, much larger than even the largest estimate of the truncation of the thin disc ($R_{\rm in}\ll 10^3 R_g$) in Marcel et al..

Third, we observe peaks/drops in $\cal C$ coinciding with both hard-soft and soft-hard state transitions. Given that $R_{\rm in}\ll R_{\rm opt,in}$ throughout all outbursts in our sample (as discussed above), it is clear that these abrupt changes in  $\cal C$ are not simply a consequence of $R_{\rm in}$ changing suddenly during the outburst.

Evidence for this type of behaviour has been (i) to an extent, observed previously in BH-LMXBs XTEJ1817$-$330 and XTEJ1859+226, and (ii) recently predicted by \citet{dubus2019}, who consider the impact a thermally-driven wind would have on BH-LMXB light-curves, in the context of the disc-instability model.  
By fitting an irradiated disc model to broadband SEDs observed throughout the 2006 outburst of XTE J1817$-$330, \cite{gierlinski2009} found that while soft state observations were consistent with a near constant fraction of bolometric X-ray luminosity being reprocessed in the outer disc, the reprocessed fraction increased by a factor $\sim6$ as the source transitioned into the hard state. These authors suggest that their observations favour direct illumination of the outer disc by the central X-ray source, and explain the apparent increase in reprocessed fraction as the source transitions to the hard state as a consequence of a change in disc albedo during the state transition, though they did not model this in detail. Note, however, that this study did not take into account the jet contribution in the optical as we do in this paper.

Similarly, through broadband SED modelling during the 1999$-$2000 outburst of XTEJ1859+226, 
\cite{kimura2019} find evidence for a decrease in the reprocessed fraction as the source dims from the soft state towards (but not quite reaching) the hard state.  Unlike \cite{gierlinski2009}, these authors did calculate the expected reprocessed fraction from direct illumination, and found that it was smaller than required. 
They suggested another source of irradiation of the outer disc, in addition to direct illumination.  \cite{kimura2019} consider the idea, originally suggested by \cite{begelman1983}, that the disc could also be irradiated by a corona/wind existing above the disc, by effectively scattering a portion of the central X-ray luminosity back onto the outer disc. They estimated the  strength of the wind irradiation as being similar to that required by the data. 
Figure \ref{fig:hs_seds} shows a comparison of 
this SED fitting method in XTEJ1859+226 with some of our data for GX339-4. We fit the soft state SED with the 
 {\sc optxrplir} model for black hole mass of $7.8M_\odot$, distance of $8$~kpc, $R_{out}=10^{5.4}R_g$ 
 and inclination of $30^\circ$ with 
 ${\cal C}\approx 2\times 10^{-3}$, similar to that derived for XTEJ1859+226 and rather smaller than the 
 value ${\cal C}\approx 7\times 10^{-3}$ derived on the same date by our method with the Marcel et al. analysis (see Figure \ref{fig:lc_outburst1}). However, their exists a number of additional biases that may be affecting our determination of $\cal C$ (up to a factor 3 to 5), comparable to the systematic uncertainty in our method (see Appendix A), most importantly from inclination. Changing the inclination to $60^\circ$ (as assumed by using the angle average disc flux) increases ${\cal C}\approx 10^{-2}$. 
 
Figure \ref{fig:hs_seds} also shows that the 
soft state V band flux is somewhat higher than the {\sc optxrplir} model fit to the H,J,I data. This could indicate that the reddening correction is overestimated as even a single temperature blackbody underestimates the V band flux. Our model fits only to the V band flux, so this (along with inclination) would also lead to a larger ${\cal C}$. Conversely, the V band data from both the bright and dim low/hard states are well fit by ${\cal C}\approx 2\times 10^{-3}$ for inclination of $30^\circ$, which is the origin of the decrease in ${\cal C}$ for the low/hard state. 

We note also that the {\sc optxrplir} model takes into account 
disc colour corrections due to electron scattering in the disc photosphere. This becomes important when the material is ionised i.e. for UV and X-ray local temperatures, so does not affect the optical spectrum so should not affect the calculation of $\cal{C}$, but leads to a shift in the spectrum above 10~eV
 (see Fig. 3 of \citealt{kimura2019}).

It is also clear from Figure \ref{fig:hs_seds} that the amount of irradiation is similar in GX339$-$4 and XTE J1859+226. However, the disc outer radius is larger in GX339$-$4, which works to extend the region over which irradiation dominates in the disc. XTE J1859+226 has not been re-scaled as the distance estimate used is also 8kpc. It has slightly smaller best fit mass of $6.9M_\odot$ which is why this spectrum has higher $L/L_{\rm Edd}\sim 0.2$ than the $0.1$ for the soft state of GX339$-$4 shown in Figure \ref{fig:hs_seds}.

\citet{dubus2019} used the analytic estimates of scattering in a thermal-radiative wind to predict the time-series evolution of $\cal C$, for a number of model BH-LMXB lightcurves with a range of orbital periods. They find complex variability in $\cal C$ throughout the outbursts, as the wind responds to changes in luminosity and spectral shape. However, there is always a sudden drop (resp. rise) in $\cal C$ when going from the hard to soft state (resp. soft to hard). This is due to the sudden change in spectral shape, which makes a sudden change in Compton temperature, so a sudden change in launch radius of the wind (see \citealt{done2018}).

Given the observed (i) complex profiles in $\cal C$, (ii) correlations between $\cal C$ and accretion state, (iii) variations in $\cal C$ (sometimes up orders of magnitude) on timescales of days to weeks, and (iv) the typically large derived values for truncation radius of the optically emitting part of the disc ($R_{\rm opt,in}\gtrsim10^5 R_g$),
we first consider a scattering origin for the X-ray irradiation in GX339$-$4.

\subsection{Irradiation Heating via a Thermally Driven Wind}\label{sec:th_wind}

We make use of the thermally-driven (Compton heated) wind prescription from \cite{done2018}. These authors have combined analytical \citep{begelman1983} and numerical \citep{woods1996} thermal wind models to predict disc wind observables (e.g., column density, ionization state, mass loss rates, wind launching radii, and velocity) as a function of changing spectral shape and luminosity ($L_{\rm X}$) throughout outburst. We briefly present the basic idea behind this (Compton heated) thermal wind model below and refer the reader to \cite{begelman1983}, \cite{woods1996}, and \cite{done2018} for further details on the model and to \cite{kimura2019} for an example of application of the model to BH-LMXB XTEJ1859+226.

\subsubsection{Deriving Observational Properties of the Wind Throughout an Outburst Cycle}

During outburst, the surface of the disc is heated to the Compton temperature ($T_{\rm IC}$), 
which only depends on the X-ray irradiating spectrum. This X-ray irradiation results in the formation of a corona above the disc. The scale height of this corona is controlled by the ratio of sound speed ($c_{\rm IC}^{2}=kT_{\rm IC}/\mu$; where $\mu$ is mean particle mass) of the gas, to the escape velocity ($v_{\rm esc}\sim GM/R$), in the disc. 
If $c_{\rm IC}\geq v_{\rm esc}$ the gas will escape in a wind, at a launch radius defined by the Compton radius,
\begin{equation}
R_{\rm IC}=\frac{GM_1}{c_{\rm IC}^2}\approx 10^{12} \left(\frac{M_1}{10 M_\odot}\right) \left(\frac{10^7\rm\,K}{T_{\rm IC}}\right)\rm\,cm,
\end{equation}
where $M_1$ is the black hole mass and $T_{\rm IC}$ is the Compton temperature of the impinging irradiation.
Otherwise the material will form a type of static corona above the disc \citep{begelman1983}.

Unlike the static corona, the wind is expanding. Thus, the rate at which the material in the wind region of the disc ($R>R_{\rm IC}$) is heated, and thus the condition for which such a wind is launched, will depend both on $L_{\rm X}$ and the irradiating spectrum. The boundary condition between this thermal wind and the corona atmosphere above the disc can be shown to follow \citep{woods1996,done2018},
\[ R_{\rm launch}=\begin{cases} 
      0.2R_{\rm IC}& L>L_{\rm crit} \\
      0.2\left( \frac{L}{L_{\rm crit}} \right)^{-1}R_{\rm IC} & L<L_{\rm crit}, 
   \end{cases}
\]
where,
\begin{equation}
L_{\rm crit}\approx 0.09 \ \left(\frac{10^7\rm\,K}{T_{\rm IC}}\right)^{1/2} L_{\rm Edd},
\end{equation}
is the critical luminosity which is enough to launch the wind at $R_{\rm IC}$ (i.e., the luminosity that will heat the gas to a temperature $kT_{\rm IC}$ as it reaches a scale height $\sim R$, thus allowing the material to escape).

In principle, the condition for launching such a wind is not only dependent on $L_{\rm X}$ and the irradiating spectral shape, but also on the underlying irradiation geometry (defined via the $\cal C$ parameter). \cite{begelman1983} found that this Compton wind, in addition to acting as a mechanism for which mass can be removed from the system, may also act as an effective medium to scatter some of the central X-ray luminosity onto the disc, thus providing a viable irradiation geometry.
The irradiation geometry must allow the X-ray flux to irradiate the outer disc. Simple prescriptions, using a radial profile for disc height (e.g., \citealt{kim1999,dubus1999}), show that point source irradiation alone is insufficient to irradiate the outer disc regions. In this situation, the cooler outer region has a smaller scale height than the hotter, inner region, thus is shadowed from the central X-ray source. Therefore, X-ray scattering in a wind above the disc is an attractive solution.

The fraction of intrinsic X-ray flux scattered in this Compton heated wind ($C_{\rm wind}$) can be derived by integrating over the wind column density predicted by \cite{done2018}, yielding \citep{dubus2019},
\begin{equation}
{C_{\rm wind}}=\int_0^1 \int_{R_{\rm in}}^{R_{\rm out}} \eta \sigma_T n_{\rm w} (1-\mu) d\mu dr \approx \frac{\eta \sigma_T \dot{M}_{\rm w}}{8\pi R_{\rm in} v_{\rm w} m_I},
\label{eq:estC}
\end{equation}
with $\eta$ is accretion efficiency (as defined in Section \ref{sec:xraydata}), $n_{\rm w}$ the wind density, $v_{\rm w}$ the mass-weighted wind outflow rate, $m_I$ is the mean ion mass per electron and $\mu=\cos i$, where $i$ is binary inclination (see Equation 6 of \citealt{done2018} for details).

We have applied this thermal wind prescription to GX339$-$4 using the: (i) bolometric luminosity as a function of time, $L_{\rm bol}(t)$, obtained by applying a distance estimate $D$ to the 3$-$200 keV flux estimated from fitting all available RXTE/PCA spectra (see Section \ref{sec:xrayspectra}), (ii) outer disc radius ($R_{\rm out}$) computed using the defined set of binary orbital parameters (see Section \ref{sec:discsize} and Table \ref{tab:binary_params}), and (iii) Compton temperature as a function of time, $T_{\rm IC}(t)$, computed from all available RXTE/PCA spectra (see Section \ref{sec:xrayspectra}), for the outbursts in our sample occurring between 2002$-$2012.

\begin{figure*}
\begin{minipage}[t]{0.63\textwidth}
\mbox{}\\[-\baselineskip]
  \includegraphics[width=\textwidth,height=1.7\linewidth]{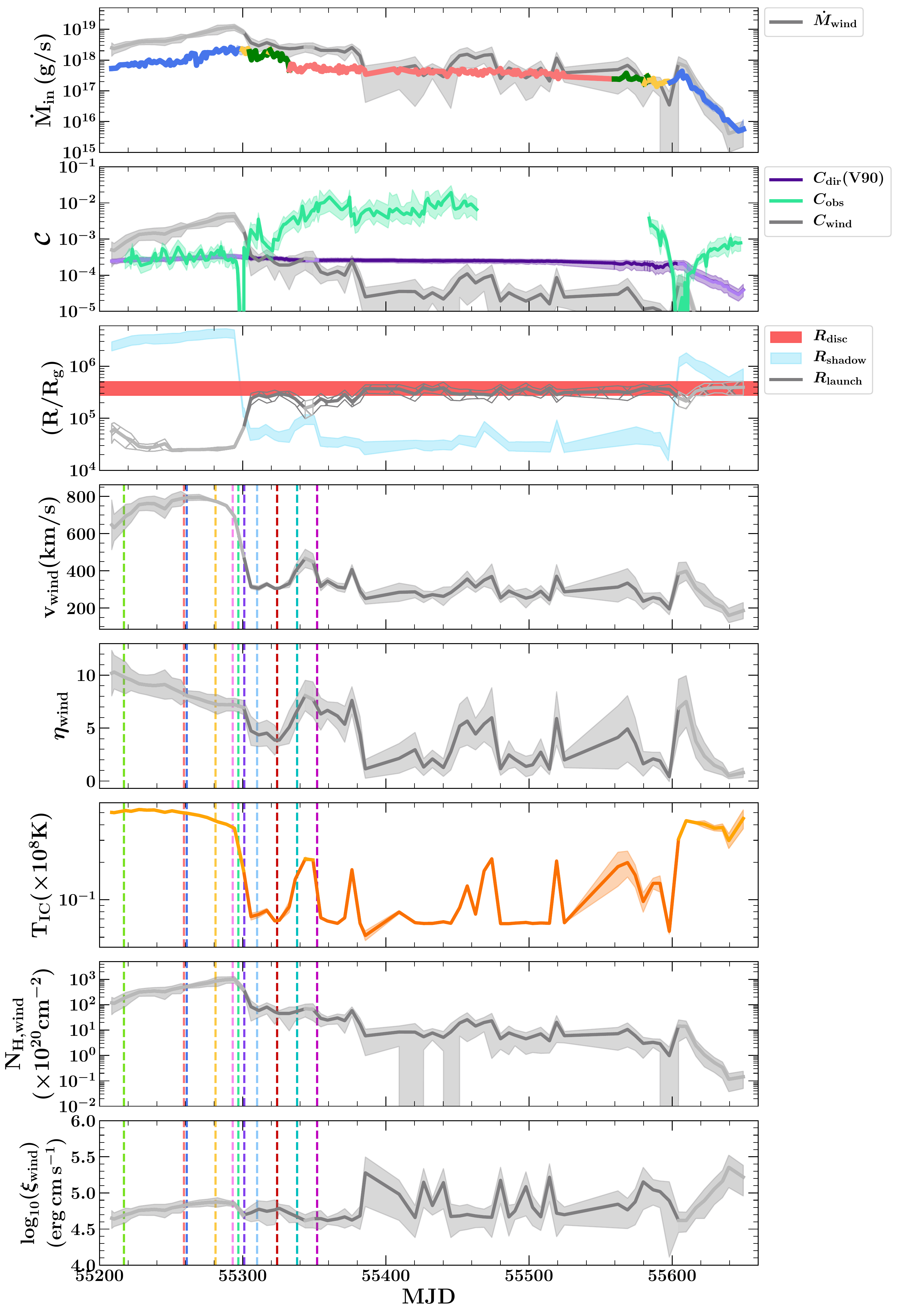}
\end{minipage}\hfill
  \begin{minipage}[t]{0.35\textwidth}
  \mbox{}\\[-\baselineskip]
  \caption{Derived thermally-driven (Compton-heated) wind properties for the 2009$-$2011 outburst of GX339$-$4. 
  \textit{(top)} Compares mass loss rate in the wind ($\dot{M}_{\rm wind}$; grey line), to the central mass-accretion rate ($\dot{M}_{\rm in}$; multi-coloured line) onto the black hole. The shaded coloured regions here represent the uncertainty in $\dot{M}_{\rm in}$ (see \citealt{marcel2019} for detailed discussion).
  This panel is colour coded by accretion state as defined by Marcel et al.: red (soft state), yellow (intermediate state; referred to as hard-intermediate state in Marcel et al.), blue (hard state), and green (soft-intermediate state).
  \textit{(second)} Compares the fraction of X-rays intercepted and reprocessed in the outer disc ($\cal C$) as a function of time, computed from the observed X-ray and optical lightcurves (see Section \ref{sec:model_disc0} and Appendix \ref{sec:model_disc}; green line), via scattering in the wind (see Section \ref{sec:th_wind}; grey line), and via a direct central source of irradiation (\citealt{vrtilek1990} (V90); purple line). See Section \ref{sec:wind_role} for details.   
  \textit{(third)} Displays the launch radii ($R_{\rm launch}$; grey line), and outer disc radii ($R_{\rm disc}$; shaded red region) computed from the chosen set of binary orbital parameters (see Table \ref{tab:binary_params} and Section \ref{sec:discsize}), and the radii at which the shadow cast by the inner attenuation zone ends ($R_{\rm shadow}$; see Section \ref{sec:th_wind} for discussion), as a function of time. 
  The remaining panels display the:
  \textit{(fourth)} wind velocity ($v_{\rm wind}$; in km/s),
  \textit{(fifth)} wind efficiency $\eta_{\rm wind}$ (defined by $\dot{M}_{\rm wind}/\dot{M}_{\rm in}$),
  \textit{(sixth)} Compton temperature ($T_{\rm IC}$; in units of $10^8$ K) computed by integrating over the best-fit RXTE/PCA spectrum (see Section \ref{sec:xrayspectra}),
  \textit{(seventh)} column density of the wind ($N_{\rm H,wind}$; in units of $10^{20} {\rm cm^{-2}}$), and
 \textit{(bottom)} log of the ionization of the wind, $\log_{10}(\xi_{\rm wind})$, as functions of time.
   The shaded coloured regions in all panels represent the $1\sigma$ confidence interval propagated for each parameter. The lighter coloured lines/shaded regions in each panel mark times when the outer (irradiated) disc is in the shadow of the inner attenuation zone (i.e., $R_{\rm shadow}>R_{\rm launch}$).
   The vertical coloured dashed lines in the fourth through bottom panels mark the epochs for which we have simulated X-ray spectra produced by the wind with {\sc xstar}. See Section \ref{sec:xstar_sims} for discussion.
}
  \label{fig:wind_outburst1}%
  \end{minipage}
\end{figure*}

Figure \ref{fig:wind_outburst1} displays the derived thermal wind properties of: (i) mass loss rate ($\dot{M}_{\rm wind}$), (ii) launch radii ($R_{\rm launch}$), (iii) velocity ($v_{\rm wind}$), (iv) efficiency ($\eta_{\rm wind}$), (v) column density ($N_{\rm H,wind}$), (vi) ionization ($\xi_{\rm wind}$), and (vii) fraction of X-rays scattered in the wind ($C_{\rm wind}$), as a function of time during the 2009$-$2011 outburst of GX339$-$4. Note that, for the wind properties shown in this Figure, we assume the system has a randomly orientation towards us, averaging over all inclination angles. Here, we first compare these wind properties to the time-series evolution of central mass accretion rate, $\dot{M}_{\rm in}(t)$, derived by Marcel et al.. Then, we compare these wind properties to the reprocessed X-ray fraction ($\cal C$), derived from the X-ray and optical light curves, using the Marcel et al. derivation of $(\dot{M}_{\rm in},R_{\rm in})$ (see Section \ref{sec:time_c}). These results for the remaining 6 outbursts of GX339$-$4 in our sample considered here, occurring between 2002$-$2012, can be found in Figures \ref{fig:wind_outburst2}$-$\ref{fig:wind_outburst6} in Appendix \ref{sec:wind_app}.

\subsection{The Predicted Hard and Soft State Wind in GX339$-$4}

Simple 
thermal wind models predict that the wind exists in both the hard and soft accretion states during the outbursts of GX339$-$4 considered here. However, typically BHXBs show the X-ray spectral signatures of disc winds only exist the soft state, disappearing as the source transitions into the hard state during outburst (see e.g., \citep{miller2006,ponti2012,neilsen2013,diaztrigo2014}. In the case of thermally-driven winds, 
the absence of observed hard state wind signatures
relates to the complex response of the wind to the changing illuminating spectral shape. Firstly, the wind is launched from closer in when the spectrum hardens, and secondly, it is irradiated by a much harder spectrum. The combination of these two effects means that the wind is predicted to be 
completely ionised, so not detectable as absorption lines in X-ray spectra \citep{chakravorty2013,higginbottom2015,done2018}. This has also been shown in detail in full radiation hydrodynamic simulations as well \citep{tomaru2019a,tomaru2019b}.

\subsubsection{Simulating X-ray Spectra Produced by the Wind During an Outburst Cycle}\label{sec:xstar_sims}

We have performed simulations using the {\sc xstar} photoionization code to determine whether spectral features resulting from our predicted hard and soft state thermal wind could be observable in X-ray spectra. We explicitly make use of the {\sc xstar2xspec} routine, which creates a table model for use in {\sc xspec} by combining multiple {\sc xstar} simulations for a range of input parameters.

For a multitude of epochs, during state transitions in the 4 canonical outburst cycles for which detailed spectral information is available, we run the {\sc xstar2xspec} routine: (i) using the observed best-fit spectral model (see details below) as the input continuum SED, (ii) fixing both the density at the wind launch radius\footnote{While the \cite{done2018} wind model assumes gas density varies as $n(R)=n_0(R/R_{\rm launch})^{-2}$, we assume density remains constant with radius. This choice was made because of well-known convergence issues when the above radial dependence for density is used. See \url{https://heasarc.gsfc.nasa.gov/xstar/docs/html/xstarmanual.html} for details.} ($n_0$; computed via the thermal wind model) and the (0.0136$-$13.6 keV) luminosity ($L_0$; computed from the input continuum SED), (iii) setting the turbulent velocity at $300$ km/s and using solar abundances, and (iv) varying the wind column density ($N_{\rm H,wind}$), ionization ($\xi_{\rm wind}$), and line-of-sight velocity ($v_{\rm wind}$).

\begin{figure}
  \center
\includegraphics[width=1.\linewidth,height=1.5\linewidth]{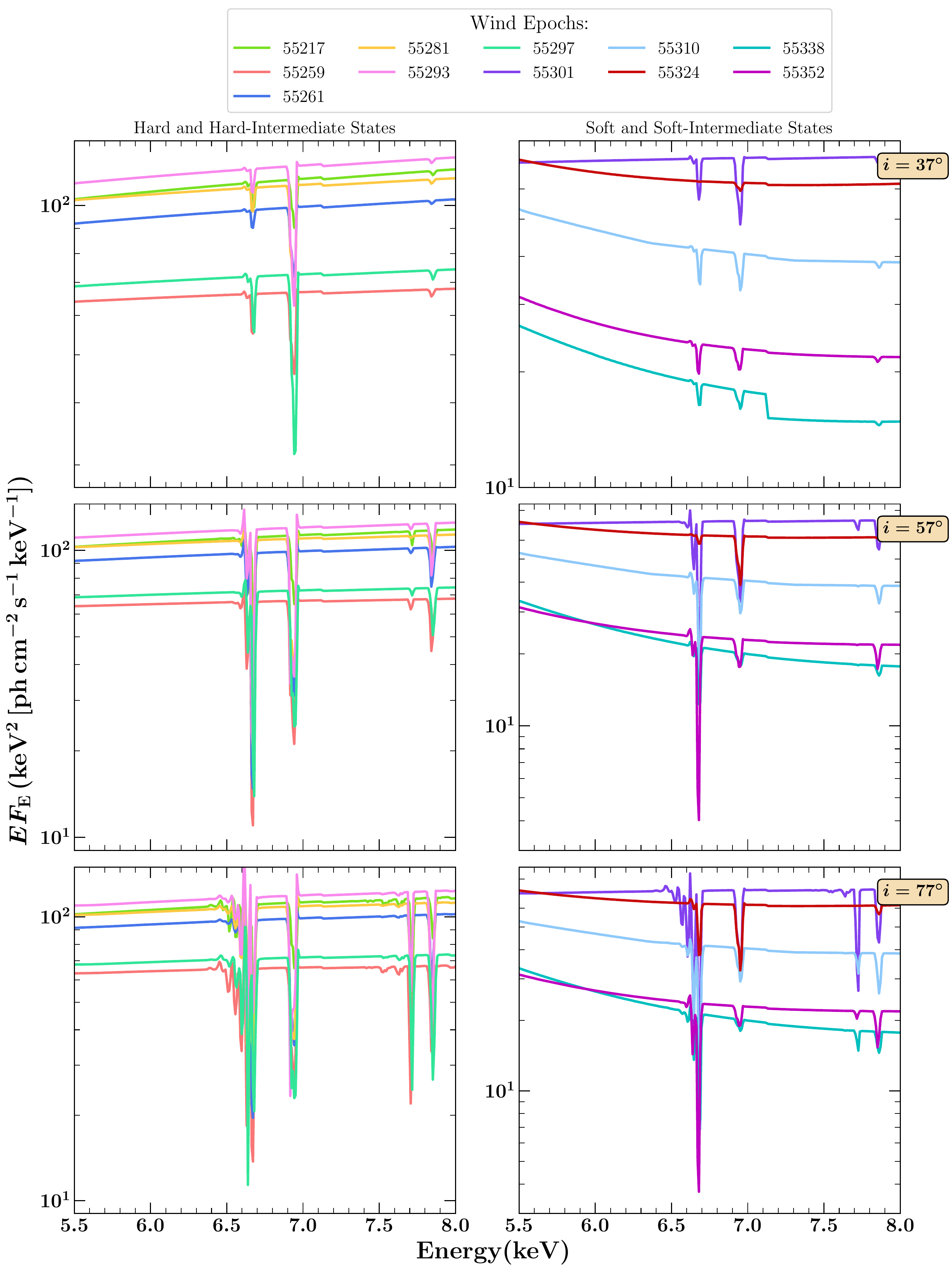}
\caption{Simulated X-ray spectra of the thermal wind via {\sc xstar} (See Section \ref{sec:xstar_sims}) for a range of epochs during the 2009--2011 outburst cycle of GX339$-$4, assuming an inclination of $i=37^{\circ}$\textit{(top)}, $i=57^{\circ}$ \textit{(middle)}, and $i=77^{\circ}$ \textit{(bottom)}. The wind epochs are split by accretion state for clarity.}
  \label{fig:xstar_mods1}%
\end{figure}

Figure \ref{fig:xstar_mods1} shows the resulting simulated X-ray spectra, for 11 individual epochs occurring during the hard-soft state transition of the 2009$-$2011 outburst of GX339$-$4. Each simulated spectra here is produced by combining the table model created from the {\sc xstar2xspec} routine with the: (i) broad-band (X-ray to UV) spectra from Swift/XRT and UVOT (see \citealt{reynolds2013} for details) to characterize the input continuum SED, and (ii) observable wind properties from the \cite{done2018} thermal wind model ($N_{\rm H,wind}$, $\xi_{\rm wind}$, $v_{\rm wind}$), derived during each particular epoch, assuming three different inclination angles in the range of $37^{\circ}<i<78^{\circ}$ (predicted from optical analysis; \citealt{heida2017}).  
Figures \ref{fig:xstar_mods2}--\ref{fig:xstar_mods4} in Appendix \ref{sec:wind_app} show these results for the hard-soft state transitions occurring during the 2002$-$2003, 2004$-$2005, and 2006$-$2007 outbursts of GX339$-$4, respectively. In this case, as broad-band Swift spectral observations are not available, we characterize the input continuum SED using the available RXTE/PCA spectra (see \citealt{clavel2016} for details).

Whether or not it would be possible to observe features from the predicted thermal wind in X-ray spectra is largely dependent on the true binary inclination of GX339$-$4. While a reliable measurement for binary inclination (e.g., via detection of ellipsoidal modulations) does not yet exist, modelling of the reflection component in X-ray spectra has provided a range of contradictory results. A number of studies making use of a combination of XMM-Newton and RXTE data, taken during the hard/intermediate states of the 2002$-$2003, 2004$-$2005, and 2009$-$2011 outbursts, uniformly favour a low inclination of $i\sim10^{\circ}-20^{\circ}$ \citep{miller2004,miller2006,reis2008,plant2015}. Such a low inclination is consistent with both the independent detection of a one-sided jet \citep{gallo2004}, and promising evidence for wind spectral features previously detected in emission during the soft state of the 2004$-$2005 outburst \citep{miller2015}. On the other hand, by: (i) analyzing the same XMM-Newton data (e.g., \citealt{donetrigo2010,basak2016}), (ii) making use of Swift and NuSTAR data taken during the 2009$-$2011, 2013, and 2014$-$2015 outbursts \citep{furst2015,parker2016,wangji2018}, and (iii) performing an independent analysis of available RXTE data in the hard state over a wide range of luminosity \citep{garcia2015}, other authors derive higher inclination estimates between $i\sim30^{\circ}-60^{\circ}$. 

It is important to note that reflection modelling is actually measuring the inclination of the inner disc, which does not have to be the same as the binary inclination\footnote{This situation can happen if, for example, the black hole spin angular-momentum axis is not aligned with the binary angular-momentum vector. Spin-orbit misalignment may be common in BHXBs (see e.g., \citealt{atri2019})}.
Regardless, the predicted absorption features (see Figure \ref{fig:xstar_mods1} and Figures \ref{fig:xstar_mods2}--\ref{fig:xstar_mods4} in Appendix \ref{sec:wind_app}) are not observed in existing hard or soft state X-ray spectra of GX339$-$4. The typical equivalent widths (EWs) of these predicted features in our simulated spectra  are EW$\gtrsim15$ eV, for the He-like FeXXV (6.7 keV) and H-like FeXXVI (6.97 keV) absorption lines, and thus would be detectable by current missions (e.g., Chandra; see \citealt{miller2006b}). Thus, we favour a low inclination of $<37^{\circ}$. If this is the case, it is unlikely that the predicted (hard and soft state) thermal wind would ever be observable (at least in absorption) in X-ray spectra of this source. 

\subsection{The Role the Wind Plays During Outburst Cycles in GX339$-$4}\label{sec:wind_role}

Taking a time-average over the 2002$-$2012 period in which our outburst sample covers, we estimate only $\sim 25\%$ of the transferred mass escapes in this thermal wind in GX339$-$4. This low wind mass-loss rate is consistent with the low $\alpha$-viscosity parameters ($\alpha\sim0.2$; \citealt{tetarenko2018}) derived from, and expected (from predictions of the DIM) decay timescales seen in, the observed X-ray light-curves of GX339$-$4.

While this thermal wind, predicted to be present in both the hard and soft accretion states, may not be a dominant mechanism for mass loss in this system, it does play an important role in the accretion process in terms of irradiating the disc.
Using this wind as a medium in which to scatter X-rays back onto the disc, we are able to reproduce some of the features, independently predicted to be present (e.g., \citealt{kimura2019, dubus2019}), in our computed ${\cal C}(t)$ outburst profiles (see Section \ref{sec:time_c}). 
Such features include the peaks/drops in ${\cal C}$ occurring during accretion state transitions and complex (day-week timescale) variability. 
However, irradiation via scattering in such a thermal wind alone cannot fully account for the ${\cal C}$ required to reproduce
the observed X-ray and optical light-curves throughout entire outburst cycles of GX339$-$4.

\subsubsection{The Hard State Wind}\label{sec:hs_wind}
First, we find wind-driven irradiation tends to over-predict the scattering required in the bright hard states of GX339$-$4.
Multiple authors \citep{begelman1983,tomaru2019a,tomaru2019b} have suggested the possibility that a thermal wind could be intrinsically suppressed in the hard state by the larger shadow cast by the heated inner atmosphere (corona) over the inner disc. Thus, we have investigated the effect that inner corona attenuation would have on our results.

\cite{begelman1983} were able to show that it is relatively easy for the static corona to become optically thick in the radial direction. Here an inner attenuation zone forms, with the ability to cast a shadow over the disc surface, strongly affecting the illumination pattern in the outer disc, and in-turn the thermal wind properties \citep{tomaru2019a,tomaru2019b}. 
The outer disc can only be illuminated when the disc scale height increases enough that it rises above the shadow zone. For an irradiated, isothermal disc (i.e., $H(R)\propto R^{9/7}$, assuming $1/3$ of the flux thermalizes in the disc; \citealt{cunningham1976,vrtilek1990}), the radii at which the shadow zone ends can be written as \citep{tomaru2019a},
\begin{equation}
R_{\rm shadow}=3\times10^{7} \left(\frac{T_{\rm IC}}{10^8 K}\right)^{7/8} \left(\frac{M_1}{M_{\odot}} \right)^{1/2} \left(\frac{L}{L_{\rm edd}} \right)^{3/8} \Xi_{\rm h,min}^{-7/8} R_g.
\end{equation}
In Figure \ref{fig:wind_outburst1} (as well as the figures in Appendix \ref{sec:wind_app}), we compare the time-evolution of $R_{\rm shadow}$ to the outer disc radius. Given that $R_{\rm shadow}$ is typically $\gtrsim R_{\rm disc}$ during the hard states of GX339$-$4, there is considerable uncertainty over the properties of any hard state wind we derive here.

\subsubsection{The Soft State Wind}\label{sec:ss_wind}
Second, we find wind-driven irradiation tends to under-predict the scattering required in the soft states of GX339$-$4. Thus, we have considered the possibility that a hybrid source of irradiation could exist in this system. Such a configuration would involve irradiation occurring via a combination of (i) scattering in a thermal wind (${\cal C}_{\rm wind}$), and (ii) direct irradiation from a central source of the form \citep{king1997,dubus1999},
\begin{equation}
    {\cal C}_{\rm dir}=\eta (1-a)\frac{H}{R}\left( \frac{d\ln H}{d \ln R}-1 \right),
    \label{eq:c_direct}
\end{equation}
where $\eta$ is accretion efficiency, and $a$ and $H(R)$ are the X-ray albedo and scale height of the disc as a function of radius, respectively. 
To demonstrate this possibility, in the second panel of Figure \ref{fig:wind_outburst1} (as well as Figures \ref{fig:wind_outburst2}$-$\ref{fig:wind_outburst6} in Appendix \ref{sec:wind_app}), we also plot the theoretical estimate of ${\cal C}_{\rm dir}$ computed for a isothermal disc \citep{cunningham1976}, in a LMXB system (\citealt{vrtilek1990}; V90).
However, we caution that this direct irradiation is highly sensitive to the evolution of the disc shape during the outburst. The outer disc is easily shadowed, or the strength of irradiation diminished by the convex shape of the disc, when self-consistent calculations are carried out  \citep{meyer1982,cannizzo1995,dubus1999,kim1999}.

In doing so, we find that direct irradiation is insufficient to account for the measured $\cal C$ when the disc enters the soft state and the thermal wind is not dense enough to scatter enough light (Figure \ref{fig:wind_outburst1}). This remains puzzling. One possibility is that our thermal wind model underestimates the wind density in this state. We consider this unlikely given the good agreement generally found between the analytic estimates and more elaborate numerical simulations (see Section \ref{sec:intro}). Other possibilities include: overestimating the optical contribution from the disc (Section \ref{sec:emission_mechs}), leading to a higher effective $\cal C$ because of contributions from the hotspot, companion or jet (although not in soft state), or a change in albedo or $H/R$ as the disc responds differently to the soft X-ray spectrum.

Given our results, we favour two possible explanations: we are overestimating $\cal C$  in this source due to seeing it at a low inclination, which also means we underestimate the thermal wind as gravitational redshift reduces the Compton temperature in the observed spectrum compared to that seen by the disc \citep{munozdarias2013}. Alternatively, there may be an additional source of scattering onto the disc from a magnetic wind. This 
would need to be completely ionised in order to circumvent the constraints on the wind features discussed above.

\section{Summary}\label{conclusion}

While X-ray irradiation of the accretion disc is known to play a key role in regulating the outburst cycles of BHXB systems, how, and to what degree, the discs in these binary systems are irradiated remains largely unknown. The light curve profiles of BHXB outbursts encode within them distinct observational signatures of the irradiation source heating the disc in the system \citep{king1998,dubus2001,tetarenko2018b}. Accordingly, we have developed a methodology that makes use of a combination of X-ray and optical light-curves to track the evolution of physical properties of the X-ray irradiation source heating the discs in these binary systems.

By applying this methodology to $\sim15$ yrs of outburst activity in GX339$-$4, we are able to derive the evolution of the (i) fraction of the X-ray flux that is intercepted and reprocessed in the outer disc, $\cal C$, (ii) irradiation temperature at the outer disc radius, $T_{\rm irr}(R_{\rm out})$,
over 9 individual outburst cycles. 
In doing so, we find the profiles of ${\cal C}(t)$ and $T_{\rm irr}(R_{\rm out},t)$ throughout individual cycles. These time-series evolutions contain significant variability on timescales of days to weeks, along with distinct temporal features including, most notably, peaks/drops in $\cal C$ and $T_{\rm irr}$ that coincide with hard-soft state transitions.

We have first considered a scattering origin for the X-ray irradiation in GX339$-$4. The (i) observed complex outburst profiles in $\cal C$ and $T_{\rm irr}$, (ii) typically large derived values of $R_{\rm opt,in}\gtrsim10^5 R_g$, and (iii) the fact that the distinct temporal features in the BHXB outburst $\cal C$ profile have previously been associated with irradiation via a thermally-driven disc wind (e.g., \citealt{dubus2019}), are all suggestive that a disc wind may play a role in irradiating the disc in this system.

Making use of the thermally-driven (Compton-heated) wind model of \cite{done2018}, we have: (i) predicted the time-series evolution of key observational properties of this wind, namely, mass loss rate, launch radii, velocity, column density, ionization, and fraction of X-rays scattered in the wind ($C_{\rm wind}$), and (ii) simulated the X-ray spectra produced by such a wind using the {\sc xstar} photoionization code, for a multitude of epochs during individual state transitions, throughout seven individual outburst cycles.

Contrary to X-ray spectral observations of BHXBs, which typically show disc wind signatures present only in the soft accretion state \citep{miller2006,ponti2012,neilsen2013,diaztrigo2014}), the simple analytic approximation of \cite{done2018} predicts that the wind exists in both the hard and soft accretion states. We 
make detailed photo-ionisation models and find that both hard and soft state outflows predict features in the X-ray spectra, in conflict with the observations, unless 
the source inclination is lower than $\sim 40^\circ$. 
A low inclination is suggested by both continuum fitting (the very low disc temperature: \citep{munozdarias2013}) and X-ray reflection spectroscopy (see e.g., \citealt{miller2004,miller2006,reis2008,plant2015}).

Our findings also suggest this hard and soft state wind is not an efficient mechanism to remove significant amount of mass from the system. In fact, we estimate typically only $\sim25$\% of the transferred mass (from companion star to disc) is lost via such a wind. Nonetheless, the wind 
may still play an important role as 
a mechanism for irradiation heating.

Overall, we find that the strength of irradiation ($\cal C$) required to
 account for the observed X-ray and optical light curves of GX339$-$4 
 is higher than what a combination of thermal wind and direct irradiation can explain. X-ray to infrared global spectral fits of irradiated disc models to individual snapshot observations taking into account all the photometric information can provide a more accurate evaluation of {\cal C}, and help disentangle the disc and jet contributions, at single points during an outburst. Using this SED fitting method, has allowed us to evaluate additional sources of uncertainty in our method for deriving {\cal C}. For example, Figure \ref{fig:hs_seds} shows such an SED fit to a soft state observation: taking into account colour corrections, and allowing for a lower extinction, can reduce the value of {\cal C} by a factor 3 to 4. Even then, neither X-rays scattered in the thermal wind, nor direct X-rays from a central source (when a thermal wind is not dense enough to scatter enough light), can fully explain the magnitude of $\cal C$ during this soft state. 

This is puzzling, as full radiation-hydrodynamic simulations of thermal-radiative winds have been extremely successful in matching to the observed absorption features in high inclination, large disc systems
such as the BHXB H1743$-$322 \citep{tomaru2019a,tomaru2019b}. It may be that there are substantial differences between the approximate analytic models for thermal winds used here, and the results of full radiation hydrodynamic simulations. Alternatively, it could also be that we underestimate the thermal wind in this source due to seeing it at a low inclination.
We will explore these effects in a later work.  
Tailored radiation-hydrodynamic simulations to GX339$-$4 will enable us to fully assess the scattered flux (via a thermal-radiative wind) illuminating the disc in the system. Thus, allowing us to determine whether scattering and direct illumination in a thermal-radiative wind can really produce the observed heating of the outer disc, or whether additional mechanisms, such as scattering in a magnetically-driven outflow, need to be considered as well.

\section*{Acknowledgements}
The authors would like to thank the anonymous referee for their insightful comments that improved the quality of this manuscript.
BET thanks Mark Reynolds for providing his comprehensive broadband Swift spectral results, and Tolga Dincer for providing additional SMARTS data between 2013$-$2015, for GX339$-$4. BET also thanks Jon Miller for his assistance with, and helpful discussion regarding, the {\sc xstar} simulations and X-ray reflection modelling.  
BET acknowledges support from the University of Michigan through the McLaughlin Fellowship. GD and MC acknowledge support from the Centre National d'Etudes Spatiales (CNES). This research has made use of (i) data, software, and/or web tools obtained from the High Energy Astrophysics Science Archive Research Center (HEASARC), a service of the Astrophysics Science Division at NASA Goddard Space Flight Center (GSFC) and of the Smithsonian Astrophysical Observatory's High Energy Astrophysics Division, (ii) data supplied by the UK Swift Science Data Centre at the University of Leicester, (iii) data from the SMARTS 1.3m telescope, (iv) the SVO Filter Profile Service (\url{http://svo2.cab.inta-csic.es/theory/fps/}), supported by the Spanish MINECO through grant AyA2014-55216, and (v) NASA's Astrophysics Data System (ADS).

\begin{landscape}
\begin{table}
	\centering
	\caption{Correlation Between X-ray Irradiation, Accretion State, and Outburst Phase}
	\medskip
	\label{tab:Cirrinfo}
	\begin{tabular}{lcccccccc} 
		\hline
		Outburst ID&Average $\cal C$&\multicolumn{4}{c}{\underline{Average $\cal C$ per Accretion State}} & \multicolumn{3}{c}{\underline{Average $\cal C$ per Outburst Phase}} \\[0.3cm]
		&&Soft State&Hard State&(Hard-)Intermediate State&Soft-Intermediate State&
		Rise&Plateau&Decay\\
		\hline
		\multicolumn{9}{c}{Prescription 1: $\cal C$ derived using $\dot{M}_{\rm in}$ computed from observed X-ray lightcurves and $R_{\rm in}$ estimated from X-ray reflection spectroscopy}\\
		\hline
        
2002-2003 &$(1.3\pm0.1)\times10^{-3}$&$(2.7\pm0.5)\times10^{-3}$&$(1.9\pm0.2)\times10^{-3}$&$(8.0\pm1.1)\times10^{-4}$&$\cdots$&
        $(3.3\pm0.7)\times10^{-3}$&$(9.5\pm1.0)\times10^{-4}$&$(2.0\pm0.2)\times10^{-3}$\\[0.07cm]
        
        2004-2005 &$(2.4\pm0.1)\times10^{-3}$&$(5.4\pm0.3)\times10^{-3}$&$(1.9\pm0.1)\times10^{-3}$&$(5.4\pm0.7)\times10^{-3}$&$\cdots$&
        $(4.1\pm0.4)\times10^{-2}$&$(2.1\pm0.1)\times10^{-3}$&$(3.1\pm0.3)\times10^{-3}$\\[0.07cm]
        
        2006 &$(1.4\pm0.1)\times10^{-2}$&$\cdots$&$(1.4\pm0.1)\times10^{-2}$&$\cdots$&$\cdots$&
        $\cdots$&$\cdots$&$(1.4\pm0.1)\times10^{-2}$\\[0.07cm]
        
        2006-2007 &$(1.5\pm0.1)\times10^{-3}$&$(1.2\pm0.1)\times10^{-3}$&$(1.7\pm0.1)\times10^{-3}$&$(1.0\pm0.5)\times10^{-3}$&$\cdots$&
        $(2.5\pm0.6)\times10^{-3}$&$(1.3\pm0.1)\times10^{-3}$&$(3.3\pm0.3)\times10^{-3}$\\[0.07cm]
        
        2008 &$(1.1\pm0.1)\times10^{-2}$&$\cdots$&$(1.1\pm0.1)\times10^{-2}$&$\cdots$&$\cdots$&
        $(9.6\pm1.2)\times10^{-3}$&$\cdots$&$(1.3\pm0.1)\times10^{-2}$\\[0.07cm]
        
        2009 &$(2.3\pm0.2)\times10^{-2}$&$\cdots$&$(2.3\pm0.2)\times10^{-2}$&$\cdots$&$\cdots$&
        $(2.4\pm0.5)\times10^{-2}$&$\cdots$&$(2.3\pm0.2)\times10^{-2}$\\[0.07cm]
        
        2009-2011 &$(5.9\pm0.2)\times10^{-4}$&$(9.6\pm0.8)\times10^{-4}$&$(5.5\pm0.2)\times10^{-4}$&$(6.8\pm1.6)\times10^{-4}$&$\cdots$&
        $(6.2\pm0.4)\times10^{-4}$&$(5.6\pm0.5)\times10^{-4}$&$(5.9\pm0.4)\times10^{-4}$\\[0.07cm]
        
        2013&$(1.3\pm0.1)\times10^{-2}$&$\cdots$&$(1.3\pm0.1)\times10^{-2}$&$\cdots$&$\cdots$&
        $(1.7\pm0.3)\times10^{-2}$&$\cdots$&$(1.1\pm0.2)\times10^{-2}$\\[0.07cm]
        
        2014-2015 &$(1.9\pm0.2)\times10^{-4}$&$(1.9\pm0.2)\times10^{-3}$&$(1.4\pm0.2)\times10^{-4}$&$(1.3\pm0.1)\times10^{-3}$&$\cdots$&
        $\cdots$&$(1.8\pm0.2)\times10^{-4}$&$(4.0\pm0.8)\times10^{-4}$\\[0.07cm]
		\hline
		\multicolumn{9}{c}{Prescription 2: $\cal C$ derived using ($\dot{M}_{\rm in}$, $R_{\rm in}$) computed by Marcel et al.)}\\
		\hline
        2002-2003 &$(6.6\pm0.5)\times10^{-4}$&$(2.9\pm1.2)\times10^{-3}$&$(5.6\pm0.9)\times10^{-4}$&$(6.4\pm1.3)\times10^{-4}$&$(1.8\pm0.4)\times10^{-3}$&
        $(1.6\pm0.2)\times10^{-3}$&$(5.2\pm0.6)\times10^{-4}$&$(7.5\pm1.0)\times10^{-4}$\\[0.07cm]
        
        2004-2005 &$(1.2\pm0.05)\times10^{-3}$&$(5.0\pm0.6)\times10^{-3}$&$(9.8\pm0.6)\times10^{-4}$&$(1.8\pm0.4)\times10^{-3}$&$(2.8\pm0.4)\times10^{-3}$&
        $(1.9\pm0.2)\times10^{-3}$&$(1.1\pm0.05)\times10^{-3}$&$(1.4\pm0.1)\times10^{-3}$\\[0.07cm]
        
        2006 &$(3.4\pm0.2)\times10^{-3}$&$\cdots$&$(3.4\pm0.2)\times10^{-3}$&$\cdots$&$\cdots$&
        $\cdots$&$\cdots$&$(3.4\pm0.2)\times10^{-3}$\\[0.07cm]
        
        2006-2007 &$(4.3\pm0.2)\times10^{-4}$&$(1.8\pm0.5)\times10^{-3}$&$(3.2\pm0.3)\times10^{-4}$&$(6.3\pm1.5)\times10^{-4}$&$(1.2\pm0.1)\times10^{-3}$&
        $(1.1\pm0.2)\times10^{-3}$&$(3.9\pm0.2)\times10^{-4}$&$(5.9\pm0.6)\times10^{-4}$\\[0.07cm]
        
        2008 &$(4.5\pm0.4)\times10^{-3}$&$\cdots$&$(4.5\pm0.4)\times10^{-3}$&$\cdots$&$\cdots$&
        $(4.4\pm0.5)\times10^{-3}$&$\cdots$&$(5.2\pm0.6)\times10^{-3}$\\[0.07cm]
        
        2009 &$(2.7\pm0.1)\times10^{-3}$&$\cdots$&$(2.7\pm0.1)\times10^{-3}$&$\cdots$&$\cdots$&
        $(1.4\pm0.2)\times10^{-3}$&$\cdots$&$(3.9\pm0.3)\times10^{-3}$\\[0.07cm]
        
        2009-2011 &$(2.5\pm0.1)\times10^{-4}$&$(5.7\pm0.5)\times10^{-3}$&$(2.5\pm0.1)\times10^{-4}$&$(6.2\pm1.1)\times10^{-4}$&$(1.4\pm0.1)\times10^{-3}$&
        $(2.8\pm0.2)\times10^{-4}$&$(7.0\pm0.6)\times10^{-4}$&$(1.9\pm0.1)\times10^{-4}$\\[0.07cm]

\hline
\multicolumn{9}{p{0.9\columnwidth}}{\hangindent=1ex NOTE. -- The time-averaged $\cal C$ is computed here using an iterative weighted mean technique (see \citealt{tetarenko2016} for details on this statistical method). This technique allows one to take into account the asymmetric ($1\sigma$) uncertainties in $\cal C$.} 
	\end{tabular}
\end{table}

\end{landscape}




\bibliographystyle{mnras}
\bibliography{refs_GX339.bib}




\appendix

\section{The Relationship Between $\dot{M}$ and Absolute Magnitude for Irradiated Accretion discs}\label{sec:model_disc}

This relationship depends on: (i) the size of the thin disc in the system, (ii) how temperature varies with radius in this disc, $T_{\rm eff}(R)$, and (iii) the optical bandpass which one observes the system in.

\subsection{Disc Size}\label{sec:discsize}

The size of a BH-LMXB disc, at any point during outburst, can be characterized by its inner ($R_{\rm in}$) and outer ($R_{\rm out}$) radius. The inner disc radius ($R_{\rm in}$) will evolve inward/outward (``truncation'') as a BH-LMXB system transitions through the typical pattern of accretion states during outburst \citep{done2007,done2010}. Thus, to compute $R_{\rm in}$, an assumption on how the inner disc radius varies as a function of central mass accretion rate, $R_{\rm in}(\dot{M}_{\rm in}$), is required (see Section \ref{sec:inner_disc} for details).

Realistically, outer disc radius ($R_{\rm out}$) must increase in outburst because of the angular momentum transported outwards. Typically this increase is expected to be a factor of $\sim2-3$ at most (see e.g., \citealt{dubus2001}). 
We instead approximate $R_{\rm out}$ as being constant during outburst. Given that we expect other sources of uncertainty, relevant to real BH-LMXB systems (e.g., distance, inclination, extinction), to dominate here, we believe the uncertainty caused by this assumption will have no significant effect on our results. To compute $R_{\rm out}$, we follow the procedure of \cite{tetarenko2018}. They have established a hierarchical method to estimate $R_{\rm out}$ by sampling this quantity from a uniform distribution between the circularization radius ($R_{\rm circ}$), and the radius of the black hole Roche lobe ($R_1$) in the system. Using this method, an estimate of $R_{\rm out}$ can be obtained with prior knowledge of black-hole mass ($M_1$), binary mass ratio ($q$) and orbital period ($P_{\rm orb}$).

\subsection{Disc Temperature Profile}

Realistically, as a BH-LMXB disc cycles through periods of quiescence and outburst, the way in which temperature varies with radius, 
$T_{\rm eff} (R)$, changes as a result of (heating and cooling) fronts propagating throughout the disc (see \citealt{dubus2001}). This change ultimately causes a type of hysteresis between optical magnitude and $\dot{M}_{\rm in}$.

For simplicity, we instead assume a steady-state disc during outburst. In this case $\dot{M}$ does not vary with radius. Thus, the evolution of the disc can be thought of as a succession of steady states, where $\dot{M}=\dot{M}_{\rm in}$. Under the steady-state assumption, the temperature profile of a BH-LMXB disc can be written as a combination of viscous \citep{frank2002},
\begin{equation}
T_{\rm visc}^4=\frac{3G M_1 \dot{M}}{8 \pi \sigma R^3}\left[ 1-\left(\frac{R_{\rm in}}{R}\right)^{1/2} \right],
\end{equation}
and irradiated \citep{dubus1999},
\begin{equation}
T_{\rm irr}^4={\cal C} \frac{\dot{M} c^2}{4 \pi \sigma R^2},
\label{eq:temp_irr_eq}
\end{equation}
portions such that,
\begin{equation}
T_{\rm eff}^4(R)=T_{\rm visc}^4(R)+T_{\rm irr}^4(R).
\label{eq:temp}
\end{equation}

\subsection{Conversion of Disc Emission to Observed Flux and Absolute Magnitude}\label{sec:convert_mdot}
To convert from disc emission to an optical flux, as observed in a particular band-pass, we adapt the procedure developed by \cite{dubus2018}, for irradiated accretion discs. The optical flux in a bandpass $Y$ can be written as,
\begin{equation}
F_Y=\frac{\cos i}{D^2}\int_{R_{\rm in}}^{R_{\rm out}} 2 \pi R \omega_Y(T_{\rm eff})\sigma T_{\rm eff}^4 dR,
\label{eq:disc_flux}
\end{equation}
where $i$ is the binary inclination, $D$ is the distance to the system, $\omega_Y$ is,
\begin{equation}
\omega_Y=\frac{1}{\sigma T_{\rm eff}^4}\frac{\int f_Y B_{\nu} d\nu}{\int f_Y d\nu},
\end{equation}
and $f_Y$ is the filter transmission curve. The absolute magnitude, in band-pass $Y$, can be computed via,
\begin{equation}
M_Y=-2.5 \log_{10}\left[ \left(\frac{D}{10\, \rm{pc}} \right) \left( \frac{F_Y}{Z_Y} \right) \right],
\end{equation}
where $Z_Y$ is the filter zero point in the Vega system.

In this way, the absolute magnitude, in a particular bandpass $Y$ ($M_Y$), depends only on $\dot{M}_{\rm in}$, $M_1$, $R_{\rm in}$, $R_{\rm out}$, and $\cal C$. Thus, through numerical integration of Equation \ref{eq:disc_flux}, taking $T_{\rm eff}(R)$ in the functional form given in Equation \ref{eq:temp}, it is possible to reconstruct $\cal C$ from an optical light curve, given the estimate of $\dot{M}_{\rm in}$ derived from a simultaneous X-ray light curve.

Note that in this paper, we make exclusive use of the OIR filters used with SMARTS/ANDICAM (see Section \ref{sec:oirdata}). The wavelength range, effective wavelength, zero point, and transmission curves for these filters have been obtained from the SVO filter service\footnote{\url{http://svo2.cab.inta-csic.es/theory/fps/}}.

\subsection{Application to Synthetic Model Lightcurves}\label{apply_model_lcs}

Before applying the above methodology to observational data, we first tested it on synthetic model light-curves. This is necessary to verify the validity of, and quantify the uncertainty produced by, use of the steady state disc assumption. Model $\dot{M}_{\rm in}$ and optical (V and I band) lightcurves were computed using the numerical scheme of \cite{hameury1998}, adapted to include irradiation heating \citep{dubus1999} and inner disc evaporation \citep{menou2000}.
The parameters of black hole mass ($4M_{\odot}<M_1<15M_{\odot}$), disc size ($3\times10^{10}\,\rm{cm}<R_{\rm out}<1\times10^{12}\,\rm{cm}$), $\alpha$-viscosity ($0.1<\alpha_h<1.0$),

\begin{figure}
  \center
  \includegraphics[width=\columnwidth,scale=0.5]{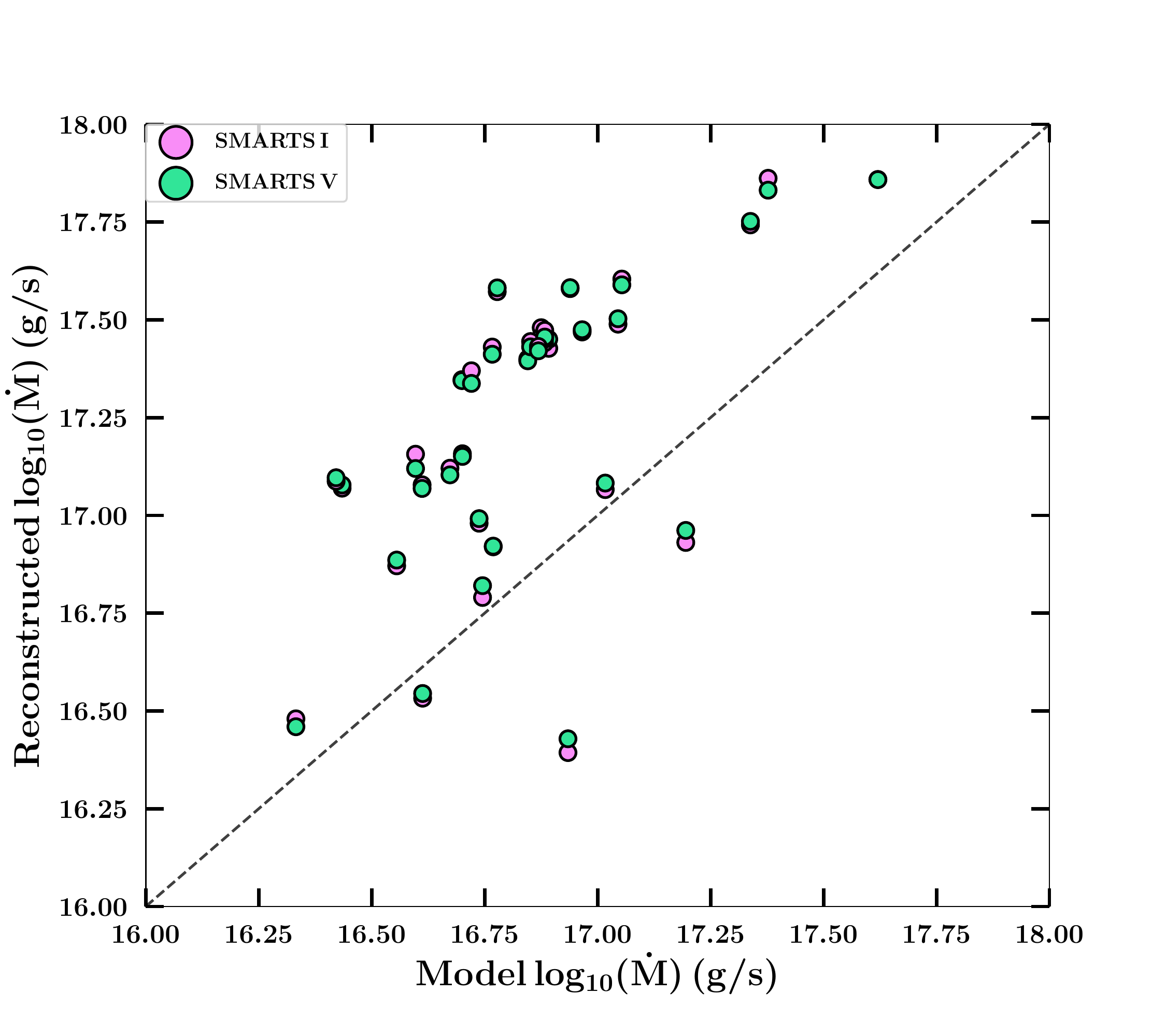}
  \caption{Correlation plot comparing the model $\dot{M}_{\rm in}$, averaged over a full outburst cycle, to the time-averaged reconstructed $\dot{M}_{\rm in}$ in both the V (green circles) and I (pink circles) SMARTS/ANDICAM filters, for our set of 46 disc models. The black dotted line represents the 1-to-1 line on the plot.}%
  \label{fig:mdot_compare}%
\end{figure}

\noindent and reprocessed X-ray fraction ($0.005< {\cal C} < 0.1$)\footnote{Note, using ${\cal C}$ smaller than a few $10^{-3}$ causes the light curves to behave like that of a non-irradiated disc (see e.g., \citealt{dubus2001,dubus2019}).}, were varied to produce a set of 46 individual model lightcurves covering the parameter space valid for BH-LMXB systems.

For each model, we (i) reconstruct $\dot{M}_{\rm in}$ from the V and I band model light curves using the model parameters: $M_1$,$R_{\rm out}$, and $\cal C$, and assuming they were ``observed'' with SMARTS/ANDICAM, (ii) use our methodology (Sections \ref{sec:discsize}--\ref{sec:convert_mdot}) along with the model optical and $\dot{M}_{\rm in}$ light curves to reconstruct $\cal C$ throughout a model outburst, (iii) analyze the relationship between $\dot{M}_{\rm in}$ and $M_V$, and $\dot{M}_{\rm in}$ and $M_I$, comparing the steady-state assumption to the hysteretic pattern of disc evolution computed from the code, and (iv) compute a time-averaged $\cal C$, and time-averaged reconstructed $\dot{M}_{\rm in}$, to compare to the model $\cal C$ and model $\dot{M}_{\rm in}$ averaged over the entire outburst cycle.

\begin{figure}
  \center
  \includegraphics[width=\columnwidth,scale=0.5]{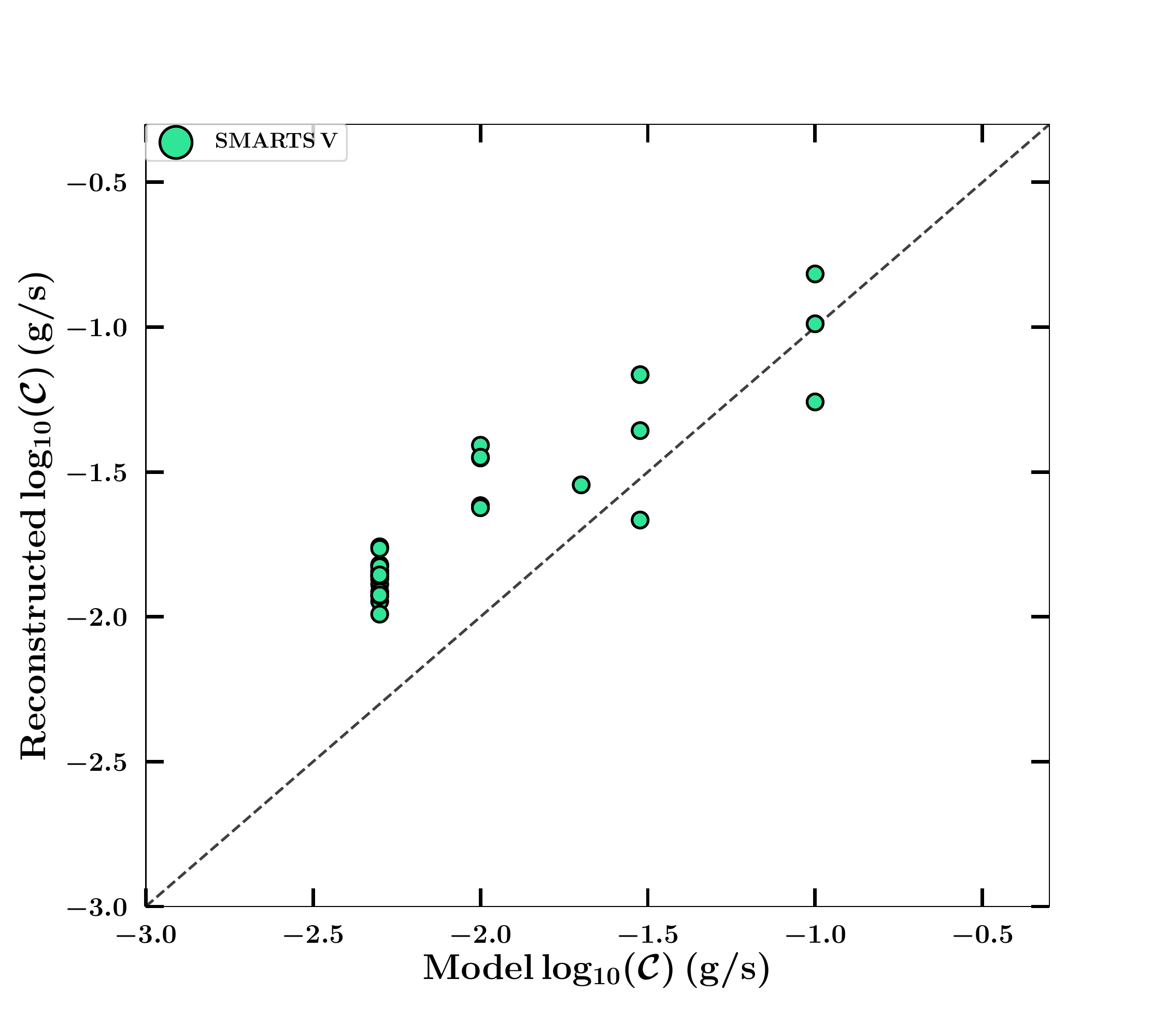}
  \caption{Correlation plot comparing the model $\cal C$ to the reconstructed $\cal C$ (using the V-band light-curve) averaged over a full outburst cycle, for our set of 46 disc models. The black dotted line represents the 1-to-1 line on the plot.}%
  \label{fig:c_compare}%
\end{figure}

\begin{figure*}
  \center
  \includegraphics[width=0.92\linewidth,height=.6\linewidth]{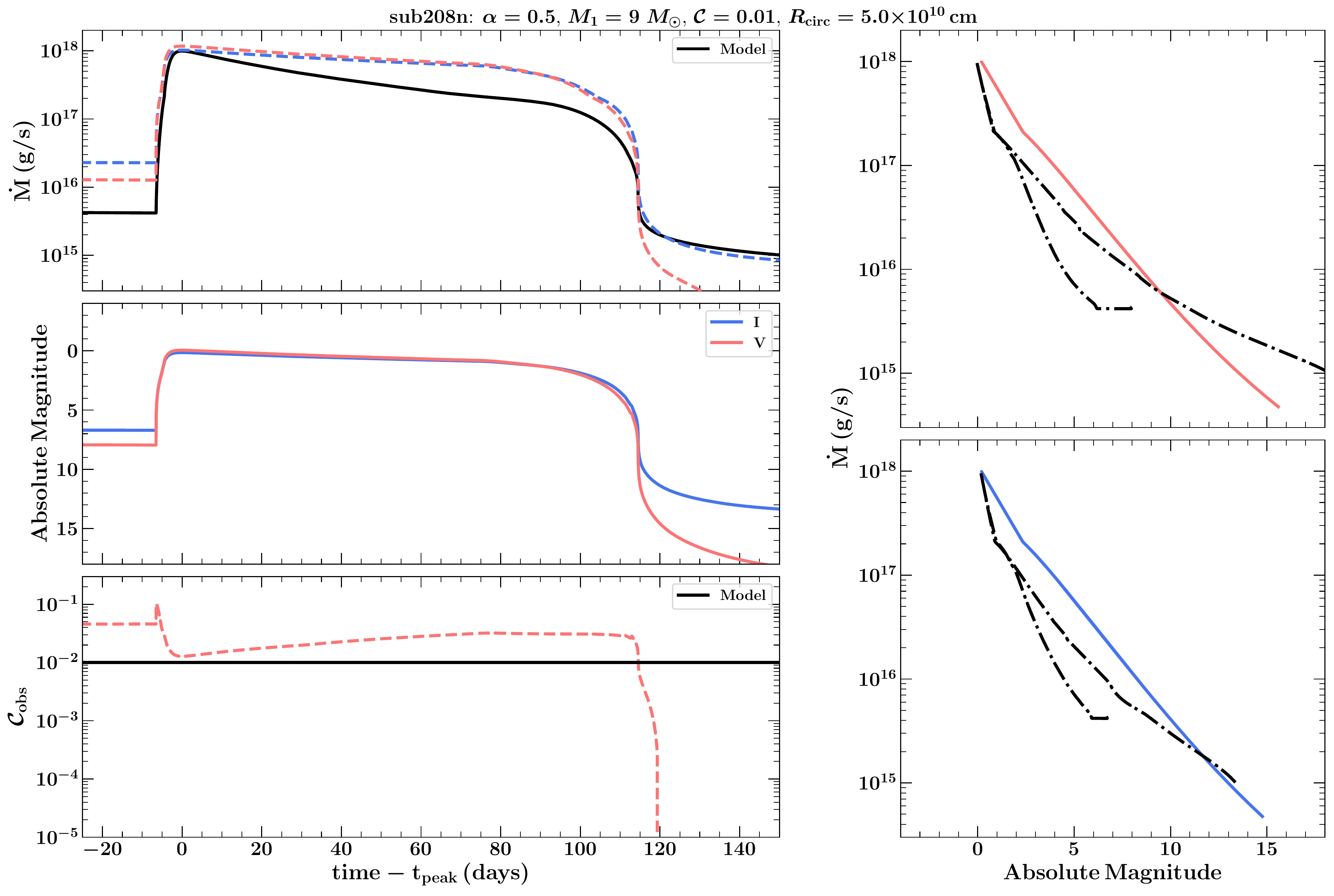}\hfill
  \includegraphics[width=.92\linewidth,height=.6\linewidth]{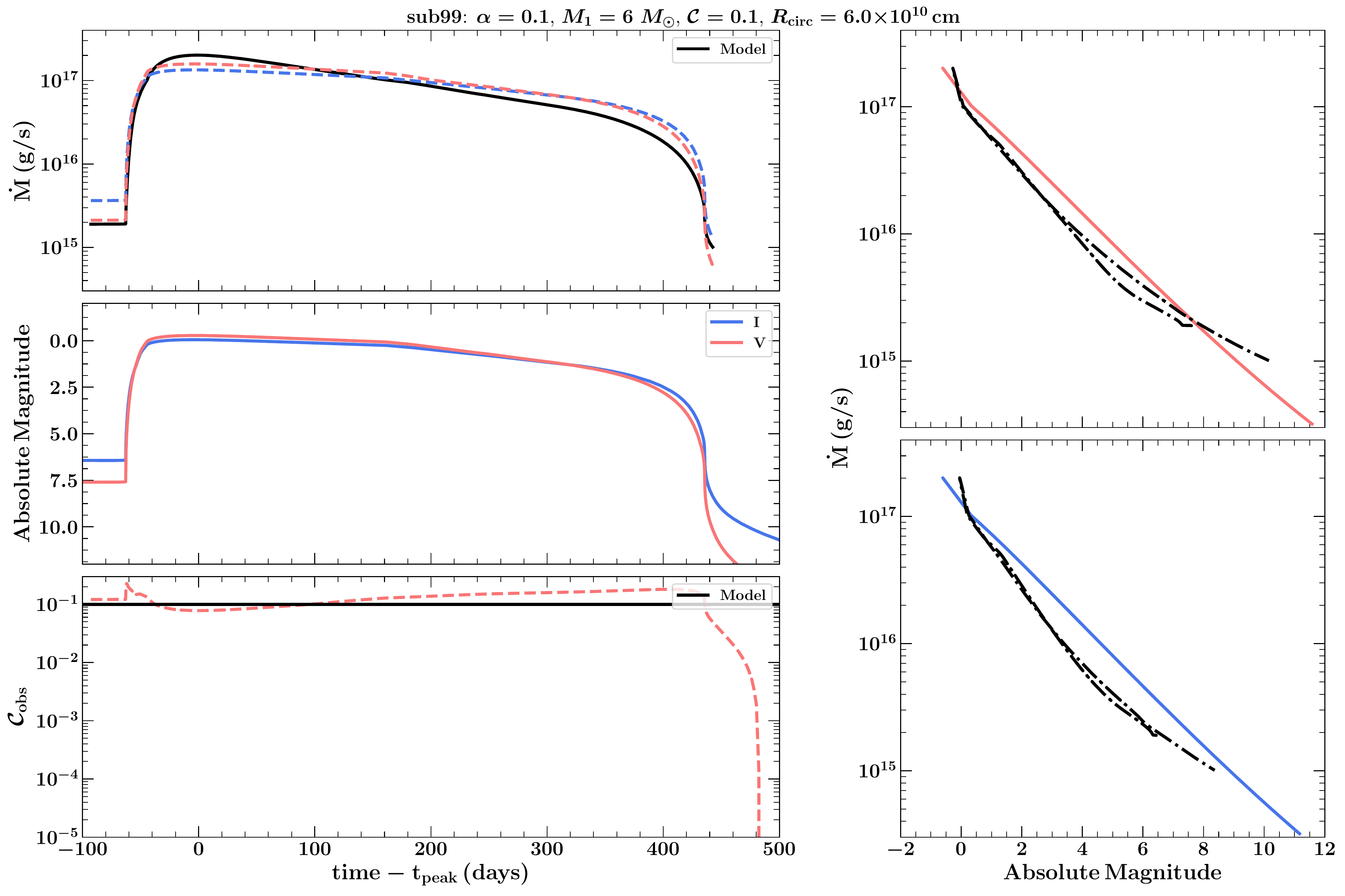}
  \caption{Example model lightcurve analysis. \textbf{Left Panel:} The model and reconstructed $\dot{M}$ \textit{(top)} , optical lightcurves \textit{(middle)}, and $\cal C$ \textit{(bottom)}, for a parameter set  ($M_1$, $\alpha_h$, $\cal C$, and $R_{\rm circ}$). 
  The solid lines represent the output of the numerical DIM code (see Section \ref{apply_model_lcs}). The dotted lines represent the parameters reconstructed using the numerical $\dot{M}-M_V$ and $\dot{M}-M_I$ relationships built from the methodology outlined in Section \ref{sec:model_disc}.
  \textbf{Right Panel:} Compares the $\dot{M}-M_V$ \textit{(top)} and $\dot{M}-M_I$ \textit{(bottom)} relationships, computed from the disc code evolution (dot-dashed black lines) to the steady state assumption (solid red and blue lines).}%
  \label{fig:model_test1}%
\end{figure*}

\begin{figure*}
  \center
  \includegraphics[width=0.92\linewidth,height=.6\linewidth]{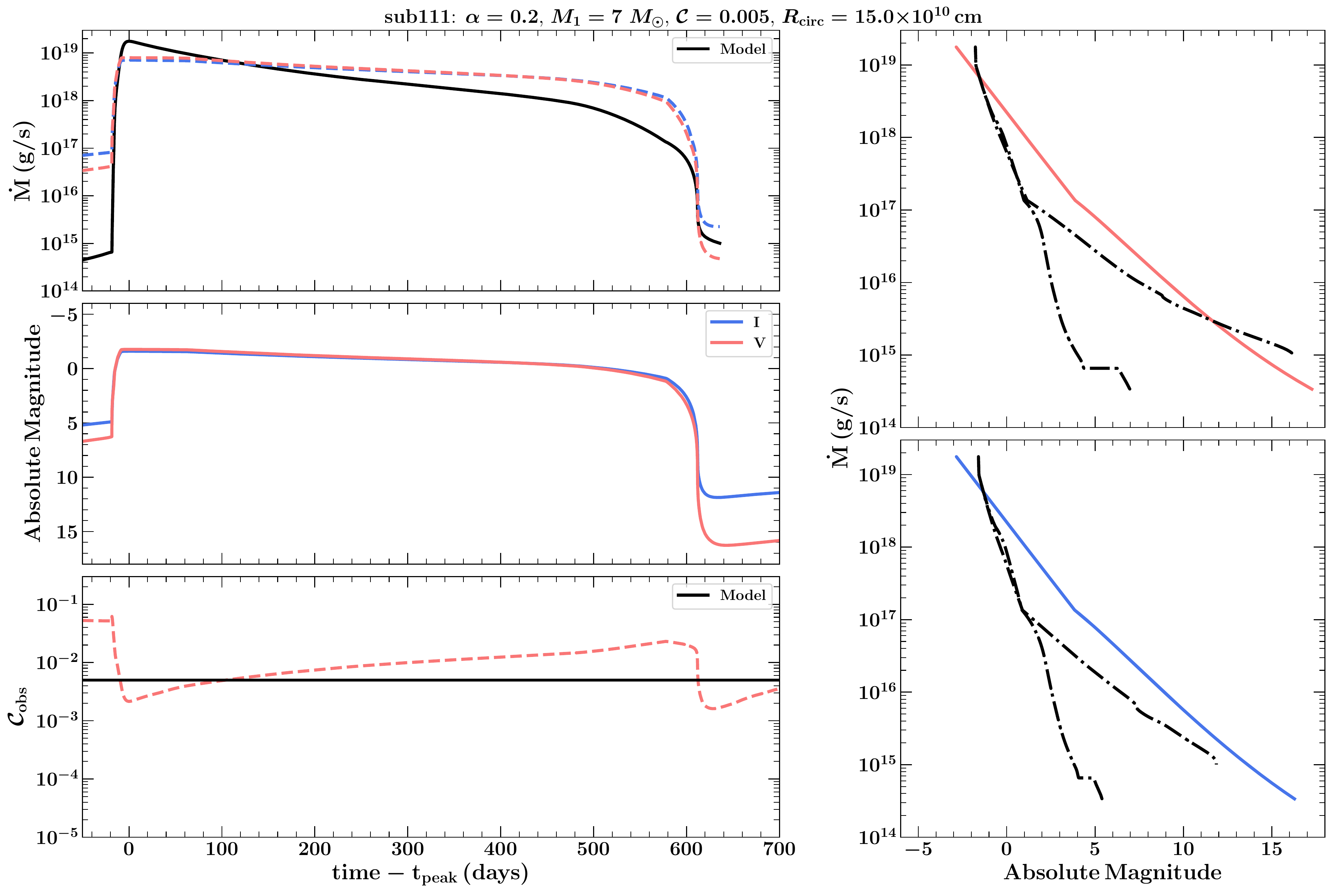}\hfill
  \includegraphics[width=.92\linewidth,height=.6\linewidth]{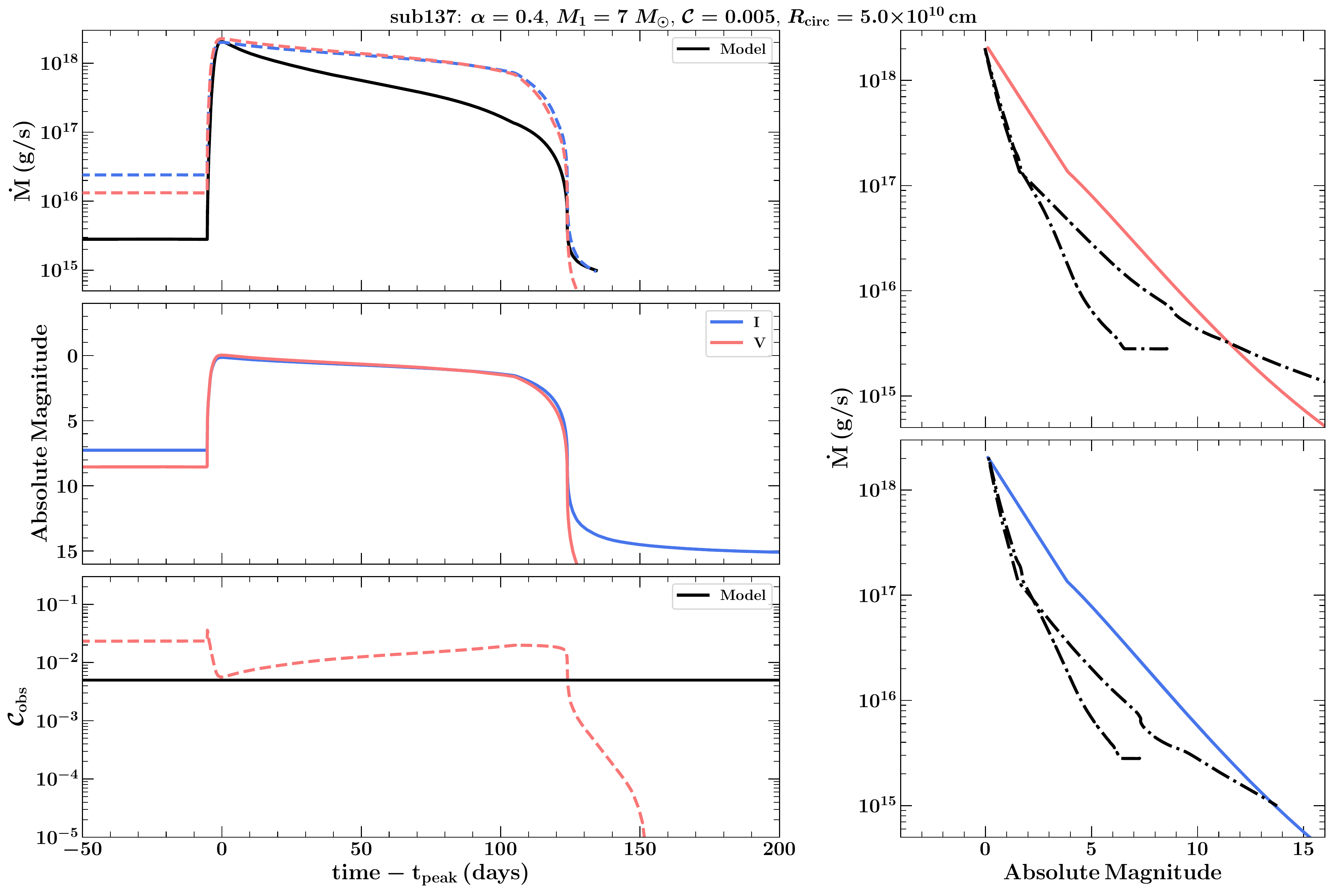}
  \caption{Same caption as Figure \ref{fig:model_test1}.}%
  \label{fig:model_test2}%
\end{figure*}

\begin{figure*}
  \center
  \includegraphics[width=0.92\linewidth,height=.6\linewidth]{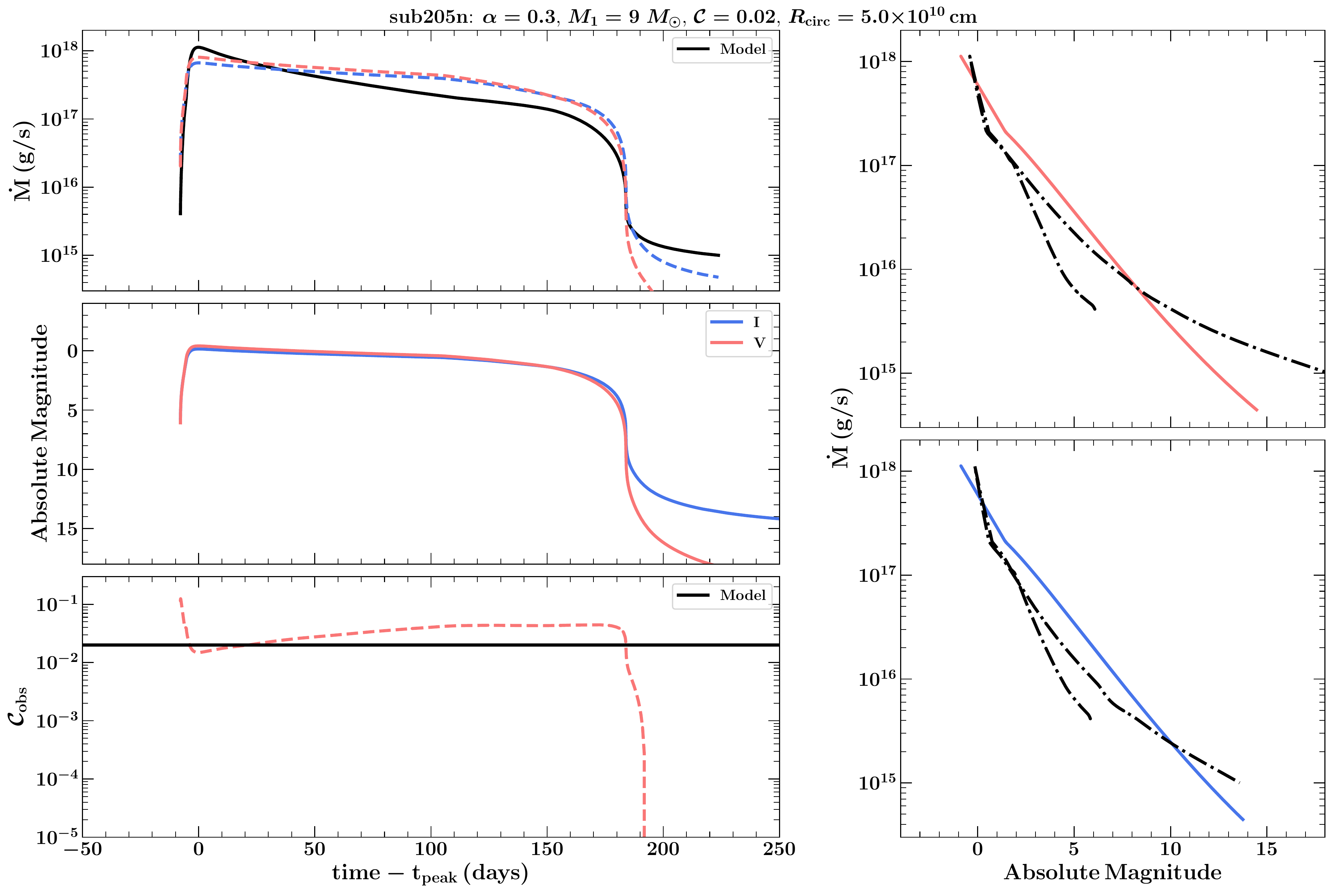}\hfill
  \includegraphics[width=.92\linewidth,height=.6\linewidth]{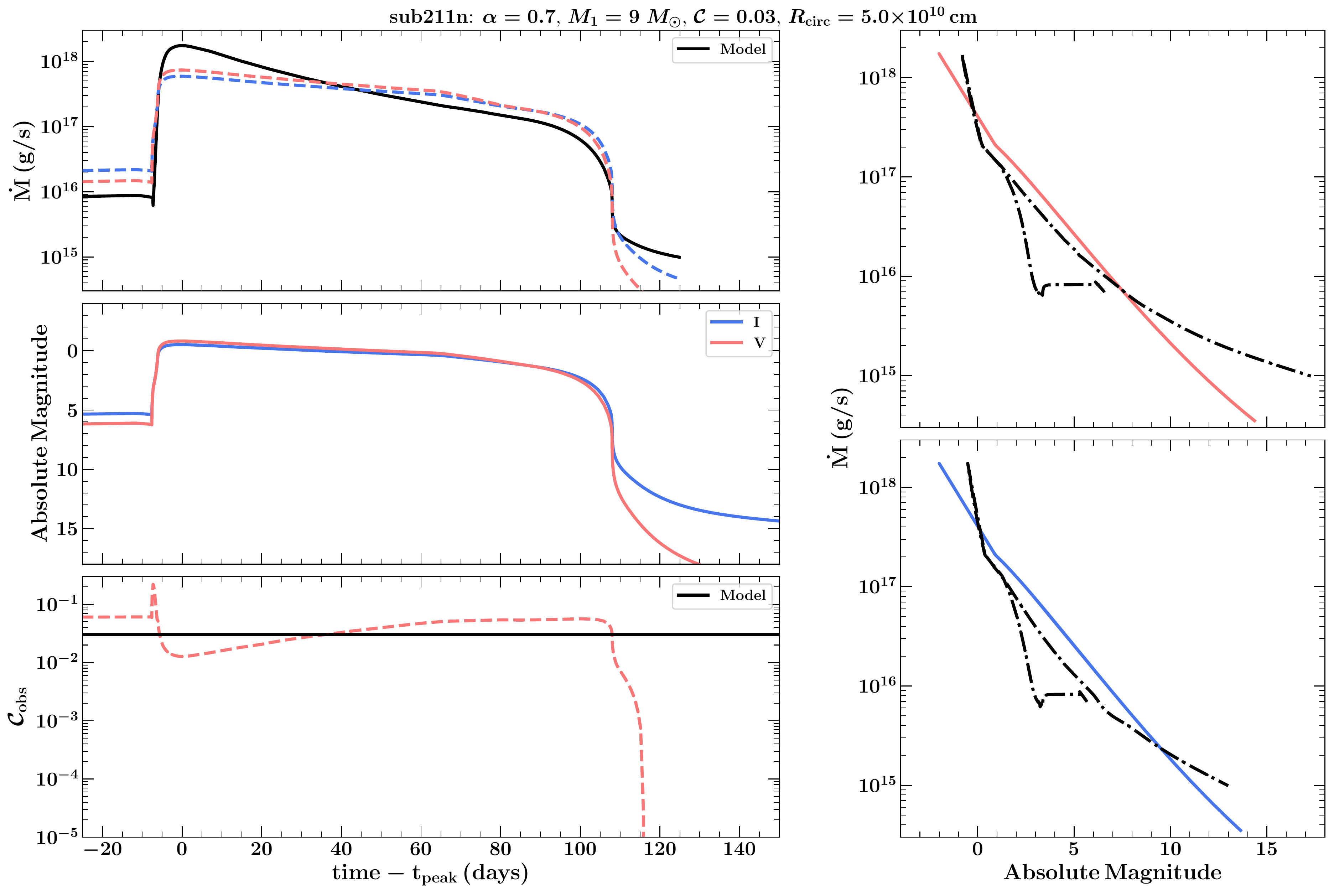}
  \caption{Same caption as Figure \ref{fig:model_test1}.}%
  \label{fig:model_test3}%
\end{figure*}

Figures \ref{fig:mdot_compare} and \ref{fig:c_compare} compare: the time-averaged model $\dot{M}_{\rm in}$ to reconstructed $\dot{M}_{\rm in}$ (in SMARTS/ANDICAM V and I filters), and the model $\cal C$ to the time-averaged reconstructed $\cal C$ (using the model V-band light curve), for the set of 46 disc models, respectively. 
Figures \ref{fig:model_test1}--\ref{fig:model_test3} display example model runs, for a representative set of disc models, which use various combinations of ($M_1$, $\alpha_h$, $\cal C$, and $R_{\rm circ}$).
Overall, we find that: (i) the reconstructed $\dot{M}_{\rm in}$ tends to be overestimated, typically by a factor of 3 or less, with a few outlying models, and (ii) our methodology can reproduce the model $\cal C$ to within a factor of 3 in all model runs.
We find this level of uncertainty to be acceptable, as other sources of error, not taken into account here, occurring in real BH-LMXB systems (e.g., distance, inclination, extinction), will dominate. 

\clearpage

\section{Analysis of OIR Emission Processes During an Outburst Cycle}\label{sec:oir_app}

\begin{figure*}
\subfloat
  {\includegraphics[width=0.96\linewidth,height=.5\linewidth]{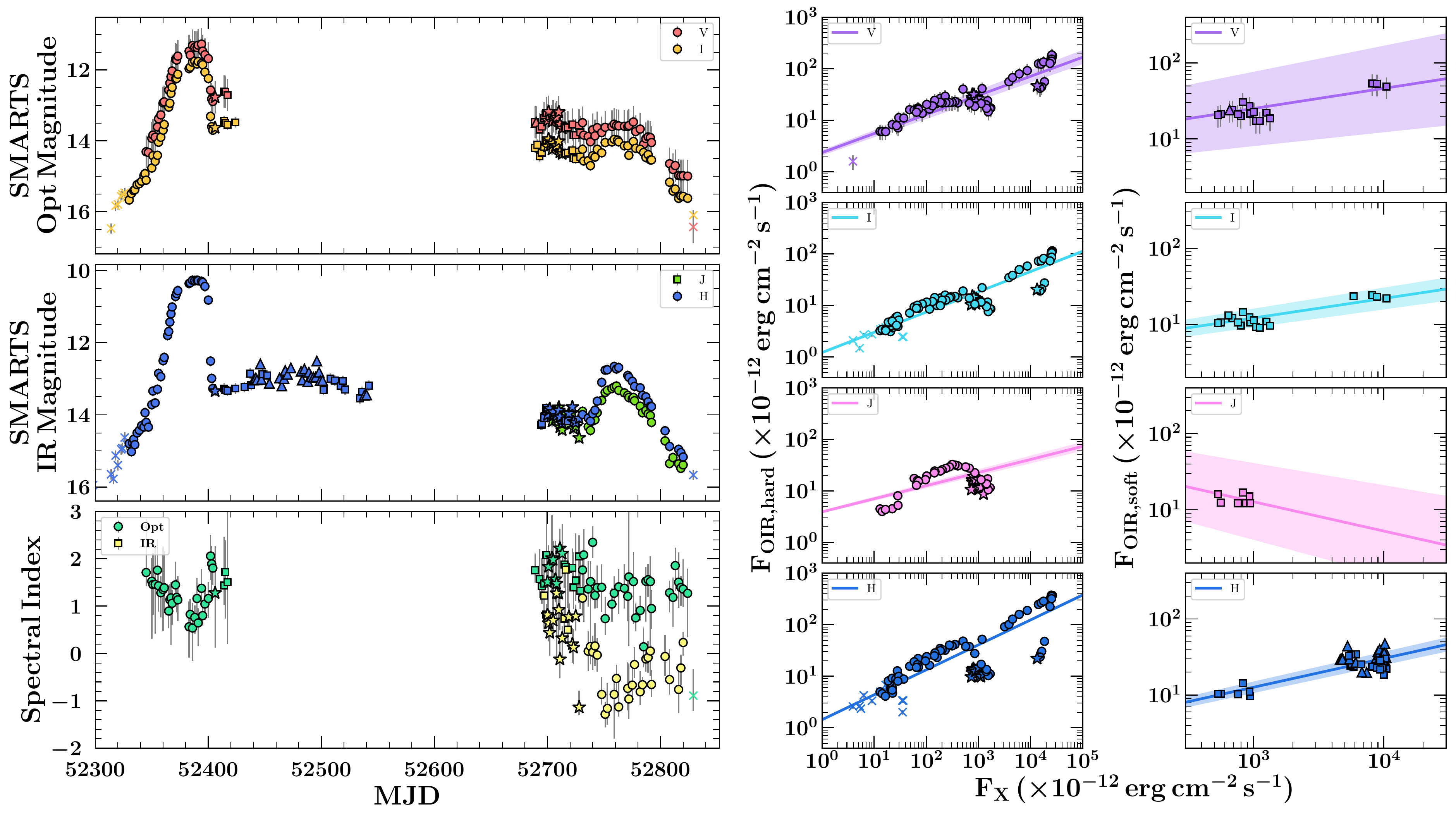}}\hfill
\subfloat
  {\includegraphics[width=0.96\linewidth,height=.5\linewidth]{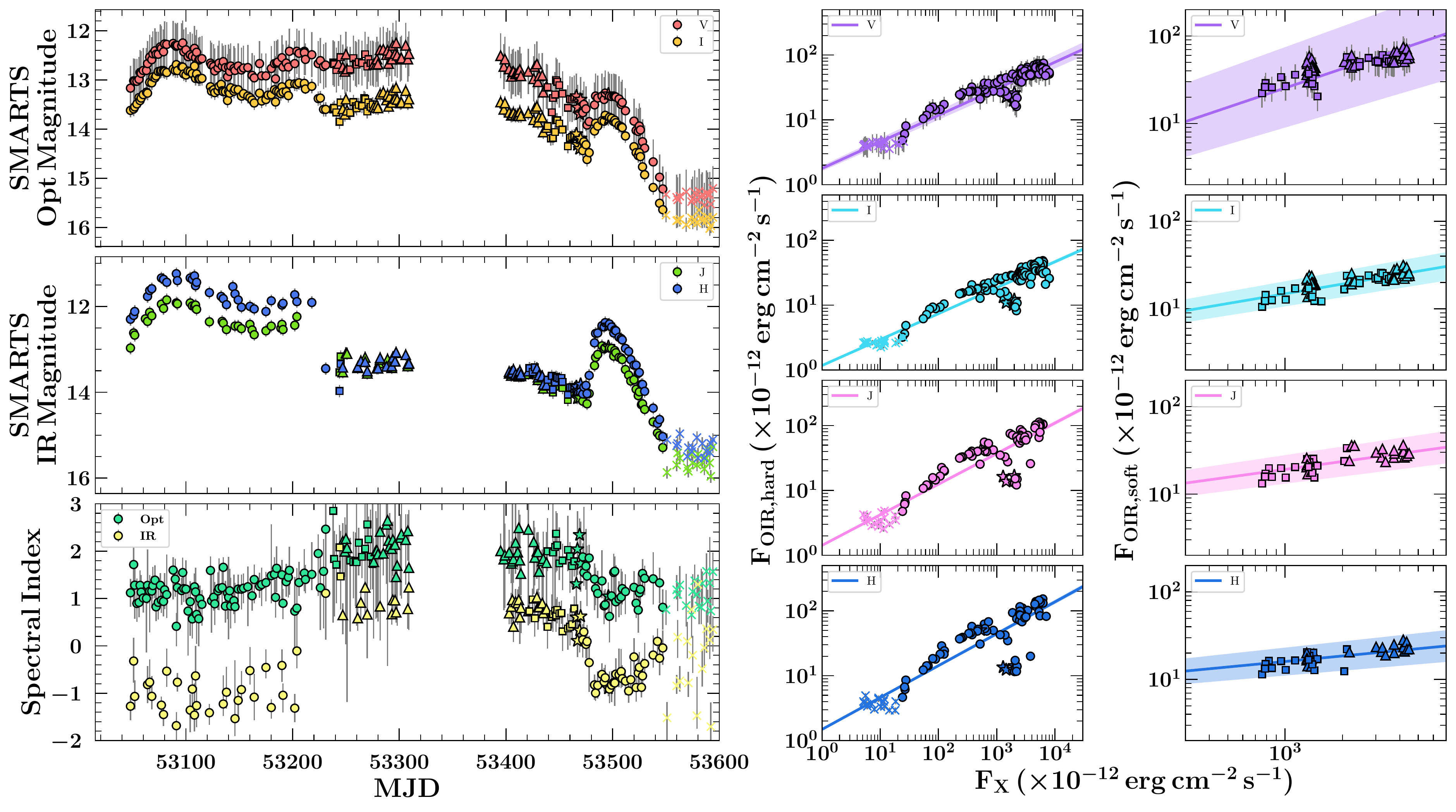}}  
  \caption{Analysis of the emission processes present during the \textit{(top)} 2002$-$2003 and \textit{(bottom)} 2004$-$2005 outbursts of GX339$-$4. Caption is the same as Figure \ref{fig:mw_outburst1}.}
  \label{fig:mw_outburst2}%
\end{figure*}

\begin{figure*}
\subfloat
  {\includegraphics[width=0.96\linewidth,height=.5\linewidth]{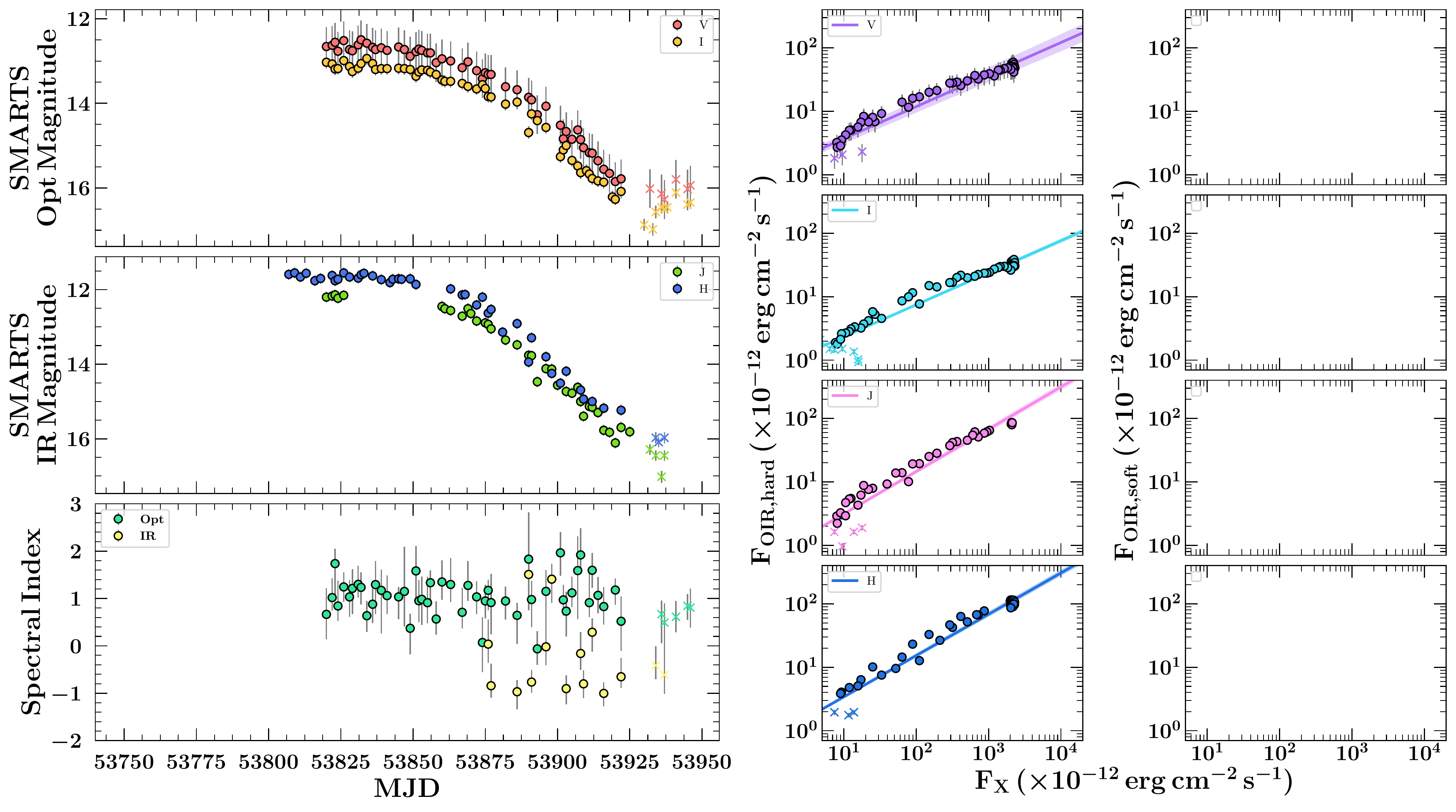}}\hfill
  
\subfloat
  {\includegraphics[width=0.96\linewidth,height=.5\linewidth]{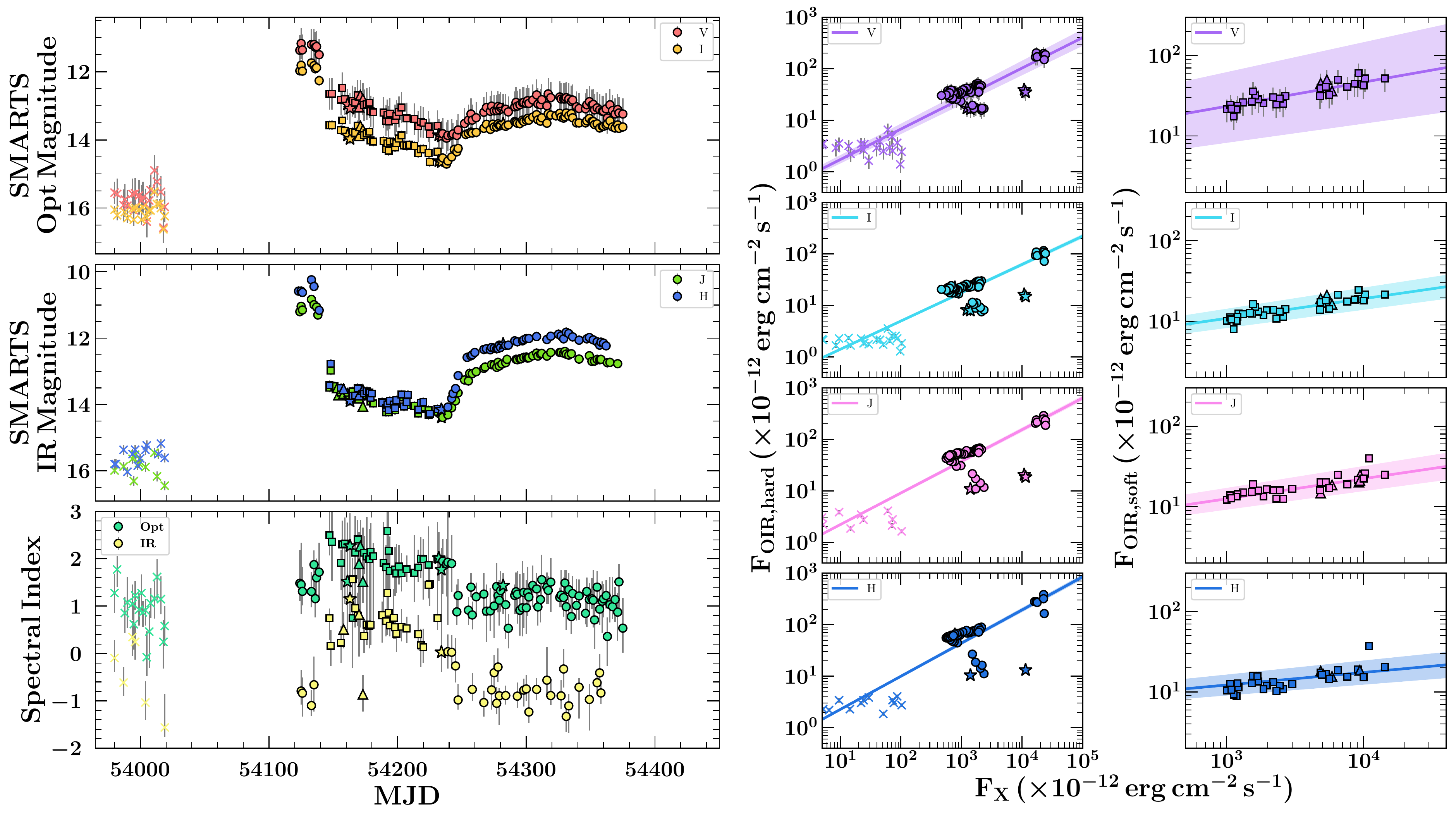}}  
  \caption{Analysis of the emission processes present during the \textit{(top)} 2006 and \textit{(bottom)} 2006$-$2007 outbursts of GX339$-$4. Caption is the same as Figure \ref{fig:mw_outburst1}. Note that no soft state correlations exist for the 2006 (failed) outburst, as GX339$-$4 remained in the hard state for the duration.}
  \label{fig:mw_outburst3}%
\end{figure*}

\begin{figure*}
\subfloat
  {\includegraphics[width=0.96\linewidth,height=.5\linewidth]{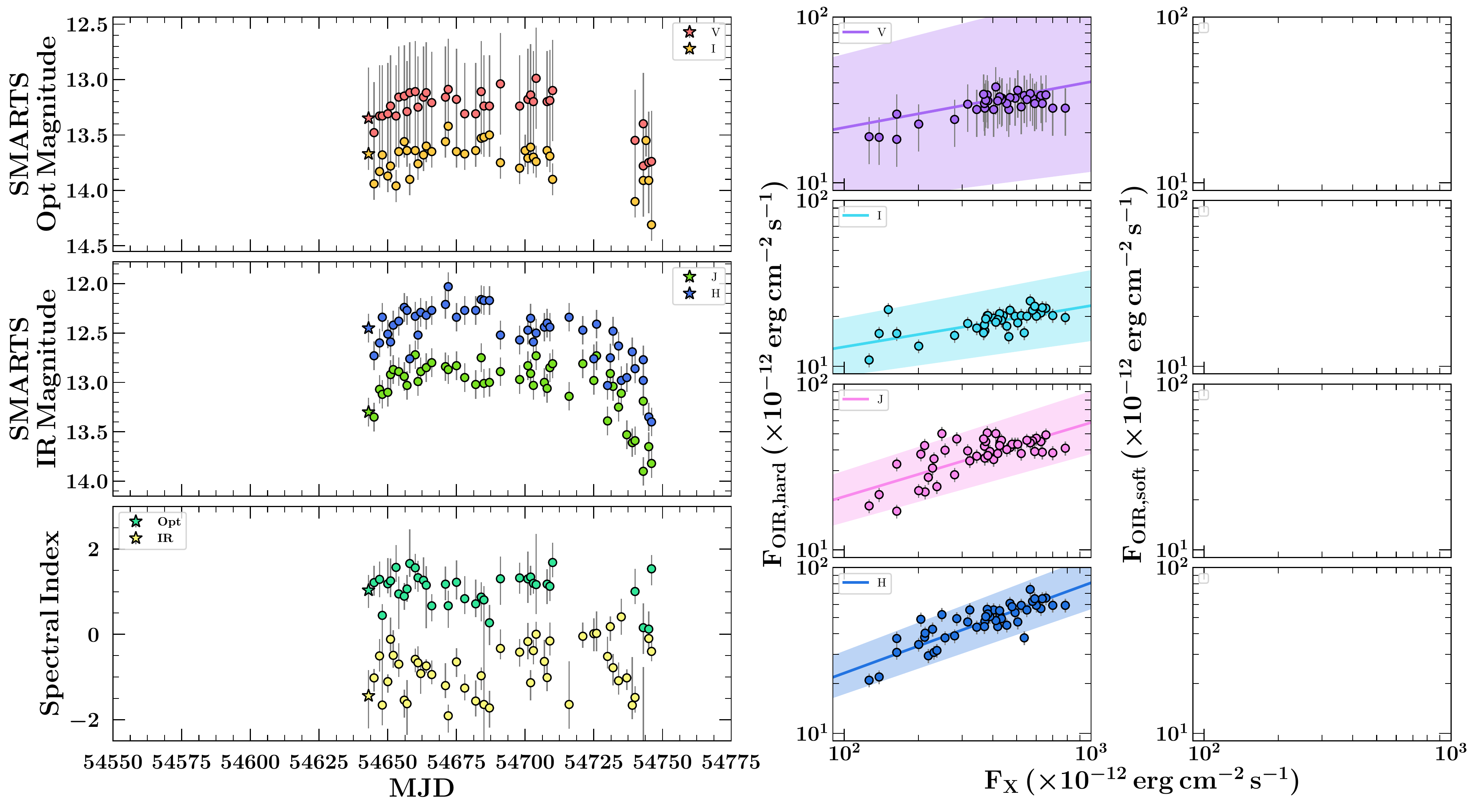}}\hfill
  
\subfloat
  {\includegraphics[width=0.96\linewidth,height=.5\linewidth]{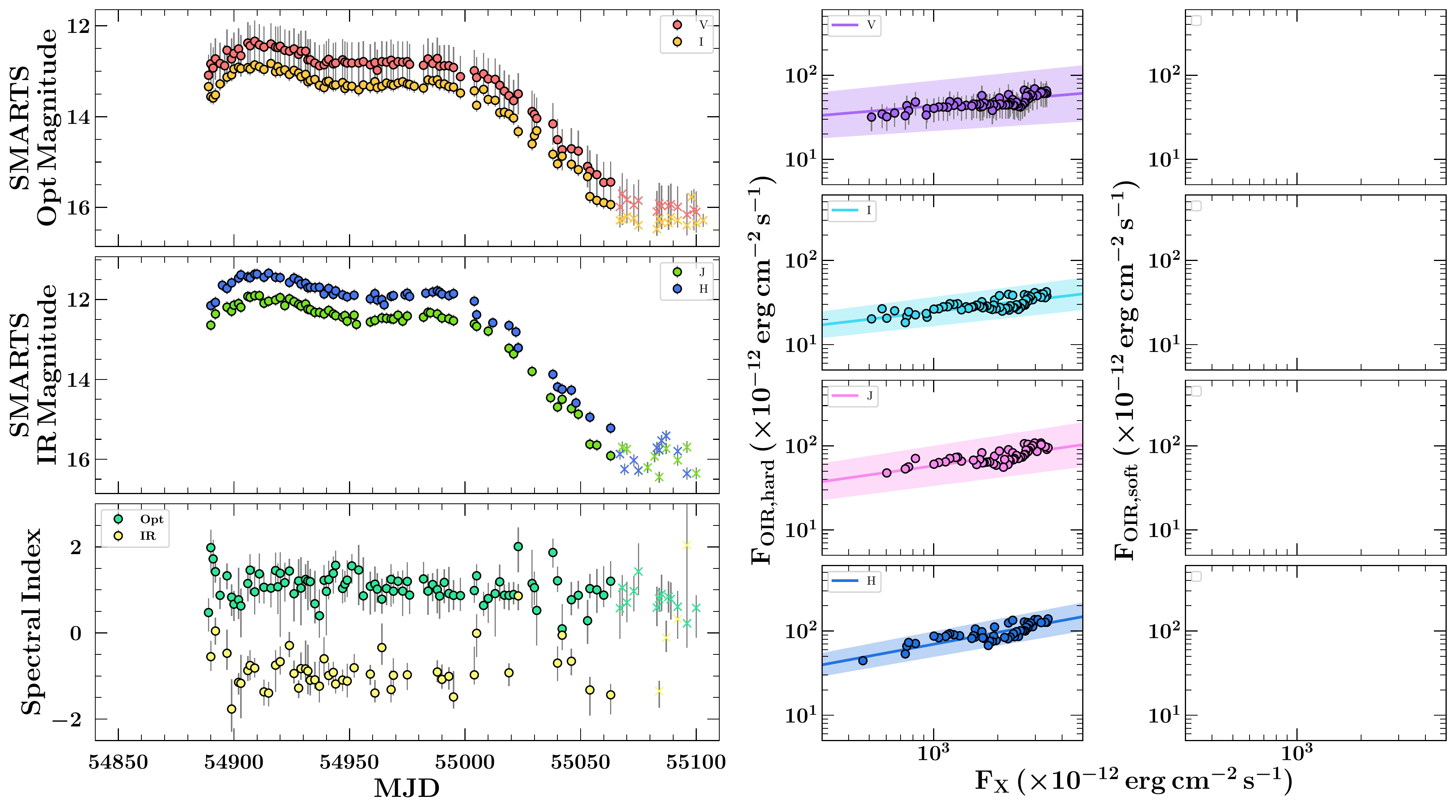}} 
  \caption{Analysis of the emission processes present during the \textit{(top)} 2008 and \textit{(bottom)} 2009 outbursts of GX339$-$4. Caption is the same as Figure \ref{fig:mw_outburst1}. Note that no soft state correlations exist for the 2008 or 2009 (failed) outbursts, as GX339$-$4 remained in the hard state for the duration.}
  \label{fig:mw_outburst4}%
\end{figure*}

\begin{figure*}
\subfloat
  {\includegraphics[width=0.96\linewidth,height=.5\linewidth]{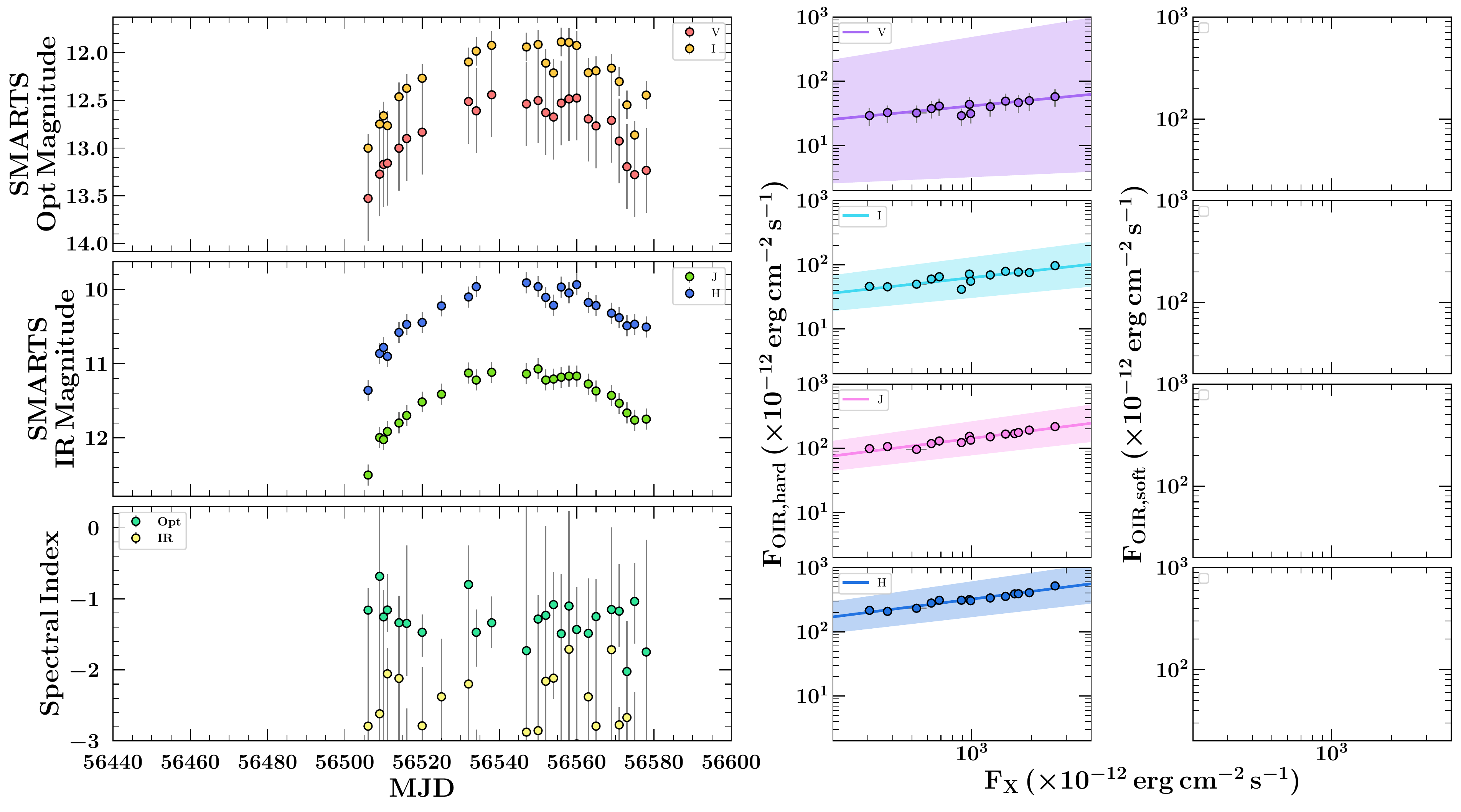}}\hfill
  
\subfloat
  {\includegraphics[width=0.96\linewidth,height=.5\linewidth]{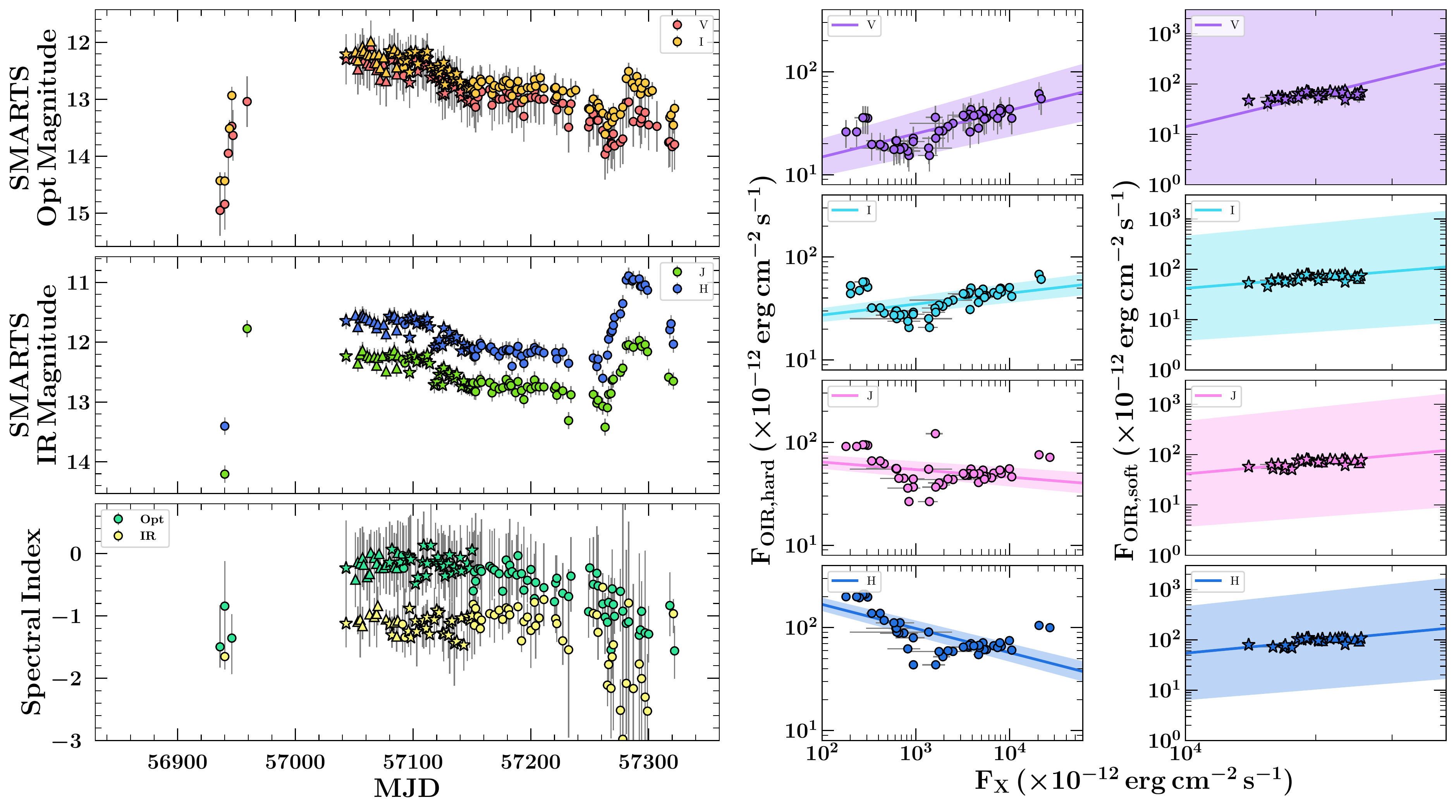}}  
  \caption{Analysis of the emission processes present during the \textit{(top)} 2013 and \textit{(bottom)} 2014$-$2015 outbursts of GX339$-$4. Caption is the same as Figure \ref{fig:mw_outburst1}. As the Marcel et al. calculations does not cover the 2013 or 2014$-$2015 outbursts of GX339$-$4, the bolometric flux estimated from the observed X-ray data is used in all correlations. Note that, no soft state correlations exist for the 2013 (failed) outburst, as GX339$-$4 remained in the hard state for the duration.}
  \label{fig:mw_outburst5}%
\end{figure*}

\clearpage

\section{Analysis of X-ray Irradiation During Outburst}\label{sec:xray_app}

\begin{figure*}
\subfloat
  {\includegraphics[width=0.96\linewidth,height=.5\linewidth]{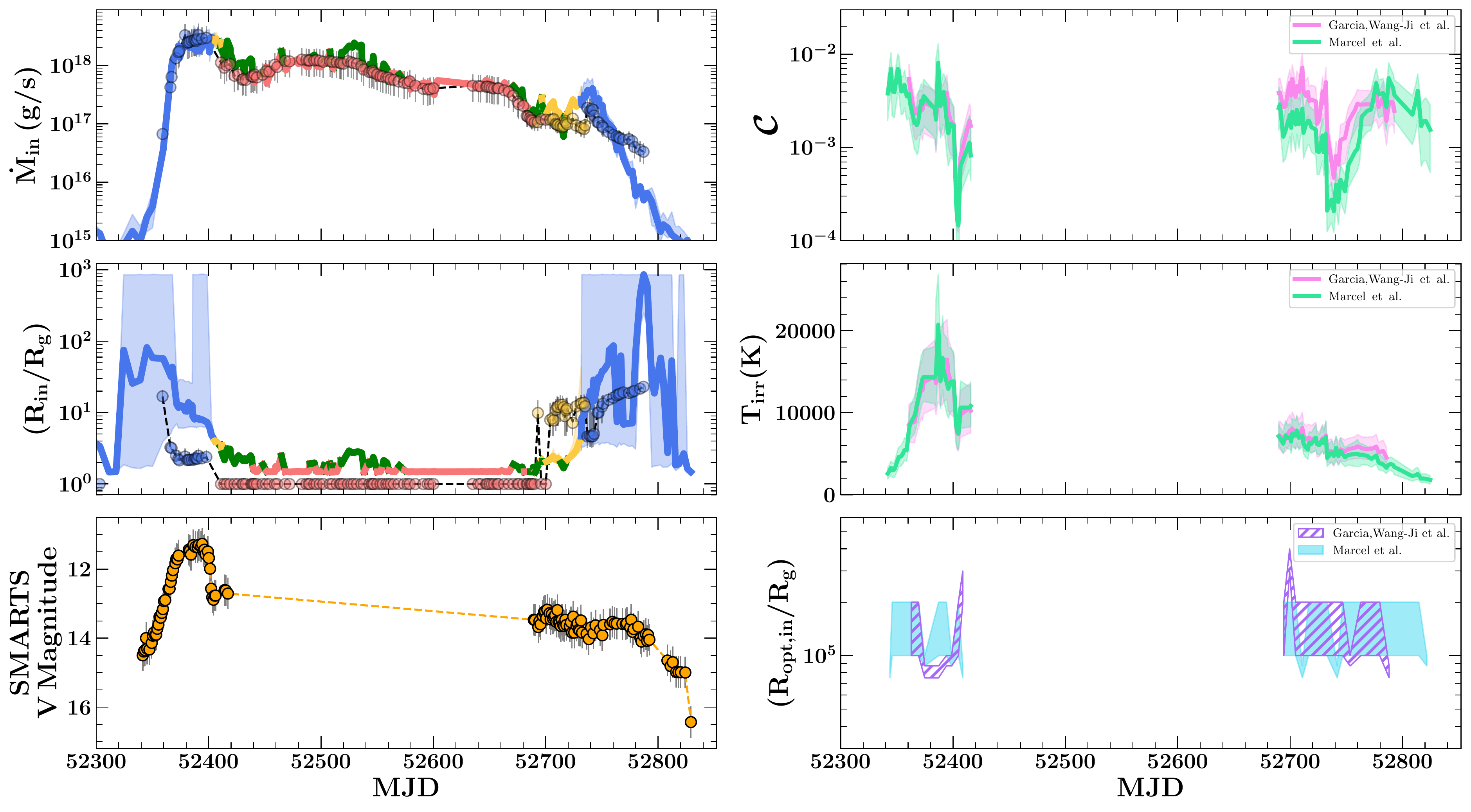}}
  
\subfloat
  {\includegraphics[width=0.96\linewidth,height=.5\linewidth]{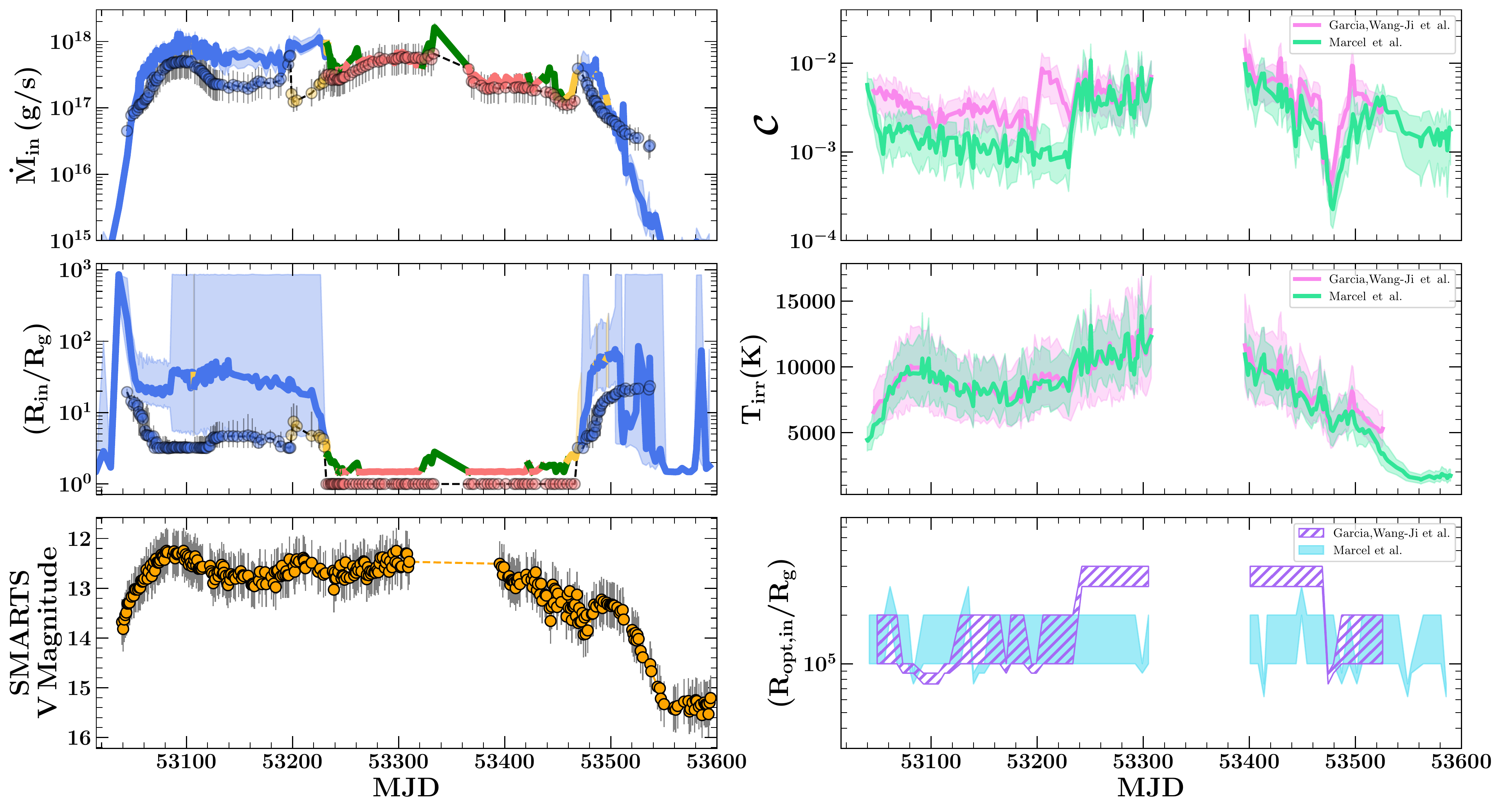}}
  \caption{The \textit{(top)} 2002$-$2003 and \textit{(bottom)} 2004$-$2005 outbursts of GX339$-$4. Caption is the same as Figure \ref{fig:lc_outburst1}.
  }%
  \label{fig:lc_outburst2}%
\end{figure*}

\begin{figure*}
\subfloat
  {\includegraphics[width=0.96\linewidth,height=.5\linewidth]{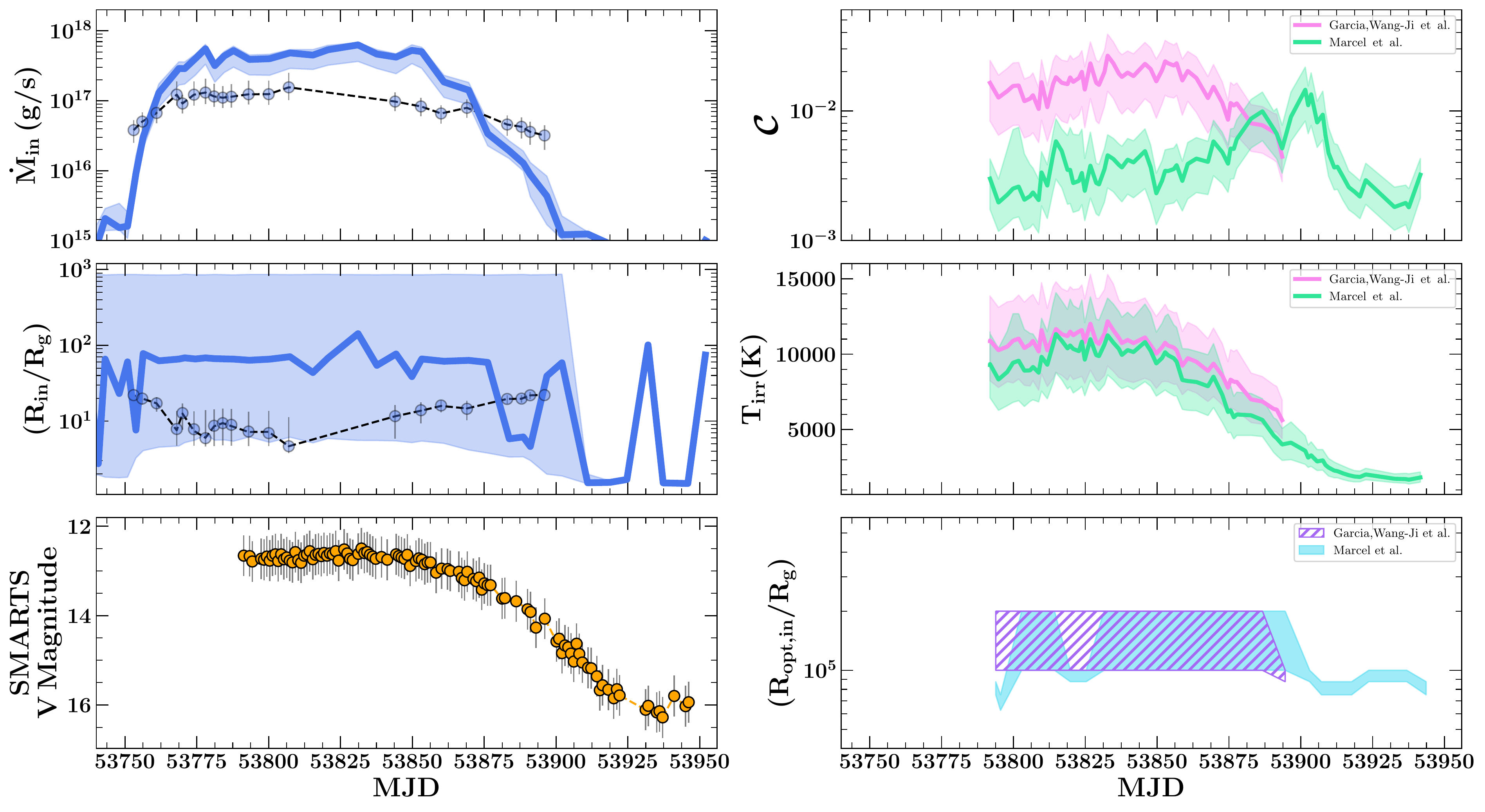}}
  
\subfloat
  {\includegraphics[width=0.96\linewidth,height=.5\linewidth]{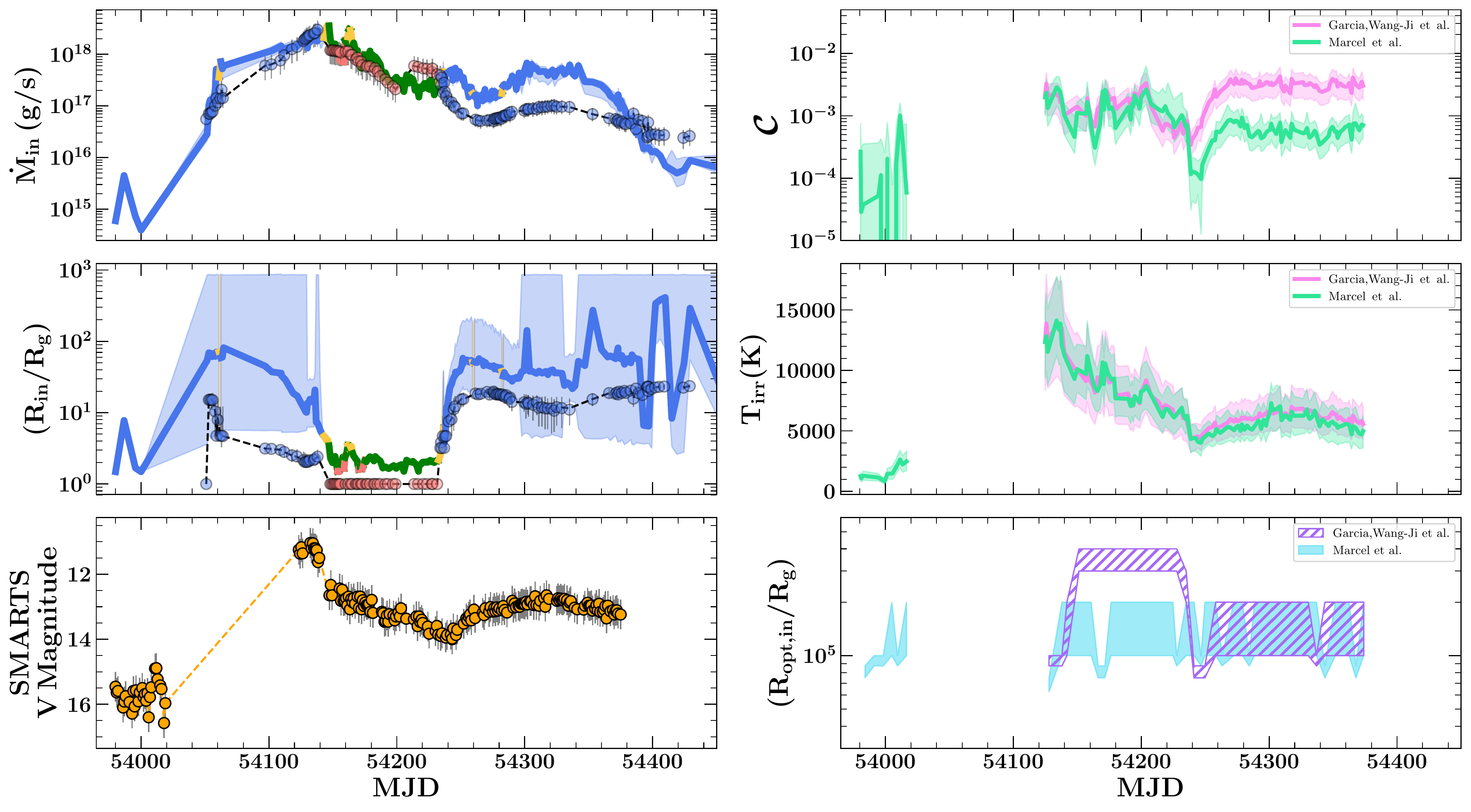}}
  \caption{The \textit{(top)} 2006 and \textit{(bottom)} 2006$-$2007 outbursts of GX339$-$4. Caption is the same as Figure \ref{fig:lc_outburst1}.
  }%
  \label{fig:lc_outburst3}%
\end{figure*}

\begin{figure*}
\subfloat
  {\includegraphics[width=0.96\linewidth,height=.5\linewidth]{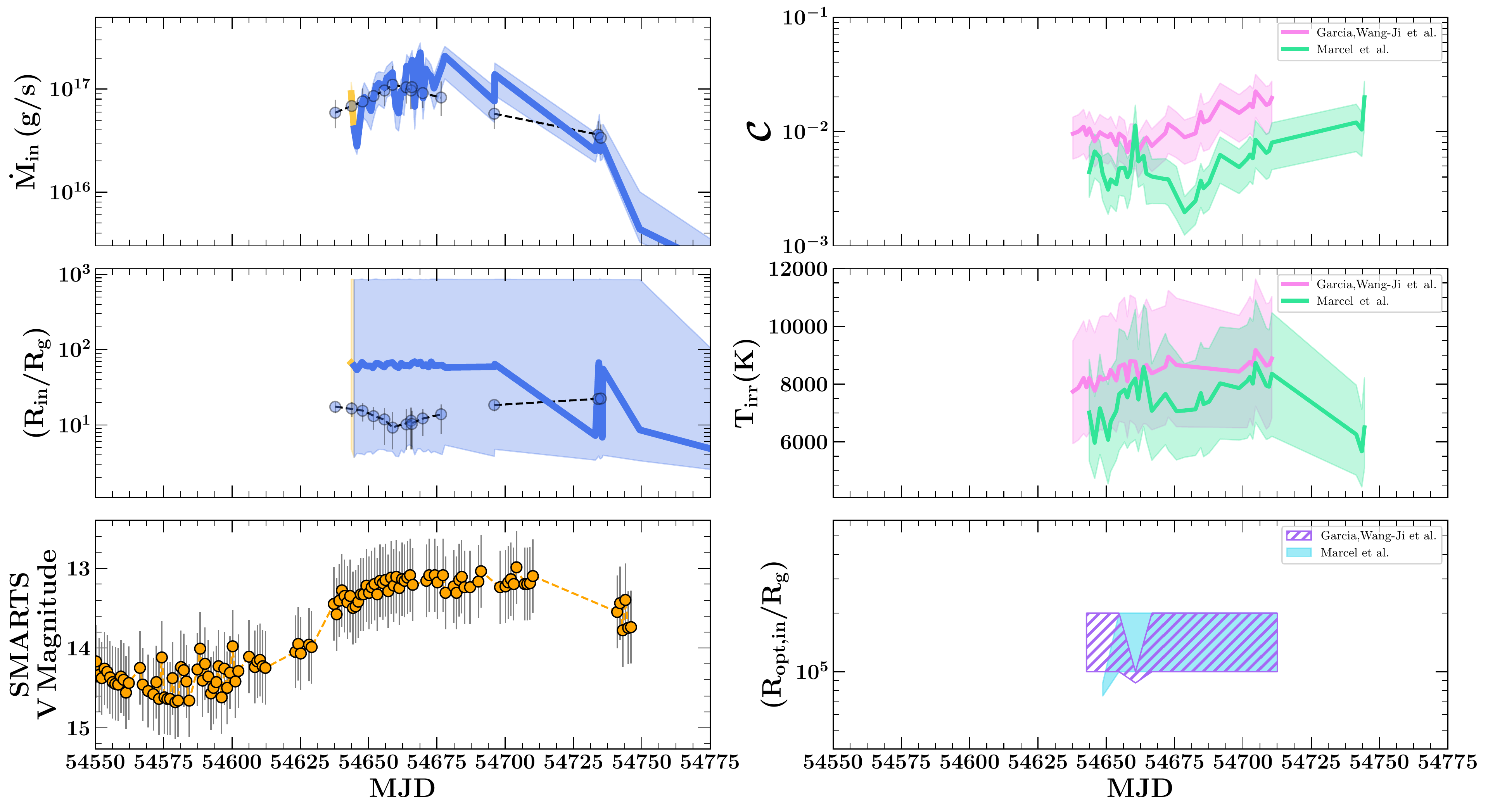}}
  
\subfloat
  {\includegraphics[width=0.96\linewidth,height=.5\linewidth]{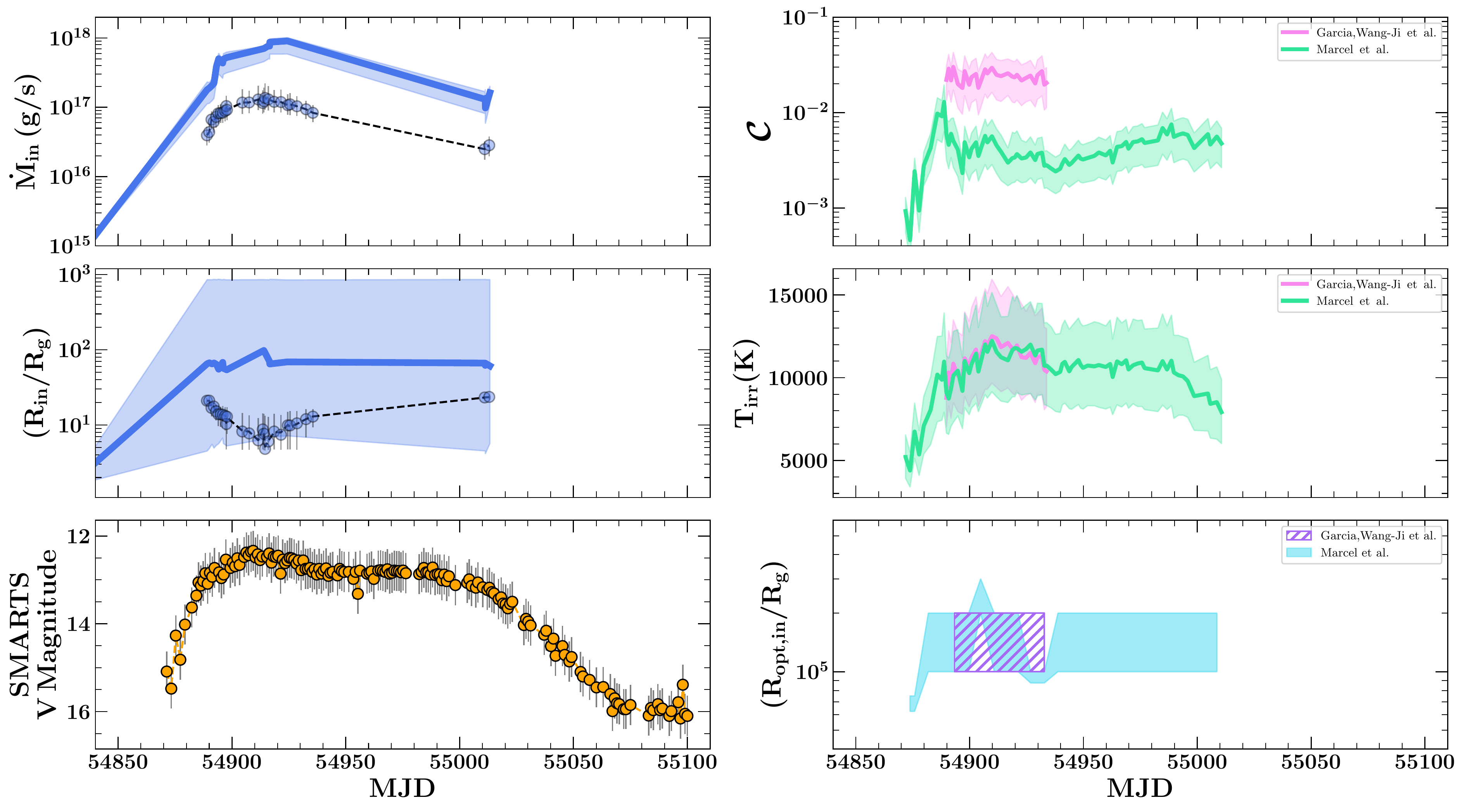}}
  \caption{The \textit{(top)} 2008 and \textit{(bottom)} 2009 outbursts of GX339$-$4. Caption is the same as Figure \ref{fig:lc_outburst1}.
  }%
  \label{fig:lc_outburst4}%
\end{figure*}

\begin{figure*}
\subfloat
  {\includegraphics[width=0.96\linewidth,height=.5\linewidth]{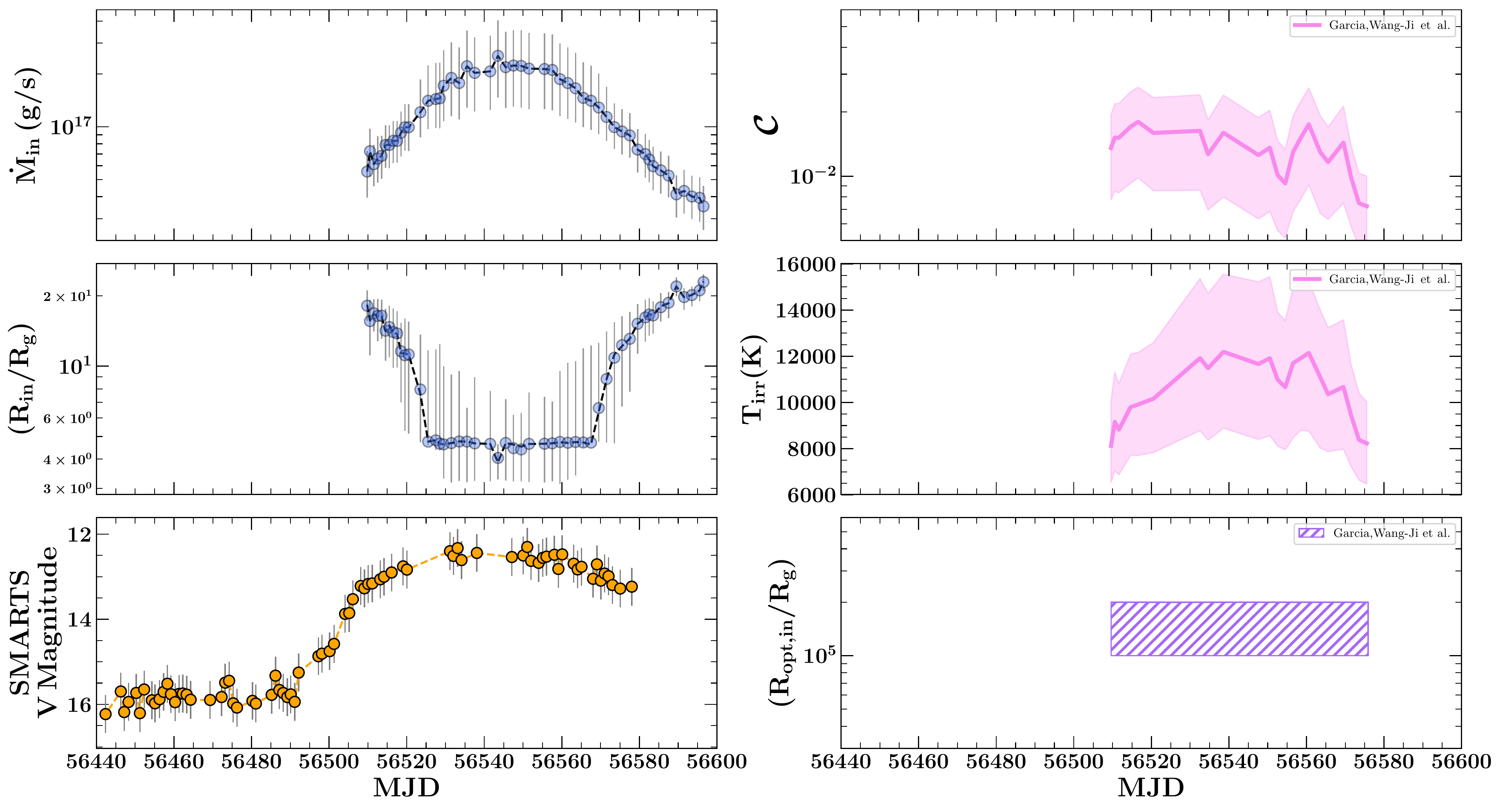}}
  
\subfloat
  {\includegraphics[width=0.96\linewidth,height=.5\linewidth]{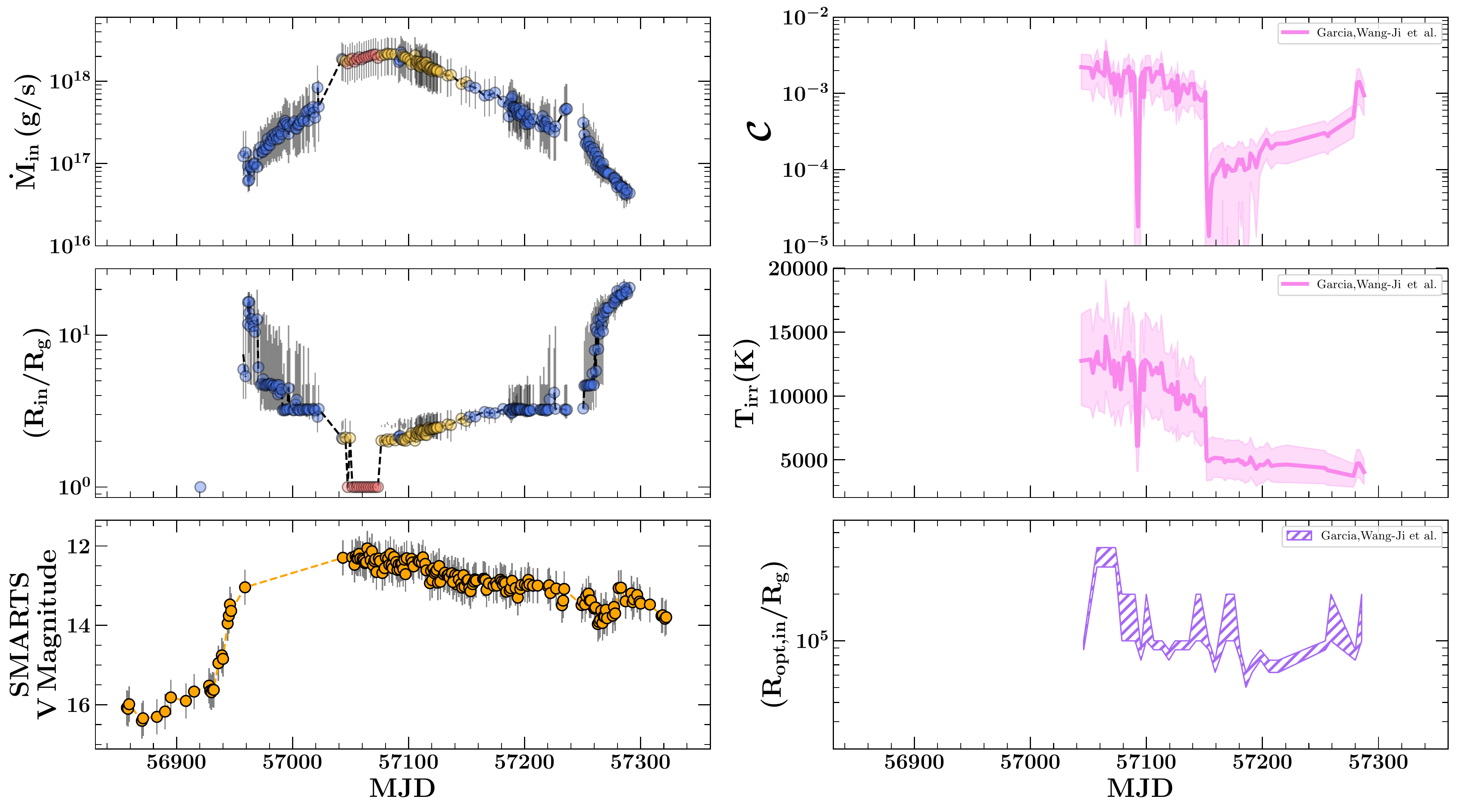}}
  \caption{The \textit{(top)} 2013 and \textit{(bottom)} 2014$-$2015 outbursts of GX339$-$4. Caption is the same as Figure \ref{fig:lc_outburst1}. Note that the Marcel et al. analysis did not cover these outbursts.
  }%
  \label{fig:lc_outburst5}%
\end{figure*}

\clearpage

\section{Thermally-driven Wind Analysis}\label{sec:wind_app}

\begin{figure*}
\begin{minipage}[t]{0.63\textwidth}
\mbox{}\\[-\baselineskip]
  \includegraphics[width=\textwidth,height=1.7\linewidth]{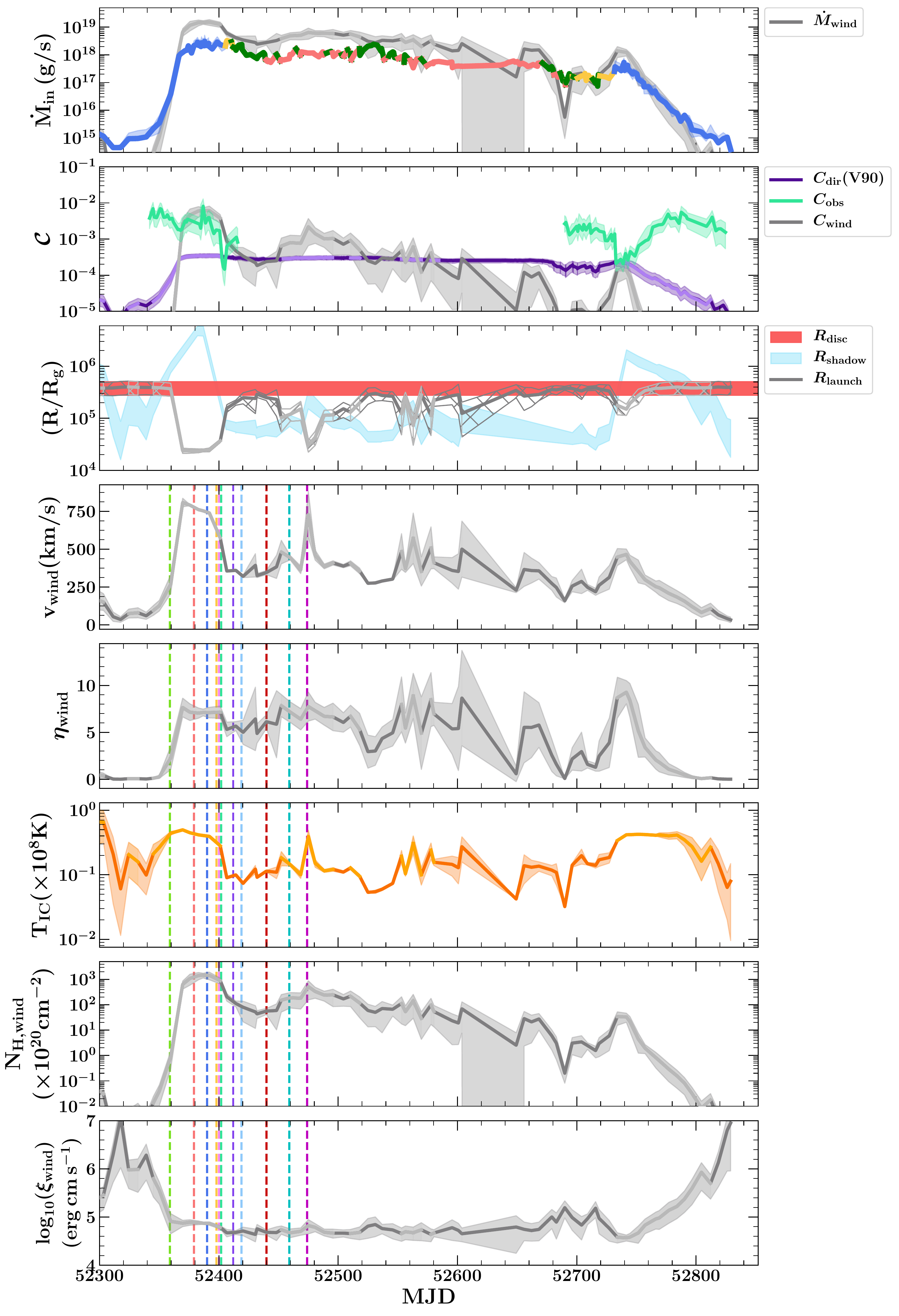}
  \end{minipage}\hfill
\begin{minipage}[t]{0.35\textwidth}
\mbox{}\\[-\baselineskip]

  \caption{Derived thermal (Compton-heated) wind properties for the 2002$-$2003 outburst of GX339$-$4. Caption is the same as Figure \ref{fig:wind_outburst1}.}%
  \label{fig:wind_outburst2}%
  \end{minipage}
\end{figure*}

\begin{figure*}
\begin{minipage}[t]{0.63\textwidth}
\mbox{}\\[-\baselineskip]
  \includegraphics[width=\textwidth,height=1.7\linewidth]{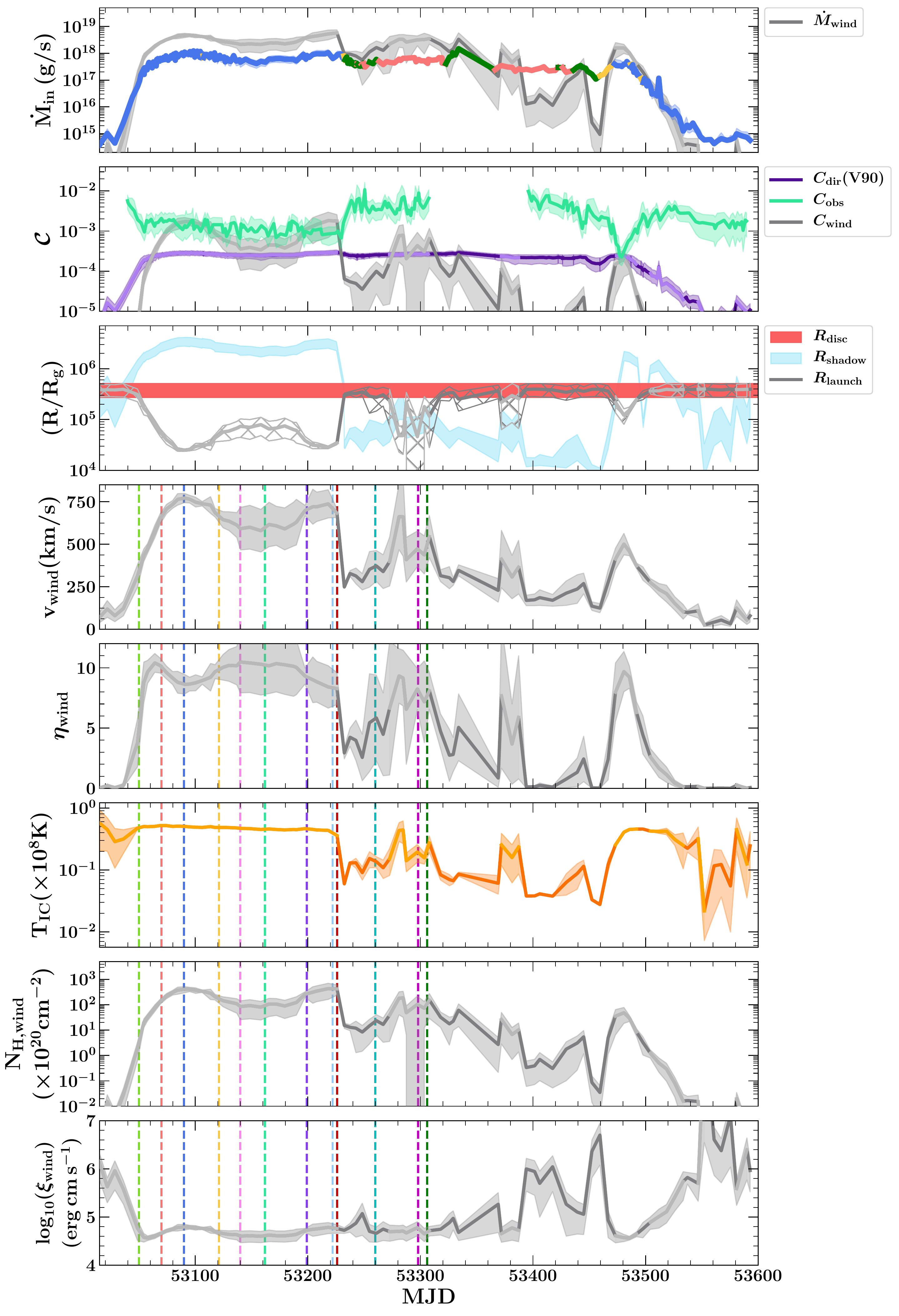}
\end{minipage}\hfill
\begin{minipage}[t]{0.35\textwidth}
\mbox{}\\[-\baselineskip]
  \caption{Derived thermal (Compton-heated) wind properties for the 2004$-$2005 outburst of GX339$-$4. Caption is the same as Figure \ref{fig:wind_outburst1}.}%
  \label{fig:wind_outburst3}%
  \end{minipage}
\end{figure*}

\begin{figure*}

\begin{minipage}[t]{0.63\textwidth}
\mbox{}\\[-\baselineskip]
  \includegraphics[width=\textwidth,height=1.7\linewidth]{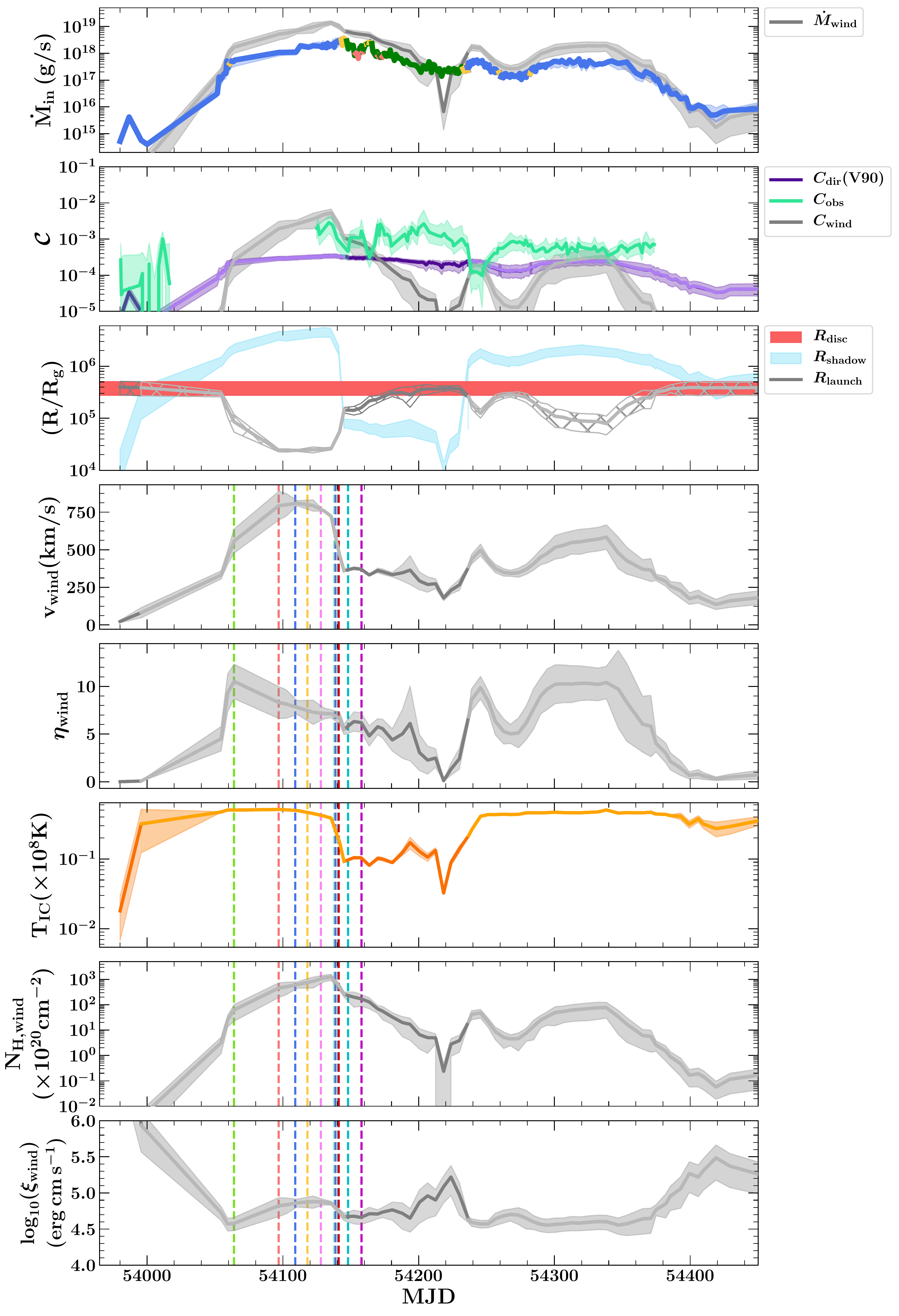}
\end{minipage}\hfill
\begin{minipage}[t]{0.35\textwidth}
\mbox{}\\[-\baselineskip]
  \caption{Derived thermal (Compton-heated) wind properties for the 2006$-$2007 outburst of GX339$-$4. Caption is the same as Figure \ref{fig:wind_outburst1}.}%
  \label{fig:wind_outburst5}%
    \end{minipage}
\end{figure*}

\begin{figure*}
\centering
\subfloat
  {\includegraphics[width=0.33\linewidth,height=0.7\linewidth]{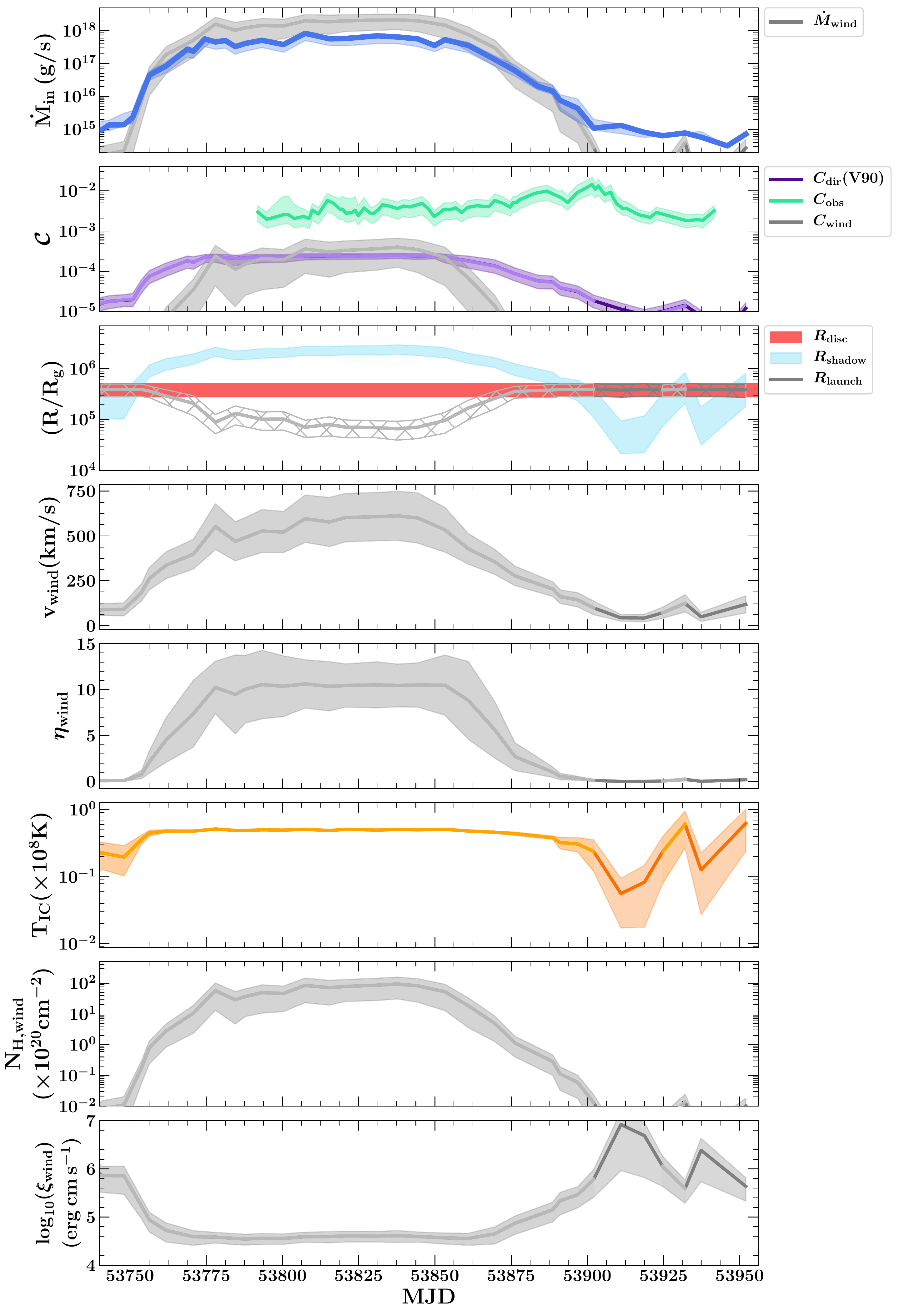}}\hfill
\subfloat
  {\includegraphics[width=0.33\linewidth,height=0.7\linewidth]{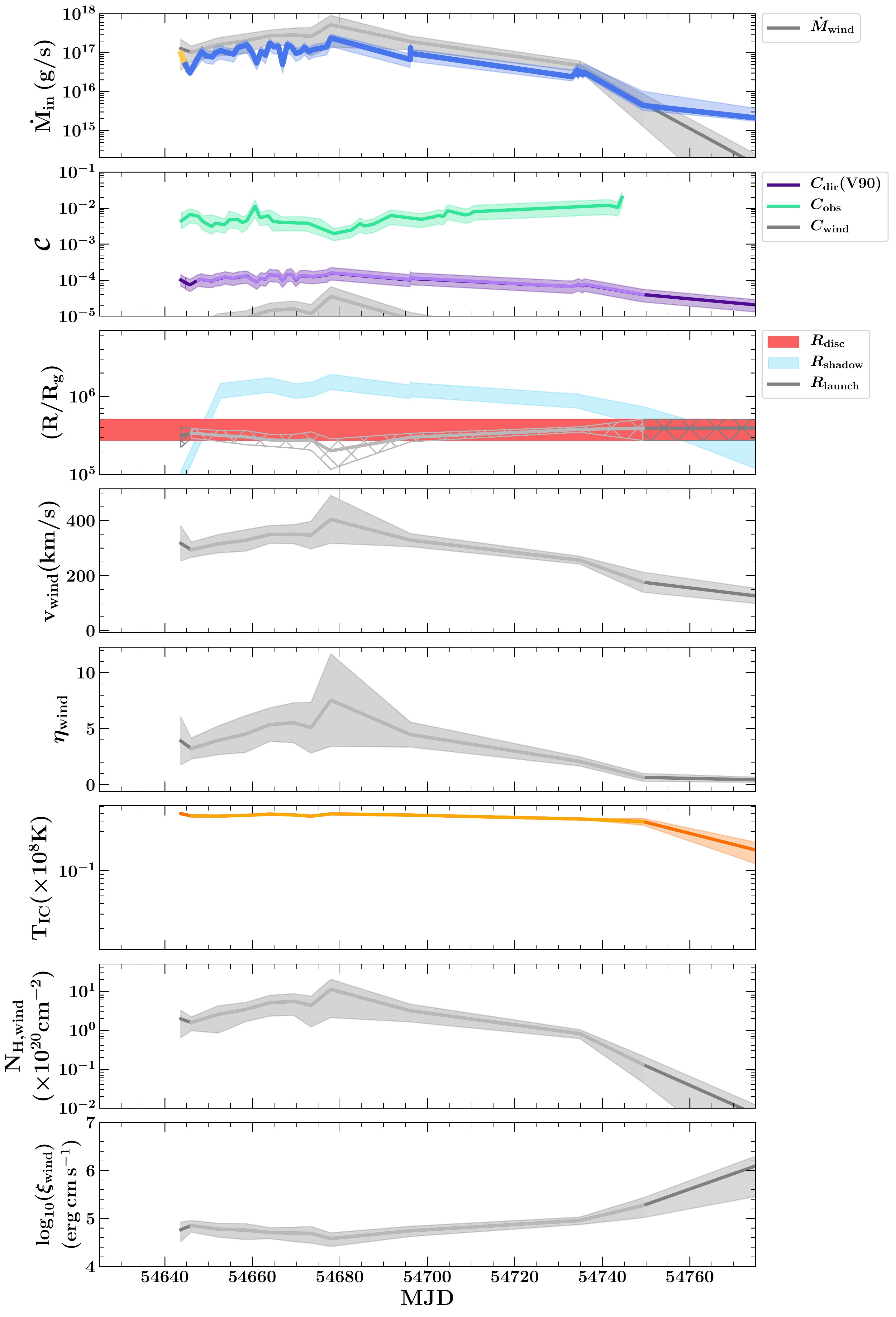}}\hfill
\subfloat
  {\includegraphics[width=0.33\linewidth,height=0.7\linewidth]{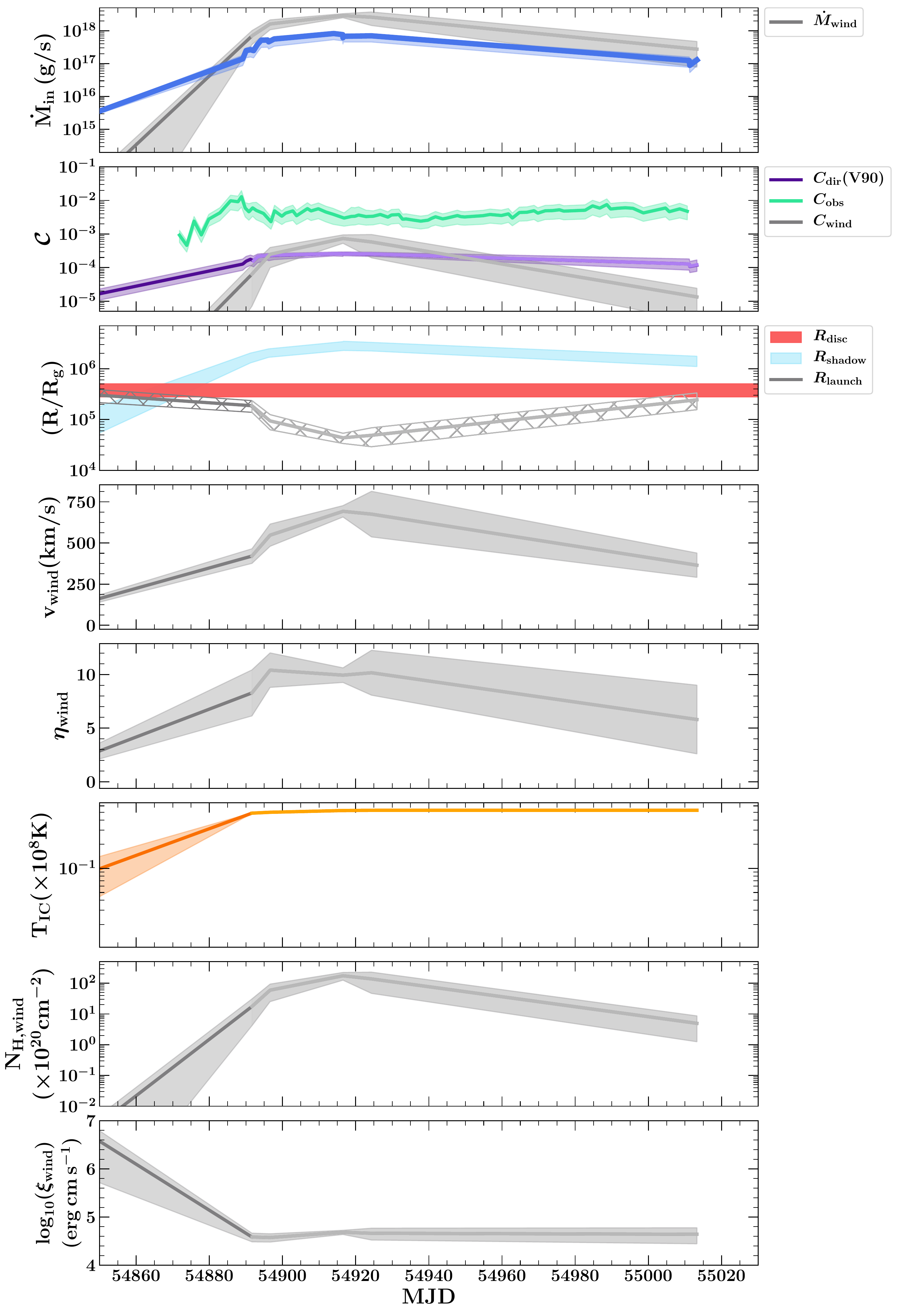}}

  \caption{Derived thermal (Compton-heated) wind properties for the failed \textit{(left)} 2006, \textit{(middle)} 2008, and \textit{(right)} 2009 outbursts of GX339$-$4. Caption is the same as Figure \ref{fig:wind_outburst1}.}%
  \label{fig:wind_outburst6}%
\end{figure*}

\begin{figure}
  \center
\includegraphics[width=1.\linewidth,height=1.2\linewidth]{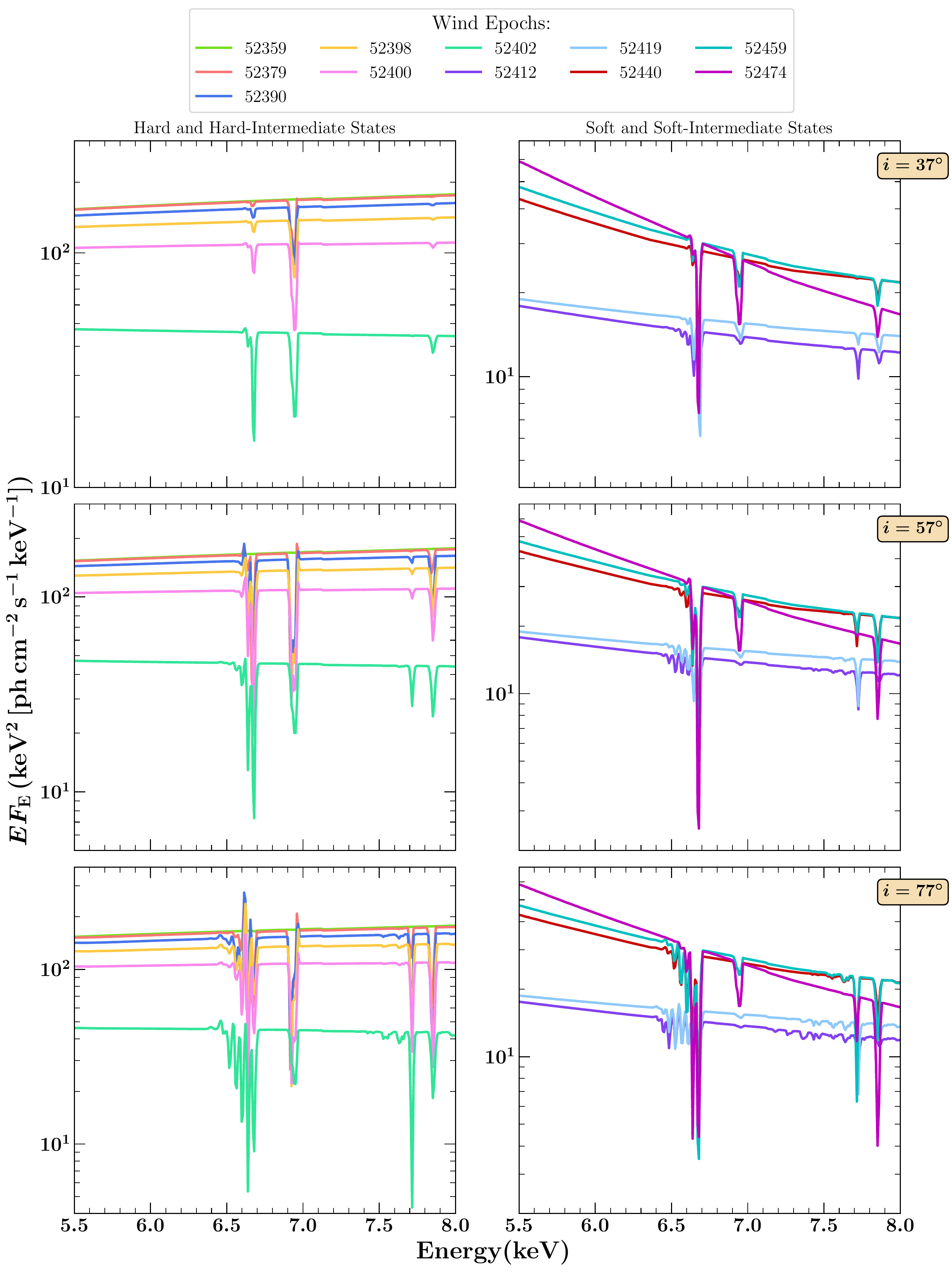}
\caption{Simulated X-ray spectra of the thermal wind via {\sc xstar} during the 2002--2003 outburst cycle of GX339$-$4. Caption the same as Figure \ref{fig:xstar_mods1}.}
  \label{fig:xstar_mods2}%
\end{figure}

\begin{figure}
  \center
\includegraphics[width=1.\linewidth,height=1.2\linewidth]{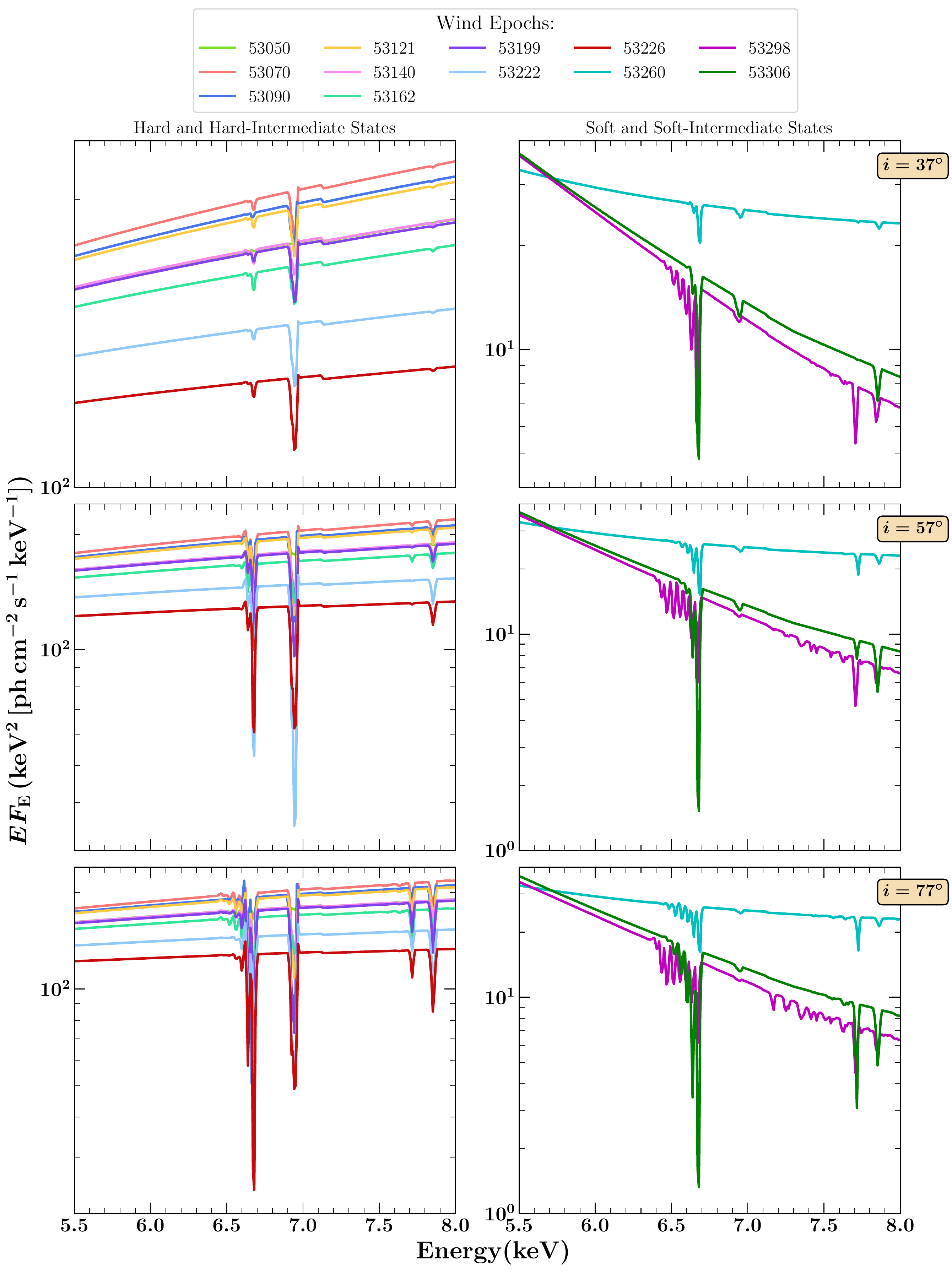}
\caption{Simulated X-ray spectra of the thermal wind via {\sc xstar} during the 2004--2005 outburst cycle of GX339$-$4. Caption the same as Figure \ref{fig:xstar_mods1}.}
  \label{fig:xstar_mods3}%
\end{figure}

\begin{figure}
  \center
\includegraphics[width=1.\linewidth,height=1.2\linewidth]{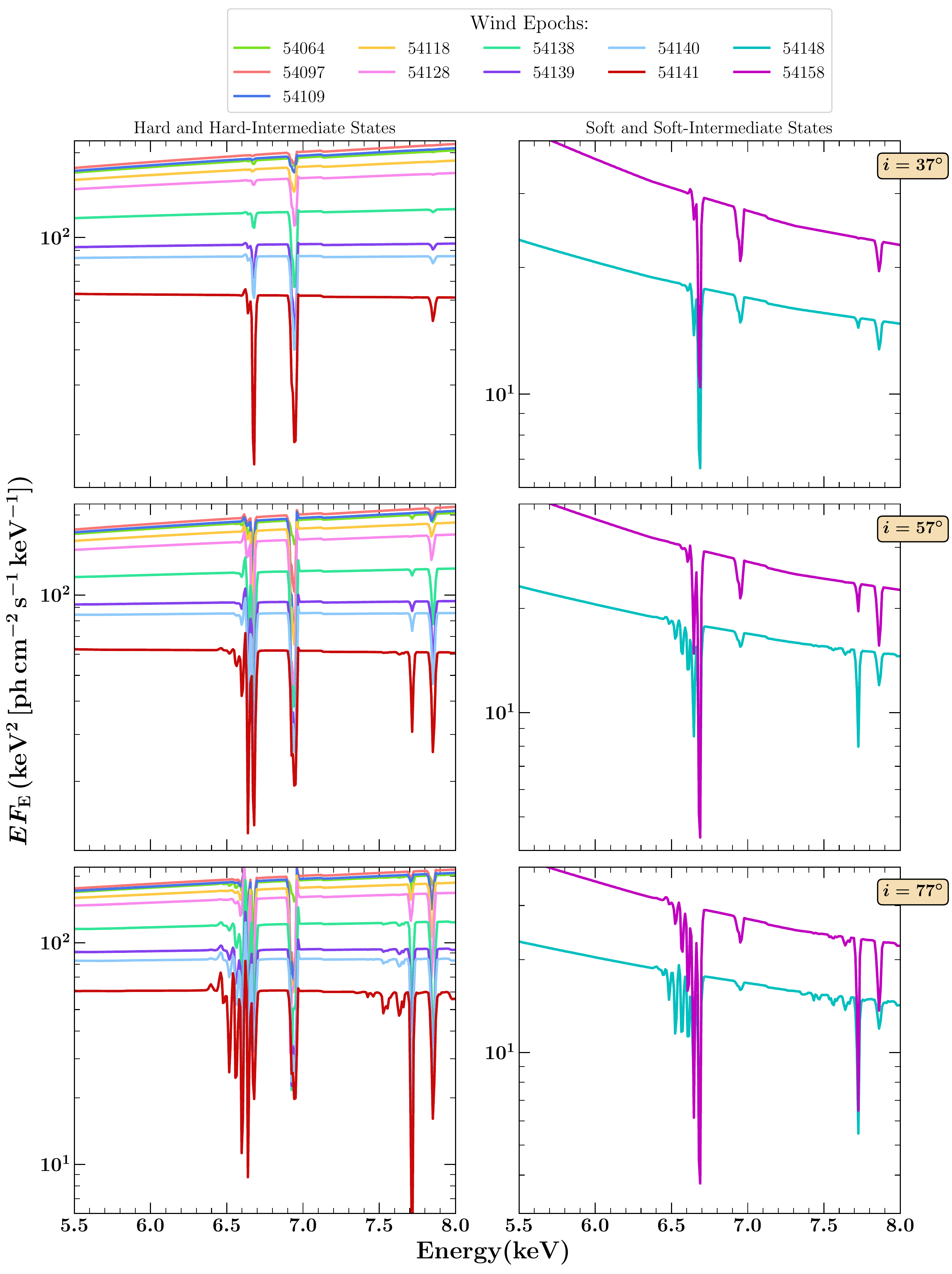}
\caption{Simulated X-ray spectra of the thermal wind via {\sc xstar} during the 2006--2007 outburst cycle of GX339$-$4. Caption the same as Figure \ref{fig:xstar_mods1}.}
  \label{fig:xstar_mods4}%
\end{figure}


\bsp	
\label{lastpage}
\end{document}